\pdfoutput=1
\documentclass[a4paper,11pt]{article}
\usepackage{jheppub}
\usepackage{afterpage}
\usepackage{subcaption}
\usepackage{lineno}
\usepackage{placeins}

\usepackage{graphicx}

\usepackage{ifthen}

\usepackage{braket}

\usepackage{verbatim}

\graphicspath{{figures/}}

\usepackage[normalem]{ulem}
\definecolor{DarkRed}{rgb}{0.5,0.0,0.0}
\definecolor{DarkGreen}{rgb}{0.0,0.5,0.0}
\definecolor{DarkBlue}{rgb}{0.0,0.0,0.5}
\definecolor{Magenta}{rgb}{1.0,0.0,1.0}
\definecolor{DarkMagenta}{rgb}{0.5,0.0,0.5}
\definecolor{Orange}{rgb}{1.0,0.5,0.0}
\definecolor{DarkOrange}{rgb}{0.8,0.3,0.0}
\definecolor{DarkCyan}{cmyk}{1.0,0.0,0.0,0.5}
\definecolor{Brown}{cmyk}{0.0,0.8,1,0.6}

\usepackage{soul}
\newcommand{\matr}[1]{\mathbf{#1}} 

\usepackage{listings}

\definecolor{lstbkgdcolor}{gray}{0.85}

\lstdefinestyle{Pstyle}{language=sh,
        xleftmargin=1.5\parindent,xrightmargin=1.5\parindent,
        columns=fixed,basicstyle=\ttfamily,basewidth=0.5em,
        frame=single,
        backgroundcolor=\color{lstbkgdcolor},
        gobble=1
        }

\lstdefinestyle{Fstyle}{language=[95]Fortran,
        xleftmargin=1.5\parindent,xrightmargin=1.5\parindent,
        columns=fixed,basicstyle=\ttfamily,basewidth=0.5em,
        frame=single,
        backgroundcolor=\color{lstbkgdcolor},
        gobble=1
        }

\lstdefinestyle{Cstyle}{language=C++,
        xleftmargin=1.5\parindent,xrightmargin=1.5\parindent,
        columns=fixed,basicstyle=\ttfamily,basewidth=0.5em,
        frame=single,
        backgroundcolor=\color{lstbkgdcolor},
        gobble=1
        }

\lstnewenvironment{snippet}{\pagebreak[2]}{}
\lstnewenvironment{Psnippet}{\pagebreak[2]\lstset{style=Pstyle}}{\pagebreak[2]}
\lstnewenvironment{Fsnippet}{\pagebreak[2]\lstset{style=Fstyle}}{\pagebreak[2]}
\lstnewenvironment{Csnippet}{\pagebreak[2]\lstset{style=Cstyle}}{\pagebreak[2]}

\newcommand{\refapp}[2][sec:]{the appendix}
\newcommand{\Refapp}[2][sec:]{The appendix}
\newcommand{\ifmulticol}[2]{%
  \ifthenelse{\lengthtest{1.9\columnwidth<\textwidth}}{#1}{#2}%
}

\newcommand{\preprintnumber}[1]{\gdef\@preprintnumber{\begin{flushright}{#1}\end{flushright}}}

\newcommand{\degree}[1]{\ensuremath{k_{\nu}^{*,\text{#1}}}} % degree
\newcommand{\ave}[1]{\ensuremath{k_{nn,\nu}^{*,\text{#1}}}} % ave degree
\newcommand{\EC}[1]{\ensuremath{CC_{EC,\nu}^{*,\text{#1}}}} %  exponential closeness
\newcommand{\HC}[1]{\ensuremath{CC_{HC,\nu}^{*,\text{#1}}}} %  harmonic closeness
\newcommand{\LC}[1]{\ensuremath{C_{\nu}^{*,\text{#1}}}} %  local clustering
\newcommand{\SC}[1]{\ensuremath{C_{s,\nu}^{*,\text{#1}}}} %  soffer clustering

\newcommand{\met}{\ensuremath{E_{\text{T}}^{\text{miss}}}}
\newcommand{\mtlmin}{\ensuremath{m_{\text{T}}^{l, \text{min}}}}
\newcommand{\ptZ}{\ensuremath{p_{\text{T}}(Z)}}
\newcommand{\dPhiZlZl}{\ensuremath{\Delta\Phi(l_{Z}^{+}, l_{Z}^{-})}}

\newcommand{\HT}{\ensuremath{H_{\text{T}}}}
\newcommand{\pt}{\ensuremath{p_{\text{T}}}}
\begin{document}
%%%%%%%%%%%%%%%%%%%%%%%%%%%%%%%%%%%%%%%%%%%%%%%%%%%%%%%%%%%%%%%%%%%%%%%%

%%%%%%%%%%%%%%%%%%%%%%%%%%%%%%%%%%%%%%%%%%%%%%%%%%%%%%%%%%%%%%%%%%%%%%%%
%######################################################################%
%#                       TITLE/ABSTRACT                               #%
%######################################################################%
%%%%%%%%%%%%%%%%%%%%%%%%%%%%%%%%%%%%%%%%%%%%%%%%%%%%%%%%%%%%%%%%%%%%%%%%

%% Article title
%\preprintnumber{ADP-20-1/T1111}

\title{Does SUSY have friends? A new approach for LHC event analysis}

%% JHEP author/abstract ================================================

%% Authors
%% Give affiliation marks to be defined below
\author[a]{Anna Mullin}
\author[a]{Stuart Nicholls}
\author[b]{Holly Pacey}
\author[b]{Michael Parker}
\author[a]{Martin White}
\author[b]{Sarah Williams}

\affiliation[a]{
  ARC Center of Excellence for Particle Physics at the Terascale \& CSSM,
  Department of Physics,
  University of Adelaide}

\affiliation[b]{
  Cavendish Laboratory. Madingley Road. Cambridge CB3 0HE, UK
}

%% Email addresses (in same order as authors)
\emailAdd{martin.white@adelaide.edu.au}
\preprint{ADP-20-1/T1111}
%% Abstract
\abstract{ %%-----------------------------------------------
We present a novel technique for the analysis of proton-proton collision events from the ATLAS and CMS experiments at the Large Hadron Collider. For a given final state and choice of kinematic variables, we build a graph network in which the individual events appear as weighted nodes, with edges between events defined by their distance in kinematic space. We then show that it is possible to calculate local metrics of the network that serve as event-by-event variables for separating signal and background processes, and we evaluate these for a number of different networks that are derived from different distance metrics. Using a supersymmetric electroweakino and stop production as examples, we construct prototype analyses that take account of the fact that the number of simulated Monte Carlo events used in an LHC analysis may differ from the number of events expected in the LHC dataset, allowing an accurate background estimate for a particle search at the LHC to be derived. For the electroweakino example, we show that the use of network variables outperforms both cut-and-count analyses that use the original variables and a boosted decision tree trained on the original variables. The stop example, deliberately chosen to be difficult to exclude due its kinematic similarity with the top background, demonstrates that network variables are not automatically sensitive to BSM physics. Nevertheless, we identify local network metrics that show promise if their robustness under certain assumptions of node-weighted networks can be confirmed.

%We show that selections placed on the network variables offer significantly greater discrimination between signal and background processes than selections on the original kinematic variables used in the definition of the network.
} %%--------------------------------------------------------

%% List of keywords

%% Arxiv number
%\arxivnumber{}
%% Generate title and abstract
\maketitle
\flushbottom

%% RevTex author/abstract ==============================================
\begin{comment} %%>>>>>>>>>>>>>>>>>>>>>>>>>>>>>>

%% Authors
%% email and affilitiations immediately after each author
\author{Martin White}
\email[]{martin.white@adelaide.edu.au}
\affiliation{
  ARC Center of Excellence for Particle Physics at the Terascale \& CSSM,
  Department of Physics,
  University of Adelaide}

%% Date
\date{\today}
 
%% insert suggested PACS numbers in braces on next line
%\pacs{}

%% insert suggested keywords - APS authors don't need to do this
%\keywords{}

%% Use the \preprint command to place your local institutional report
%% number in the upper righthand corner of the title page in preprint
%% mode.  Multiple \preprint commands are allowed.
%% Use the 'preprintnumbers' class option to override journal defaults
%% to display numbers if necessary
%\preprint{ADP-20-1/T1111}
%\preprintnumber{ADP-20-1/T1111}
%\preprint{Nordita-2014-ZZ}
%\preprint{XXXX-YY-ZZ}

%% \maketitle must follow title, authors, abstract, \pacs, and \keywords
%% when using RevTex class
\maketitle

\end{comment} %%<<<<<<<<<<<<<<<<<<<<<<<<<<<<<<<<

%%%%%%%%%%%%%%%%%%%%%%%%%%%%%%%%%%%%%%%%%%%%%%%%%%%%%%%%%%%%%%%%%%%%%%%%
%######################################################################%
%#                           BODY                                     #%
%######################################################################%
%%%%%%%%%%%%%%%%%%%%%%%%%%%%%%%%%%%%%%%%%%%%%%%%%%%%%%%%%%%%%%%%%%%%%%%%

%%%%%%%%%%%%%%%%%%%%%%%%%%%%%%%%%%%%%%%%%%%%%%%%%%%%%%%%%%%%%%%%%%%%%%%%
% INTRO ================================================================
\section{Introduction}
\label{sec:Intro}
The search for physics beyond the Standard Model (BSM) at the Large Hadron Collider (LHC) has thus far not produced any significant evidence for new phenomena, despite many analyses targeting the new particles that arise in a variety of Standard Model extensions. What this means precisely for the landscape of BSM physics models is unclear, since null results are difficult to interpret even within one BSM theory due to the large parameter space and the complexity of the particle spectra predictions. 

In fact, there are specific cases where the LHC results are known to provide \emph{no} constraint in general. For example, a recent global fit of the electroweakino sector of the Minimal Supersymmetric Standard Model (MSSM)~\cite{Athron:2018vxy} demonstrated that there is no significant general exclusion on any range of electroweakino masses. The fact that the weakly-produced supersymmetric signal is very low rate compared to the dominant Standard Model background processes means that searches need to be very heavily optimised for specific scenarios in order to provide any sensitivity for discovery. This reduces sensitivity to a large range of models that do not resemble the simplified models used for optimisation~\cite{Alwall:2008ag}. Similar arguments should apply to other sectors of supersymmetry (SUSY), and previous global fits have indeed found ample parameter space for light coloured sparticles once their decays are allowed to be more complex than those encountered in simplified models~\cite{Athron:2017yua,Athron:2017ard,Athron:2017qdc,Bagnaschi:2017tru,Costa:2017gup}.

There is clearly a strong motivation to revisit particle search techniques, and find new ways of extracting small signals from the LHC data.  All LHC analyses start from a knowledge of the reconstructed objects in each event, and their reconstructed energies and momenta. These are the \emph{attributes} of the event, and one typically searches for functions of the four-momenta of the final state particles that, within a given final state, provide effective discrimination between the signal being searched for, and the dominant SM background processes. One can then place a series of selections on these kinematic variables (or perform a machine-learning based classification) in order to find regions of the data that should contain a significant excess of signal events.

In this paper, we propose a novel approach for analysing LHC events that examines the connections between events to define new event-by-event attributes which can be analysed in the usual way. Given a set of kinematic variables, the LHC collision events form a topological structure in kinematic space, and one should expect there to be significant differences between the topological structures predicted for SM events, and those predicted for signal events. Motivated by studies of galaxy topology~\cite{Hong:2016wwh}, we build a series of network graphs from simulated LHC events, each of which uses a particular distance metric to define ``friendship'' between events based on their proximity in a chosen space of kinematic variables. For example, considering only the missing energy values reconstructed for each event, the SM events will cluster in a group at lower missing energy, each having many ``friends'' close by, while SUSY events can be expected to be few and far from the main part of the distribution, giving a small number of ``friends'' when viewed as part of a network. In defining friendship in a larger space of kinematic variables, we perform an appropriate scaling of the variables and use them to calculate a distance metric. This is then used to define connections between events that are closer than some distance $l$, which is a free parameter. A variety of local network metrics can then be calculated for each network, which serve to define new event-by-event attributes that can be used to discriminate rare signals from their dominant SM backgrounds. The analysis is complicated by the need to use local metrics that are invariant under the reweighting of the network nodes (to cope with arbitrary integrated luminosities of simulated MC events with non-trivial individual weights), and we provide a detailed solution that is applied to two supersymmetric examples based on stop quark or electroweakino production.

We note that graph networks have recently been used for a variety of applications in particle physics (usually in the context of deep learning studies), including jet tagging~\cite{Moreno:2019neq,Moreno:2019bmu,Qu:2019gqs,Henrion2017NeuralMP}, modelling kinematics within an event for classification~\cite{Abdughani:2018wrw,Choma:2018zbe}, reconstructing tracks in silicon detectors~\cite{Farrell:2018cjr}, pile-up subtraction at the LHC~\cite{Martinez:2018fwc}, investigating multiparticle correlators~\cite{Komiske:2019asc}, and particle reconstruction in calorimeters~\cite{Qasim:2019otl}. A recent work that investigates the relationship between events is the exploration of the earth-moving distance metric presented in Refs.~\cite{Komiske:2019fks,Komiske:2019jim}, although this did not involve building a network based on the metric. Our work is distinguished by the fact that it builds a graph network \emph{across} proton-proton collision events, and we investigate a number of different distance metrics, and local network metrics, for separating events.

%and show that even the simplest network metric can provide an excellent
%% excellent needs to be checked when we have results!  MAP %%
%discrimination between SM and signal events. We demonstrate that such a technique can be used for conventional optimisation with a known signal model, or can be used in a model-independent fashion with no assumption on the signal. In this way, it interpolates between conventional ``supervised'' search techniques, and unsupervised techniques that are currently a topic of active research. The decision to view events in terms of their connections rather than their attributes comes with many choices, such as the initial basis of kinematic variables, the definition of the connection between events, and the choice of network metrics that is used to provide discrimination between the signal and background processes. 

This paper is structured as follows. In Section~\ref{sec:network}, we provide a brief review of the relevant network analysis concepts, and present our method for defining connections between LHC events. We develop our electroweakino case study in Section~\ref{sec:ew}, and our stop quark example in Section~\ref{sec:stop}. A discussion of various aspects of the future applicability of our technique is contained in Section~\ref{sec:discussion}.  Finally, we present our conclusions in Section~\ref{sec:conclusions}.

\section{Network analysis}
\label{sec:network}
\subsection{Overview}
A \emph{network} is a mathematical data structure that is comprised of nodes connected by either directed or undirected edges. In the following, we will consider finite, undirected graph networks denoted by $G=(\mathcal{N},\mathcal{E})$, where $\mathcal{N}$ is the set of nodes, and $\mathcal{E}\subseteq \left\{ \{i,j\} : i \ne j \in \mathcal{N}\right\}$ is the edge list. Graphs can be described in a number of ways, and for large graphs a particularly convenient form is the \emph{adjacency matrix} which has the list of nodes as the rows and columns. Connected nodes have a 1 in the relevant adjacency matrix entry, whilst disconnected nodes have a 0. We may write this as $\matr{A}=(a_{ij})_{i,j \in\mathcal{N}}$, where $a_{ij}\in \{0,1\}$ and $a_{ij}=1$ iff $\{i,j\}\in\mathcal{E}$, and we note that the adjacency matrix will be symmetric in our case. The \emph{neighbours} of a node $\nu \in \mathcal{N}$ are those that are linked to the node by an edge, defined via:

\begin{equation}
\mathcal{N}_\nu = \left\{i\in\mathcal{N}:a_{i\nu}=1\right\} 
\end{equation}
These are also referred to as the members of $\nu$'s \emph{punctured neighbourhood}. We can define an \emph{extended adjacency matrix}  $\matr{A^+}=(a_{ij}^+)_{i,j\in\mathcal{N}}=\matr{A}+\matr{I}$ with:

\begin{equation}
a_{ij}^+=a_{ij}+\delta_{ij}
\end{equation}
where $\matr{I}$ is the identity matrix, and $\delta_{ij}$ is the Kronecker delta symbol. This then allows us to define the \emph{unpunctured neighbourhood} of $\nu$ as the set of nodes which includes both the neighbours of $\nu$, and $\nu$ itself:

\begin{equation}
\mathcal{N}^+=\left\{i\in\mathcal{N}:a_{i\nu}^+=1\right\}=\mathcal{N}_\nu\cup \left\{ \nu \right\}
\end{equation}

In our analysis, we will define the adjacency matrix for LHC events by defining a distance between events in the space of kinematic variables that are measured for each event. Assuming some distance $d_{ij}$ between the nodes in this space of variables, one can then define the adjacency matrix via:

\begin{equation}
\label{eq:adj}
a_{ij} =  \begin{cases}
    1,& \text{if } d_{ij} \le l,\\
    0,              & \text{otherwise},
\end{cases}
\end{equation}
where $l$ is a free parameter known as the \emph{linking length}. This prescription clearly leaves many choices open for how to proceed, including the choice of kinematic variables, the choice of distance metric, and the choice of linking length for a given analysis. Both the choice of distance metric and the choice of kinematic variables will change the topological structure mapped by the network, and sensible choices might lead to greater differences between the behaviour of Standard Model processes and new physics processes within the network. We show below that it is advantageous to construct networks with a variety of different distance metrics from the LHC data, and to combine variables derived from more than one network. We will also suggest some guiding principles for the selection of optimal linking lengths.

For two vectors $u$ and $v$ in the space of $n$ kinematic variables for an analysis, the distance metrics that we consider in this work are:

\begin{itemize}
\item {\bf The Euclidean distance: }$d_{\text{euc}}=\sqrt{\sum_{i=1}^n(u_i-v_i)^2}$.
\item {\bf The Chebyshev distance: }$d_{\text{cheb}}=\max|u_i-v_i|$, i.e. the maximum of the difference between similar kinematic variables for the two chosen points.  
\item {\bf The Bray-Curtis distance: }$d_{\text{bray}}=\sum_{i=1}^n\frac{|u_i-v_i|}{\sum_{j=1}^n|u_j|+\sum_{j=1}^n|v_j|}$.
\item {\bf The cityblock distance: }Also known as the Manhattan distance, the cityblock distance is given by $d_{\text{city}}=\sum_{i=1}^{n}|u_i-v_i|$. 
\item {\bf The cosine distance: }$d_{\text{cos}}=1-\frac{u\cdot v}{\sqrt{u\cdot u}\sqrt{v\cdot v}}$.
\item {\bf The Canberra distance: }$d_{\text{can}} = \sum_{i=1}^n \frac{|u_i-v_i|}{|u_i|+|v_i|}$.
\item {\bf The Mahalanobis distance: }$d_{\text{mah}}=\sqrt{(u-v)V^{-1}(u-v)^T}$, where $V^{-1}$ is the inverse of the sample covariance matrix (calculated over the entire dataset of events). 
\item {\bf The correlation distance: }$d_{\text{corr}}=1-\frac{(u-\bar{u})\cdot (v-\bar{v})}{\sqrt{(u-\bar{u})^2}\sqrt{(v-\bar{v})^2}}$, where $\bar{u}$ is the mean of the elements of the vector $u$.   
\end{itemize}

The Canberra distance metric was found to be ineffective, and we do not refer to it in the analyses below. We also note that many other possibilities exist in the literature~\cite{2006vii,Komiske:2019jim,Komiske:2019fks}, but we find that the list above offers sufficient performance whilst remaining relatively quick to evaluate. 

It is possible to have weights associated with the edges in the adjacency matrix that depart from one, leading to what is conventionally referred to as a \emph{weighted network}. In our example, we will instead have weights on the \emph{nodes} of the network. To see why these weights are necessary, consider the formation of a network that is obtained by taking all LHC events that pass some pre-selection (e.g. selection of a given final state with some basic kinematic requirements). Each event then becomes a node in the network, with connections to other events defined via the distance metric. In any real LHC analysis, one would want to compare the behaviour of Standard Model Monte Carlo simulations with the observed data. In the absence of a method for weighting the events, one would have to generate exactly the same number of events as one would expect to obtain in a given integrated luminosity of LHC data, for every relevant Standard Model process. This is clearly neither feasible nor desirable, and it does not permit the use of Monte Carlo generators with non-trivial weight assignments (such as those that arise from jet matching procedures). Instead, in a node-weighted network, the network can be populated with events whose weights are defined in the normal manner. This is straightforward from the perspective of network analysis, since network nodes are allowed to have any number of attributes assigned to them, but it complicates the calculation of network metrics as we shall see below. Although this analysis only considers weighted events from Monte Carlo simulation, these methods are also applicable to searches with data-driven background estimates that apply additional weights to appropriately selected data or Monte Carlo events.

\subsection{Network metrics}

Once a network has been defined, one can calculate a series of \emph{network metrics} that characterise the network topology. These include global metrics (defined for the network as a whole), and local metrics (which we assume to be defined for each node of the network). For a given selection of events, there is only one network that can be formed from all of the selected events. It is possible in principle to infer the presence of new physics by demonstrating that the global network metrics for the network of selected events depart from a well-modelled Standard Model expectation. However, we will instead focus on local metrics that will allow us to define attributes on an event-by-event basis. These can be substituted for the kinematic variables that are used in a regular LHC event analysis, in which searches for new physics are performed by placing selections on variables to define regions of the data where the background is expected to be small. 

The simplest example of a local network metric is the \emph{degree centrality} of a node, which is equal, for an unweighted network, to the number of other nodes that are connected to it. In a social media network, for example, the degree centrality of a given person would be equal to the number of their friends. 

In a weighted network, the definitions of both local and global metrics must be updated to take account of the fact that each node now represents a different number of effective nodes. Node-weighted network measures can be defined based on the concept of \emph{node-splitting invariance} as detailed in Ref.~\cite{nsi}, and in the following we perform calculations of node-splitting-invariant (n.s.i) local network metrics using a custom version of the {\tt pyunicorn} package~\cite{Donges:2015tta}. The full list of network measures that we utilise is as follows.

\begin{itemize}
\item {\bf The n.s.i degree: }For a given node $\nu$, this is the weighted version of the degree centrality, given by:
\begin{equation}
k^*_\nu=\frac{\sum_{i\in\mathcal{N}_\nu^+}w_i}{(W+1)},
\end{equation}

where $W=\sum_{i\in\mathcal{N}}w_i$ is the sum of the weights of all nodes in the network. In our analysis, $W$ is equivalent to the total number of events expected at the LHC for our assumed integrated luminosity.

\item {\bf The n.s.i average and maximum neighbours degree: }The average neighbours degree of a node $\nu$ represents the average size of the network region that an event linked to $\nu$ is linked to. The n.s.i measure of this quantity is given by:
\begin{equation}
k_{nn,\nu}^*=\frac{1}{(W+1)}\frac{\sum_{i\in\mathcal{N}_\nu^+}w_i k^*_i}{k^*_\nu}.
\end{equation}

One can also define an n.s.i  \emph{maximum} neighbors degree, as

\begin{equation}
k_{nnmax,\nu}^*=\max_{i\in\mathcal{N}_\nu^+}k^*_i.
\end{equation}

\item {\bf The n.s.i betweenness centrality: }The \emph{shortest path betweenness centrality} of a node $\nu$ gives the proportion of shortest paths between pairs of randomly chosen nodes that pass through $\nu$. If we label the random nodes by $a$ and $b$, we have that:
\begin{equation}
BC_\nu = \left\langle n_{ab}({\nu})/n_{ab}\right\rangle_{ab}\in [0,1],
\end{equation}
where $n_{ab}$ is the total number of shortest paths from $a$ to $b$, $n_{ab}(\nu)$ is the number of those paths that pass through $\nu$, and we have defined the average of a function of node pairs by $\left\langle h(i,j)\right\rangle_{ij}=\frac{1}{N^2}\sum_{i\in\mathcal{N}}\sum_{j\in\mathcal{N}} h(i,j)$. One can write a formal expression for this quantity by first noting that $n_{ab}$ can be written as a sum over the \emph{tuples} $(t_0,...,t_{d_{ab}})$, with $t_0=a$ and $t_{d_{ab}}=b$ ($d_{ab}$ is the number of links between the nodes $a$ and $b$ on the shortest path), where each tuple in the sum gives a contribution of 1 if every node $t_l$ in the tuple is linked to its successor $t_{l+1}$, or 0 if at least one node is not linked to its successor. Both of these conditions are met if one simply takes the product of elements of the adjacency matrix for each pair of nodes in the tuple, allowing us to write
\begin{equation}
n_{ab}=\sum_{(t_o,...,t_{d_{ab}})\in \mathcal{N}^{d_{ab}+1}, t_0=a,t_{d_{ab}}=b}\prod_{l=1}^{d_{ab}}a_{t_{l-1}t_l}.
\end{equation}

$n_{ab}(\nu)$ is given by a similar formula, except that, for some $m$ in $1...d_{ab}-1$, $t_m$ must equal $\nu$:

\begin{equation}
n_{ab}(\nu)=\sum_{m=1}^{d_{ab}-1}\sum_{(t_o,...,t_{d_{ab}})\in \mathcal{N}^{d_{ab}+1}, t_0=a,t_m=\nu, t_{d_{ab}}=b}\prod_{l=1}^{d_{ab}}a_{t_{l-1}t_l}.
\end{equation}

It is possible to make an n.s.i version of this quantity based on a \emph{weighted average} instead of an average (which would be consistent with the above formula in the limit that all node weights are 1), but the {\tt pyunicorn} package instead uses a weighted sum, giving the n.s.i betweenness centrality as:

\begin{equation}
BC^*_\nu = \left\langle n^*_{ab}(\nu)/n^*_{ab}\right\rangle^{wsum}_{ab}\in [0,W^2/w_\nu],
\end{equation}

where we have defined the \emph{weighted sum} of a function of pairs of nodes $\left\langle h(i,j)\right\rangle_{ij}^{wsum}=\sum_{i\in\mathcal{N}}\sum_{j\in\mathcal{N}} w_i h(i,j) w_j$,  and $n^*_{ab}(\nu)$ and $n^*_{ab}$ are defined below. The n.s.i betweenness centrality values obtained for our examples below do not come close to saturating the maximum value of $W^2/w_\nu$. 

Formulae for  $n^*_{ab}(\nu)$ and $n^*_{ab}$ can be derived as follows. If a node $s$ is hypothetically split into two nodes $s'+s''$, any shortest path through $s$ becomes a pair of shortest paths (one of which passes through $s'$, and the other of which passes through $s''$). In addition, a shortest path from $s''$ to some $b\ne s'$ will never meet $s'$. Thus, the betweenness centrality can be made invariant under node splitting by making each path's contribution proportional to the product of the weights of the inner nodes, but with the condition that we skip the weight $w_\nu$ in this product when calculating $n_{ab}(\nu)$. Formally, we can write a modified $n^*_{ab}$ as:

\begin{equation}
n^*_{ab}=\sum_{m=1}^{d_{ab}-1}a_{t_0t_1}\prod_{l=2}^{d_{ab}}(w_{t_{l-1}}a_{t_{l-1}t_l}),
\end{equation}

and a modified $n^*_{ab}(\nu)$ as:

\begin{equation}
n_{ab}^*(\nu)=\frac{1}{w_\nu}\sum_{m=1}^{d_{ab}-1}\sum_{(t_o,...,t_{d_{ab}})\in \mathcal{N}^{d_{ab}+1}, t_0=a,t_m=\nu, t_{d_{ab}}=b}\left(a_{t_0 t_1}\prod_{l=2}^{d_{ab}}(w_{t_{l-1}}a_{t_{l-1}t_l})\right).
\end{equation}

A geometric interpretation of the n.s.i betweenness can be obtained by considering the set of nodes of our node-weighted network $\{\mathcal{N}\}$ as a sample from a population of points $\{\mathcal{N}_0\}$. Each node $\nu$ in the network then represents some small cell $\mathcal{R}_\nu$ of points in the geometric vicinity of $\nu$ in $\{\mathcal{N}_0\}$. The n.s.i betweenness can be interpreted as an estimate of the probability density that a randomly-chosen shortest path between two randomly-chosen points in the population network passes through a specific randomly-chosen point in $\mathcal{R}_\nu$. This is ultimately a measure of the importance of the node $\nu$ in the network. 

\item {\bf The n.s.i closeness centrality: }Another measure of node importance is the \emph{closeness centrality}, which is defined for a node $\nu$ by $CC_\nu=1/\left\langle d_{\nu i}\right\rangle_i$, where $d_{\nu i}$ is the number of links on a shortest path from $\nu$ to $i$, or, if there is no path, $\infty$, and we have defined the same average over nodes that was used previously. A larger value of $CC_\nu$ for a node indicates a smaller average number of links to all other nodes in the network. The n.s.i version of this metric is given by:
\begin{equation}
CC^*_\nu = \frac{W}{\sum_{i\in\mathcal{N}}w_id^*_{\nu i}}\in[0,1],
\end{equation}
where $d^*_{\nu i}$ is an \emph{n.s.i distance function} given by:
\begin{equation}
d^*_{\nu\nu}=1~~\text{and}~~d^*_{\nu i}=d_{\nu i}~~\text{for}~~i\ne\nu.
\end{equation}

The n.s.i distance function is naively a little odd, and can be justified as follows. A weighted version of $CC_\nu$ should give us the inverse average distance of $\nu$ from other weight units or points rather than from other nodes. But for this to become n.s.i, one has to interpret each node to have a unit (instead of zero) distance to itself since, after an imagined split of a node $s\rightarrow s' s''$, the two parts $s'$ and $s''$ of $s$ have unit not zero distance. 

\item {\bf The n.s.i exponential closeness centrality: }A limitation of the closeness centrality is that it receives very low values for nodes which are very close to most of the other nodes, but very far away from at least one of them. To prevent outlying nodes from skewing the closeness calculation for ``typical'' nodes, one can use the \emph{exponential closeness centrality}, defined as $CC_{EC,\nu} = \left\langle 2^{-d_{\nu i}}\right\rangle_i$. The n.s.i measure is given by:
\begin{equation}
CC^*_{EC,\nu}= \left\langle 2^{-d^*_{\nu i}}\right\rangle_i^w\in[0,1].
\end{equation}

where we have used the notation $\left\langle g(\nu)\right\rangle_\nu^w=\frac{1}{W}\sum_{\nu\in \mathcal{N}}w_\nu g(\nu)$. 

\item {\bf The n.s.i harmonic closeness centrality: }The harmonic closeness centrality reverses the sum and reciprocal operations in the definition of closeness centrality, such that $1/d_{\nu i}$ contributes zero to the sum if there is no path from $\nu$ to $\i$. The n.s.i harmonic close centrality is given by:
\begin{equation}
CC_{HC,\nu}^* = \left\langle 1/d^*_{\nu i}\right\rangle^w_i\in[0,1].
\end{equation}
\item {\bf The n.s.i local clustering coefficient: }The local clustering co-efficient of a node $\nu$ is the probability that two nodes drawn at random  from those linked to $\nu$ are linked with each other. It is given by:

\begin{equation}
C_\nu = \frac{\sum_{i\in\mathcal{N}_\nu}\sum_{j\in\mathcal{N}_\nu} a_{ij}}{k_\nu(k_\nu-1)} = \frac{N^2\left\langle a_{\nu i} a_{ij} a_{j\nu}\right\rangle_{ij}}{k_\nu(k_\nu-1)}.
\end{equation}

The n.s.i version is given by

\begin{equation}
C_\nu^*=\frac{W^2\left\langle a_{\nu i}^+ a_{ij}^+ a_{j\nu}^+\right\rangle_{ij}^w}{k_\nu^{*2}}\in \left[ \frac{w_\nu (2k^*_\nu - w_\nu)}{k_\nu^{*2}},1 \right] \subseteq [0,1].
\end{equation}
\item {\bf The n.s.i local Soffer clustering coefficient: }An alternative form of the clustering coefficient proposed by Soffer and V\'azquez~\cite{PhysRevE.71.057101} includes a correction that reduces the impact of degree correlations:

\begin{equation}
C_{s,\nu} = \frac{N^2\left\langle a_{\nu i} a_{ij} a_{j\nu}\right\rangle_{ij}}{\sum_{i\in\mathcal{N}_\nu} (\min(k_i,k_\nu)-1)}\in[C_\nu,1] .
\end{equation}

The n.s.i version of this is given by:

\begin{equation}
C^*_{s,\nu}=\frac{W^2 \left\langle a_{\nu i}^+ a_{ij}^+ a_{j\nu}^+\right\rangle_{ij}^w}{{\sum_{i\in\mathcal{N}_\nu^+}w_i\min(k_i^*,k_\nu^*)}\in[C^*_\nu,1]}.
\end{equation}

\end{itemize}
For the case studies presented in this paper, networks of events are built for different distance metrics after specified \textit{pre-selection} criteria, which are detailed further in Sections~\ref{sec:ew} and \ref{sec:stop}. For a given choice of distance metric, events are linked in the corresponding network if their distance in the original space of kinematic variables is less than a chosen linking length $l$. The pre-selection criteria applied before the networks are constructed involve the typical combinations of particle multiplicity selections and loose kinematic selections that are encountered in SUSY searches at the LHC. Local network metrics are then used along with further requirements on kinematic variables to define search regions analogous to those used in existing LHC searches. Note that, unlike traditional analysis variables which only depend on the kinematic properties of that event, the presence of a signal in the LHC data would change the local network metrics for the background events in addition to adding signal events to the various distributions. By comparing event yields obtained from a \textit{background-only network} (which mimics the data that would be expected in the absence of a signal) to those from a \textit{signal-plus-background network} (which mimics the data that would be expected in the presence of a signal), estimates of the exclusion sensitivity are calculated for the network driven analysis. These are compared to sensitivity estimates for search regions defined only using conventional kinematic variables. 

The exclusion sensitivities are estimated using the {\tt RooStats} framework~\cite{moneta2010roostats}, which provides a calculation for the \emph{binomial significance} ($Z_{\text{bi}}$) associated with a number counting experiment that tests between signal-plus-background and background-only hypotheses, in the case that the background estimate has an uncertainty derived from an auxillary measurement~\cite{CRANMER_2006,Cousins_2008,linnemann2003measures}. Our comparative results using this significance measure are encouraging and indicate that the network-based methods can offer significant improvements in discovery potential. Since we do not have access to a full background analysis, complete simulation of the detector systems, or the systematic uncertainties available to the experimental collaborations, we do not show a full statistical analysis of the expected exclusion and discovery reach, which is left for future studies.

\section{Simulated signal and background event samples}
\label{sec:mc}
\begin{figure}[h!]
\centering
\includegraphics[scale=0.8]{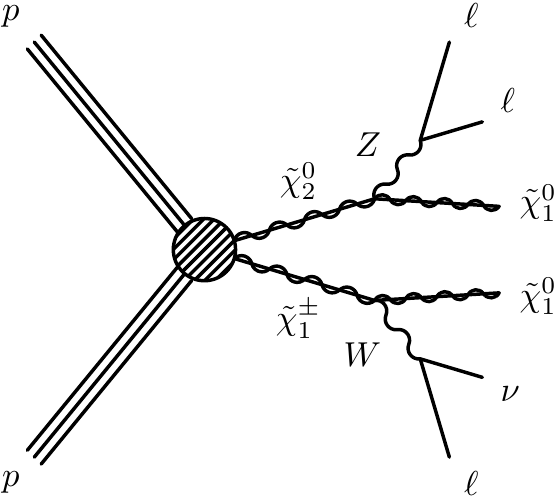} \hspace{10mm}
\includegraphics[scale=0.3]{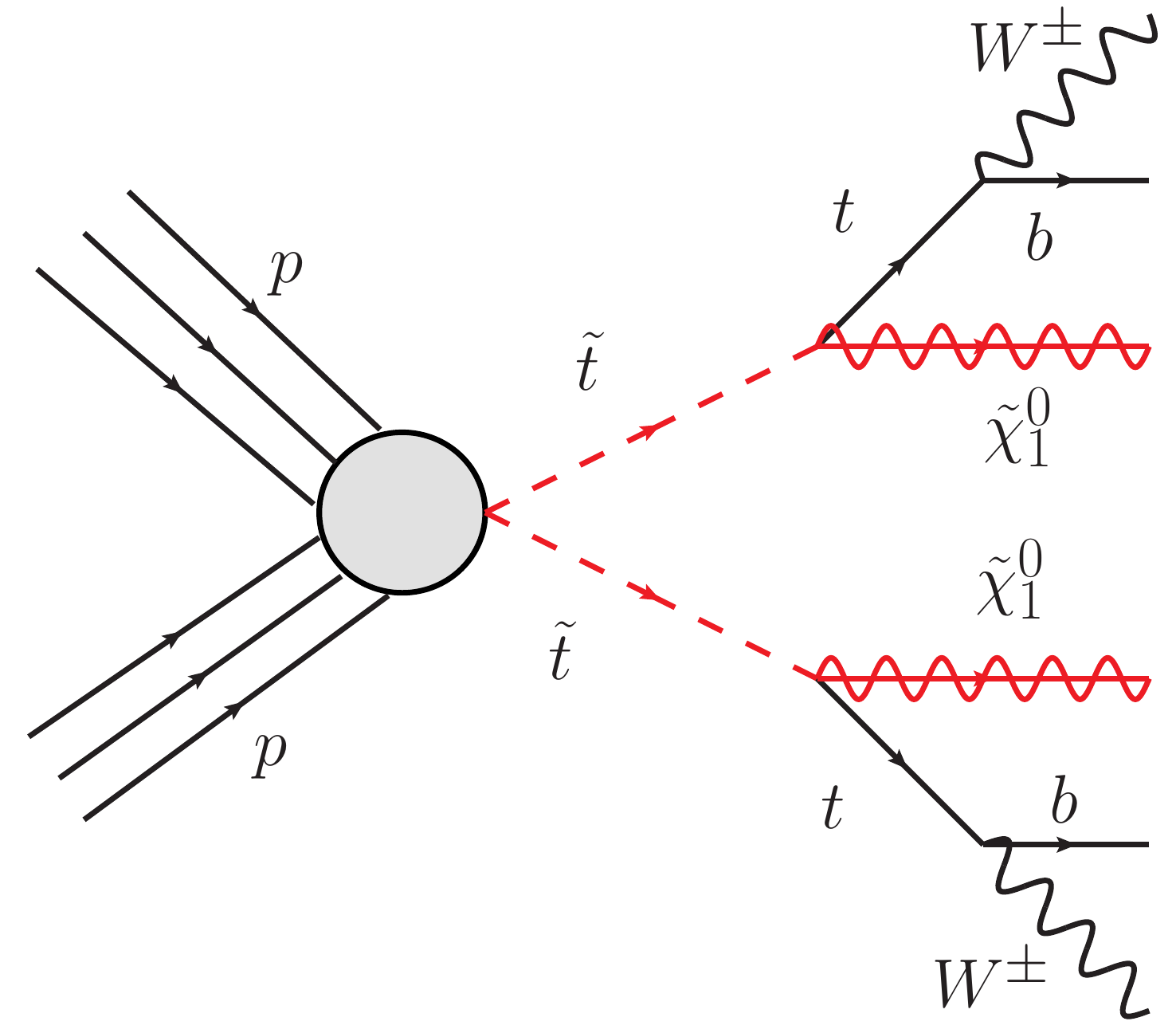}
\caption{\label{fig:feyndecay} Feynman diagrams of the simplified supersymmetric models considered in (left) the prototype electroweakino analysis and (right) the prototype stop pair analysis.}
\end{figure}

To demonstrate our method we consider two case studies: supersymmetric electroweakino and stop pair production, which will be discussed further in sections~\ref{sec:ew} and~\ref{sec:stop} respectively.  LHC searches are typically optimised on simplified models, and diagrams for the decays considered in this paper are shown in Figure~\ref{fig:feyndecay}. 

\FloatBarrier

The electroweak neutralinos ($\tilde{\chi}^{0}_{i=1,2,3,4}$) and charginos ($\tilde{\chi}^{\pm}_{i=1,2}$) are the mass eigenstates corresponding to linear combinations of the superpartners of the unbroken electroweak gauge bosons and the supersymmetric Higgs particles. The mixing of the \textit{bino}, \textit{wino}, and \textit{Higgsino} states into the electroweakino states can greatly impact the resulting phenomenology meaning that LHC searches are forced to make assumptions about the relative masses and mixings of these states.  In this paper, it is assumed that electroweakinos are produced via wino-like $\tilde{\chi}_1^{\pm}-\tilde{\chi}_2^0$ production, with subsequent decay to SM gauge bosons and the lightest neutralinos, which are assumed to be pure bino states. We consider a model with $m_{\tilde{\chi}_1^{\pm}}=m_{\tilde{\chi}_2^{0}}=400$~GeV, and $m_{\tilde{\chi}_1^0}=0$~GeV. Both gauge bosons are assumed to decay leptonically giving a 3-lepton final state, where the dominant background is SM diboson (WZ) production. This model was not excluded in the three-lepton search channel in the most recent 139 fb$^{-1}$ analysis conducted by the ATLAS collaboration~\cite{Aad:2019vvi}. The results in this paper assume 150 fb$^{-1}$ of integrated luminosity, which is comparable.

Our second example considers supersymmetric top quarks at the LHC, where it is assumed that $\tilde{t}_1\tilde{t}_1$ production dominates at the LHC, and the decay of the stop is presumed to occur with 100\% branching ratio to either a top quark and a lightest neutralino, $\tilde{\chi}_1^0$, or a $b$ quark and a lightest chargino. We consider a model with $m_{\tilde{t}_1}=500$~GeV, and $m_{\tilde{\chi}_1^0}=280$~GeV. This choice of benchmark SUSY model choice is motivated by the fact that the model was not excluded by the most recent 36.1 fb$^{-1}$ analysis conducted by the ATLAS collaboration~\cite{Aaboud:2017aeu}, and it can be expected to be challenging to observe due to the fact that the kinematic properties of the top quarks produced are rather similar to those expected in the Standard Model. We will develop a prototype analysis in the 1 lepton final state, in which one of the top quarks is assumed to decay hadronically, whilst the other decays leptonically, which is typically the most constraining final state. The dominant SM background in this final state is top quark production (including both top pair and single top production). In a real analysis, there would be a small contribution from events containing a $W$ boson produced in association with jets, but we neglect this for our proof of principle. We note that, in the final stages of our work, the ATLAS collaboration released a zero lepton stop pair production search utilising 139 fb$^{-1}$ of data which does exclude this benchmark model~\cite{Aad:2020sgw}. %However, our results remain valid in showing that, for a given set of kinematic variables in the 1 lepton case, one can improve the performance dramatically by using network measures. It remains to be seen how useful this would be in the 0 lepton case, and whether our 1 lepton network-based analysis would in fact outperform the 0 lepton analysis. 

For both case studies, the signal and dominant SM background (diboson WZ for the electroweakino case and top quark production for the stop case) are simulated with {\tt Pythia 8.240}~\cite{Sjostrand:2014zea} using the {\tt CTEQ6L1} PDF set~\cite{Pumplin:2002vw}, and an LHC detector simulation is performed using {\tt Delphes 3.4.1}~\cite{deFavereau:2013fsa,Selvaggi:2014mya,Mertens:2015kba}, using the default ATLAS detector card. Jets are reconstructed using the anti-k$_{\textup{T}}$ algorithm with a radius parameter R=0.4~\cite{Cacciari:2008gp} using the FastJet package~\cite{Cacciari:2011ma}. In normalising the electroweakino signal to the relevant integrated luminosity, we use the next-to-leading-order plus next-to-leading-log cross-section provided in Refs.~\cite{Fuks:2012qx,Fuks:2013vua}, whilst cross-section used for normalisation of the stop sample is the next-to-next-to-leading-order plus approximate next-to-next-to-leading log cross-section derived from Refs.~\cite{Beenakker:2016lwe,Beenakker:1997ut,Beenakker:2010nq,Beenakker:2016gmf}. For the SM backgrounds, the WZ sample uses the next-to-next-to-leading order cross-section presented in Ref.~\cite{Grazzini:2016swo} and the top sample uses the next-to-next-to-leading-order plus next-to-next-to-leading log $t\bar{t}$ cross-section derived from Refs.~\cite{Beneke:2011mq,Cacciari:2011hy,Baernreuther:2012ws,Czakon:2012zr,Czakon:2012pz,Czakon:2013goa,Czakon:2011xx}~\footnote{Note that this gives us an approximate weighting of the top background, since the subdominant single top contribution will be scaled by the same cross-section as the $t\bar{t}$ contribution.}.

To ensure that the extreme tails of kinematic distributions are adequately sampled, both case studies use ``sliced'' background event samples, and the stop cases uses ``sliced'' signal event samples. For the stop case study samples are generated in slices of $H_{T}$, whilst the electroweakino samples are sliced in $\hat{p}_{\textup{T}}$, the transverse momenta in the rest frame of the hard scattering process. For the electroweakino case all of the events passing the pre-selection defined in Section~\ref{sec:ew} were used to build the network, however for the stop case to speed up the network calculations for our proof of principle, we take a subsample of 10,000 events for both the signal and background, and adjust their weights accordingly. We also make use of inclusive MC samples (i.e. with no slicing) for testing, and for plotting the distance metrics, since the inclusive samples adequately describe the distributions of the distance metrics that one expects to see in LHC data, in the regions of interest.

We have performed a variety of checks that the network metrics we use are indeed safe under the reweighting of events, using both the inclusive samples, and the sliced events. These checks are presented for both of the electroweakino and stop examples in Appendix~\ref{app:nsi}. This study also considers the assumptions associated with using n.s.i. network metrics on theoretical grounds, i.e. that a weighted node can be taken to represent a group of identical nodes which possess full internal connectivity and identical external connectivity. In the subsequent sections we only present quantitative results for n.s.i. metrics that we argue are robust under these conditions; the n.s.i. degree and n.s.i. closeness, harmonic closeness and exponential closeness centralities. The betweenness centrality may not be robust under the external connectivity assumption and could thus be affected by node-weight-dependent biases. For this reason we only show qualitative results that demonstrate its potential. Improvements to methods for constructing node-weighted networks that could improve the accuracy of the second assumption are an interesting avenue for future study.

\section{Case study 1: The search for electroweakinos}
\label{sec:ew}
%\subsection{Model definition and simulation}

\subsection{Electroweakino analysis design}
\label{sec:ewanalysisdesign}
%{\bf Sarah/Holly: could you please tweak the description of the pre-selection as appropriate?}
As our first example of the application of network analysis techniques to an LHC search, we will consider the hunt for supersymmetric electroweakinos at the LHC. The first step in our analysis is to apply a pre-selection consistent with a 3 lepton electroweakino search. We require there to be exactly three light leptons (electrons or muons), with transverse momentum $p_{\text{T}}>25$~GeV and pseudorapidity $|\eta|<2.5$. We do not apply additional pseudo-rapidity or isolation requirements to the default electron/muon reconstruction provided by {\tt Delphes3}~\cite{deFavereau:2013fsa,Selvaggi:2014mya,Mertens:2015kba}. The default settings restrict electrons, photons and muons to $|\eta|<2.5$, while jets and missing transverse energy are restricted to $|\eta|<4.9$. We further require there to be no $b$-tagged jets and at most 1 non-$b$-tagged jet in the event with $p_{\text{T}}>25$~GeV and $|\eta|<2.4$, and for the dilepton invariant mass of an opposite-sign same-flavour pair in the event to satisfy $|m_{ll}-m_Z| < 10$~GeV. When investigating the potential improvements in sensitivity from including network metrics in the definition of search regions, we consider a series of variables that are typically used to discriminate electroweakino events from diboson events, choosing to focus on those whose distributions over all events show the greatest difference between our benchmark signal point and the diboson background. These are:
%
%After applying the pre-selection criteria, We have investigated a series of variables that are typically used to discriminate electroweakino events from diboson events, choosing to focus on those whose distributions over all events show the greatest difference between our benchmark signal point and the diboson background. These are:

\begin{itemize}
\item The missing transverse energy $E_{\text{T}}^{\text{miss}}$, which represents the momentum imbalance transverse to the beam direction and can be used to infer the presence of weakly interacting neutral particles escaping the detector.
\item $m_{\text{T}}^{l, \text{min}}$: defined as $\text{min}( m_{\text{T}}(l_{1},E_{\text{T}}^{\text{miss}}), m_{\text{T}}(l_{2},E_{\text{T}}^{\text{miss}}), m_{\text{T}}(l_{3},E_{\text{T}}^{\text{miss}}))$
\item $p_{\text{T}}(Z)$: the reconstructed transverse momentum of the $Z$ boson.
\item $\Delta\Phi(l_{Z}^{+}, l_{Z}^{-})$: the azimuthal angle between the two leptons associated with  the $Z$ boson, which are taken to be the same-flavour opposite sign pair whose di-lepton invariant mass is closest to the $Z$-boson mass in the case of any ambiguity (with the remaining lepton assigned to the $W$-boson).
\item $\Delta\Phi(Z, l_{W})$: the azimuthal angle between the reconstructed $Z$ boson and the lepton coming from the $W$ boson.
%\item sumpt: $p_{\text{T}}(l_1) + p_{\text{T}}(l_2) + p_{\text{T}}(l_3) + p_{\text{T}}(Jet_1)$.
\end{itemize}

\begin{figure}[htb!]
\begin{tabular}{cc}
\begin{subfigure}[b]{0.45\textwidth}
\includegraphics[width=\textwidth]{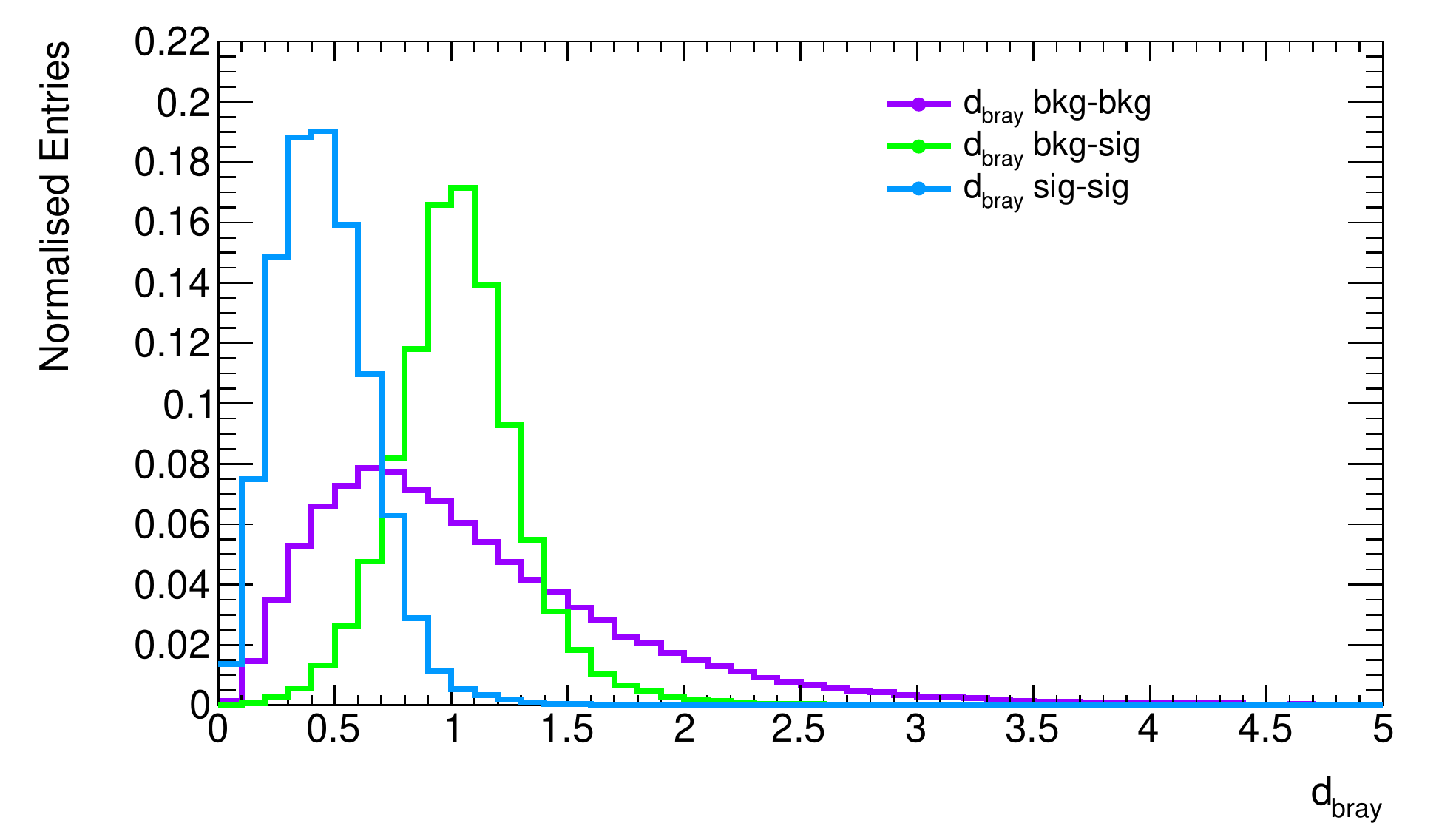}
\centering
\caption{Bray-Curtis}
\label{1}
\end{subfigure} &
\begin{subfigure}[b]{0.45\textwidth}
\includegraphics[width=\textwidth]{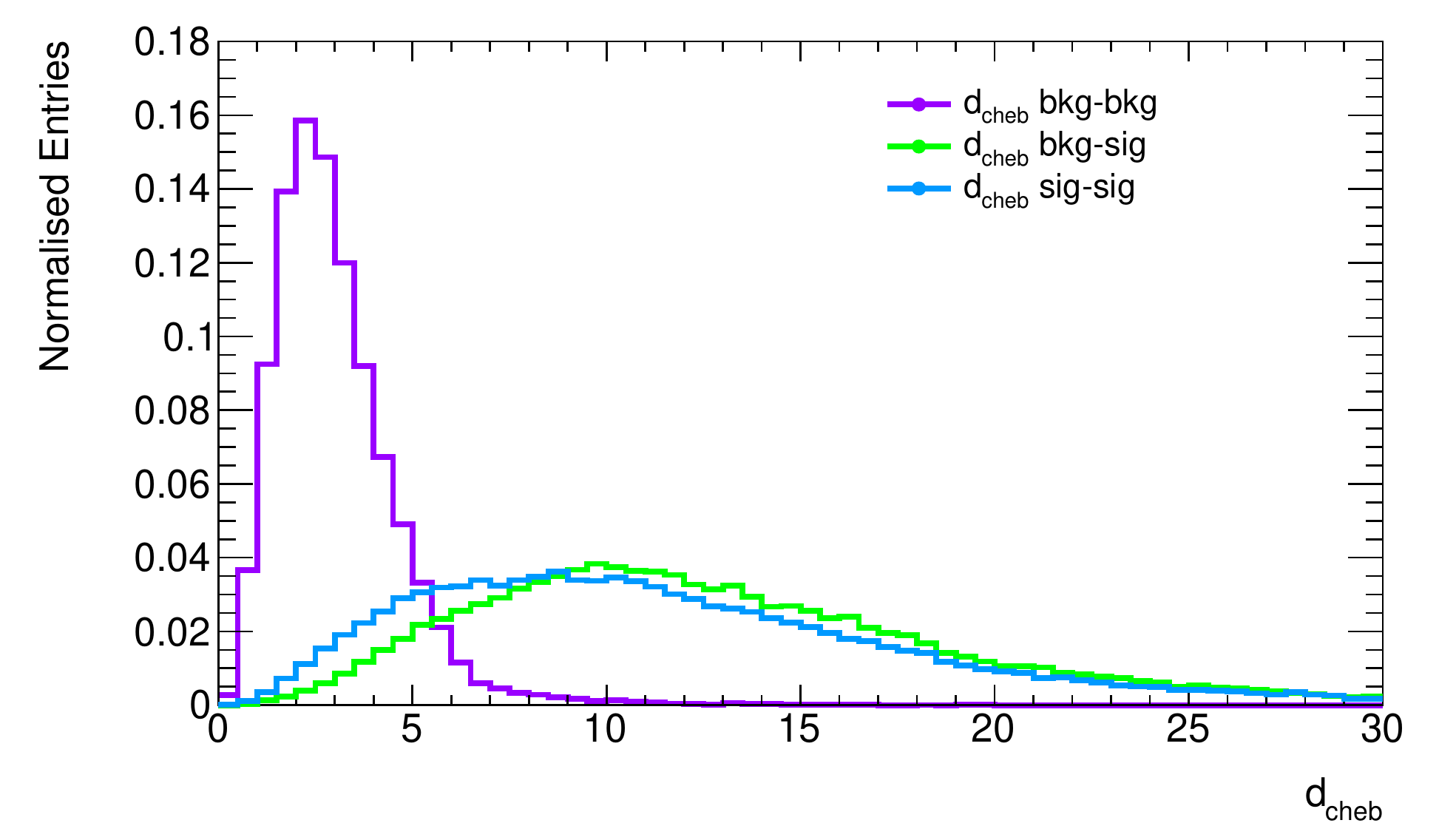}
\centering
\caption{Chebyshev}
\label{3}
\end{subfigure} \\
\begin{subfigure}[b]{0.45\textwidth}
\includegraphics[width=\textwidth]{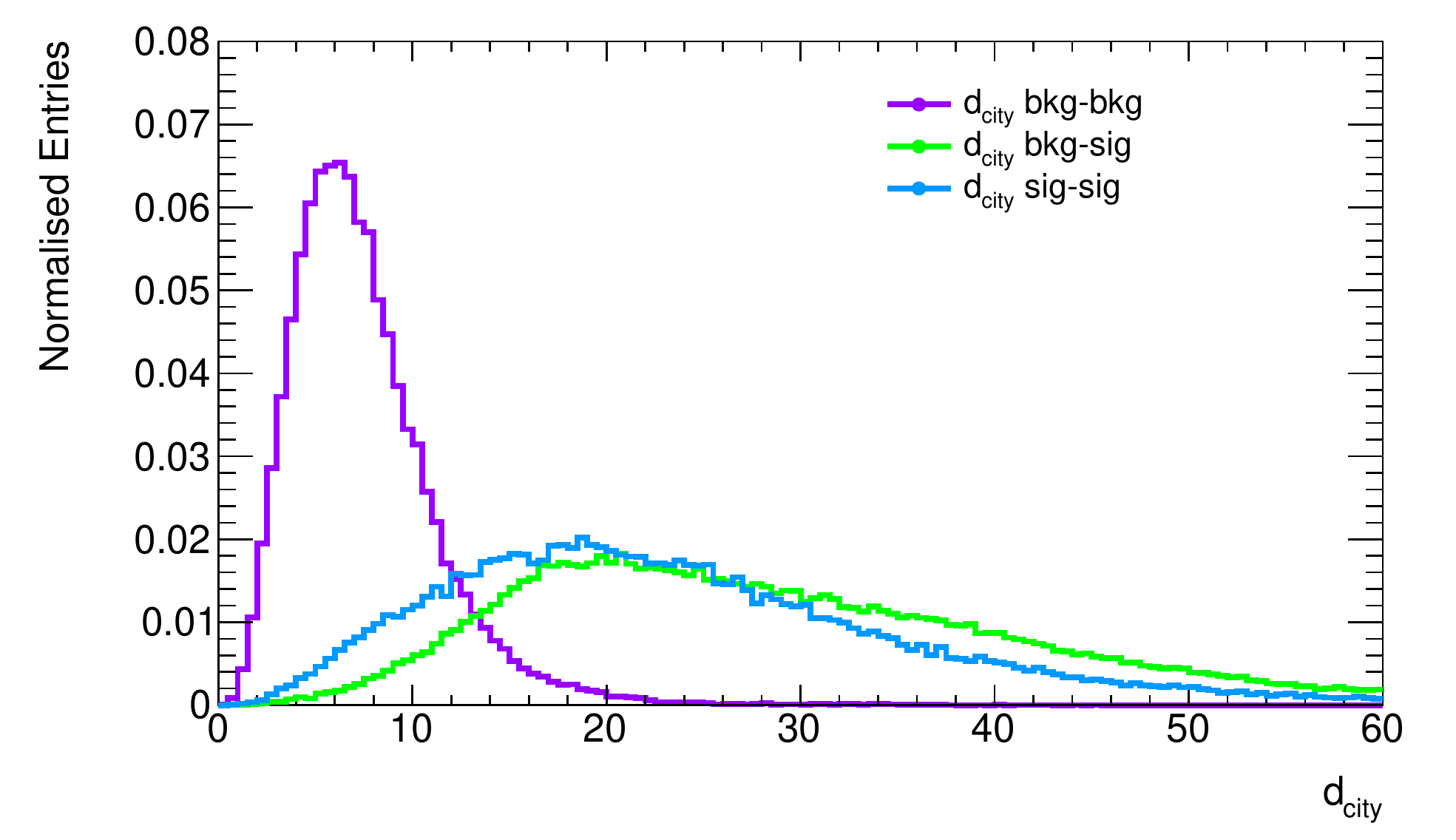}
\centering
\caption{cityblock}
\label{4}
\end{subfigure} &
\begin{subfigure}[b]{0.45\textwidth}
\includegraphics[width=\textwidth]{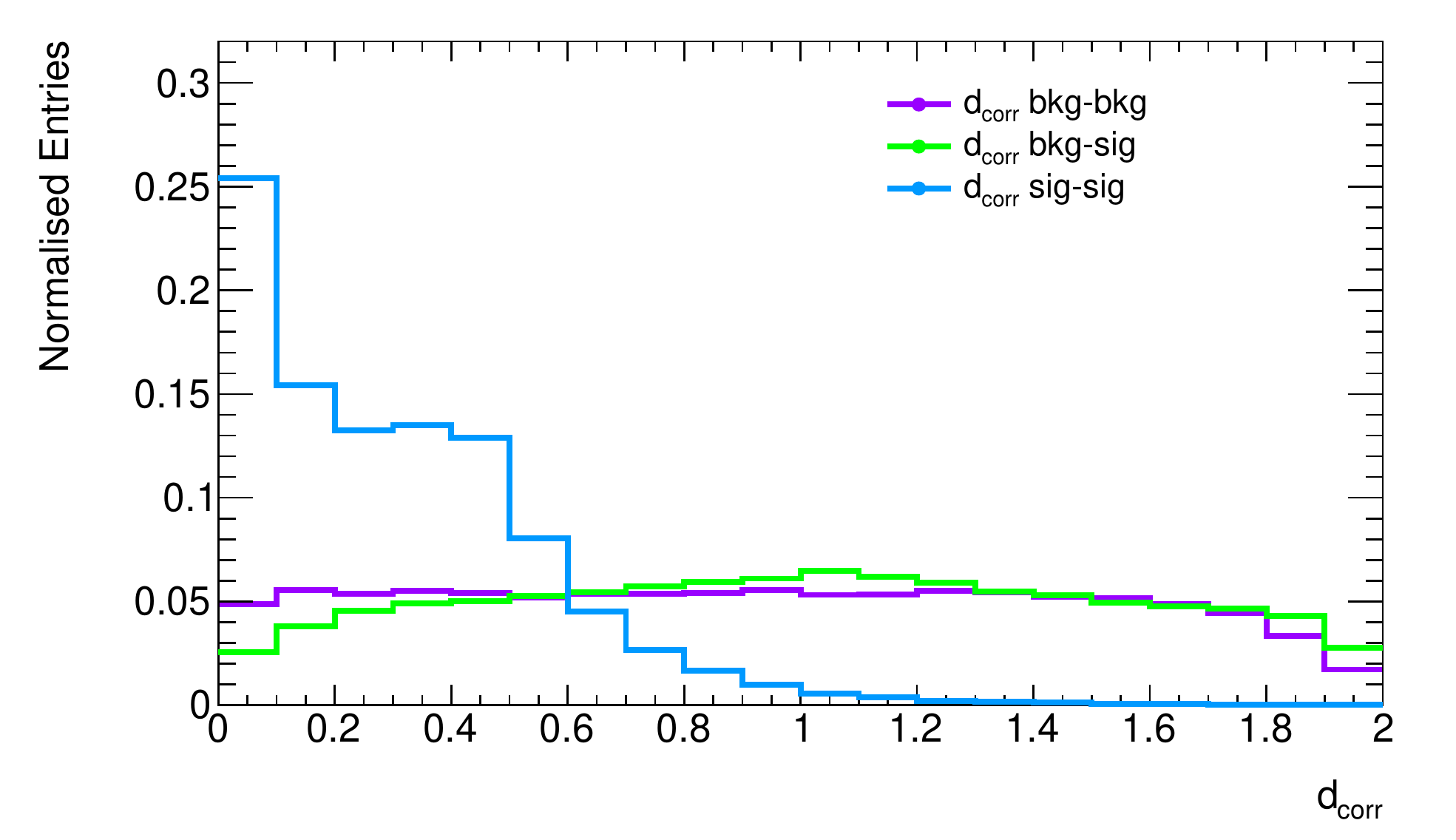}
\centering
\caption{correlation}
\label{5}
\end{subfigure} \\
\begin{subfigure}[b]{0.45\textwidth}
\includegraphics[width=\textwidth]{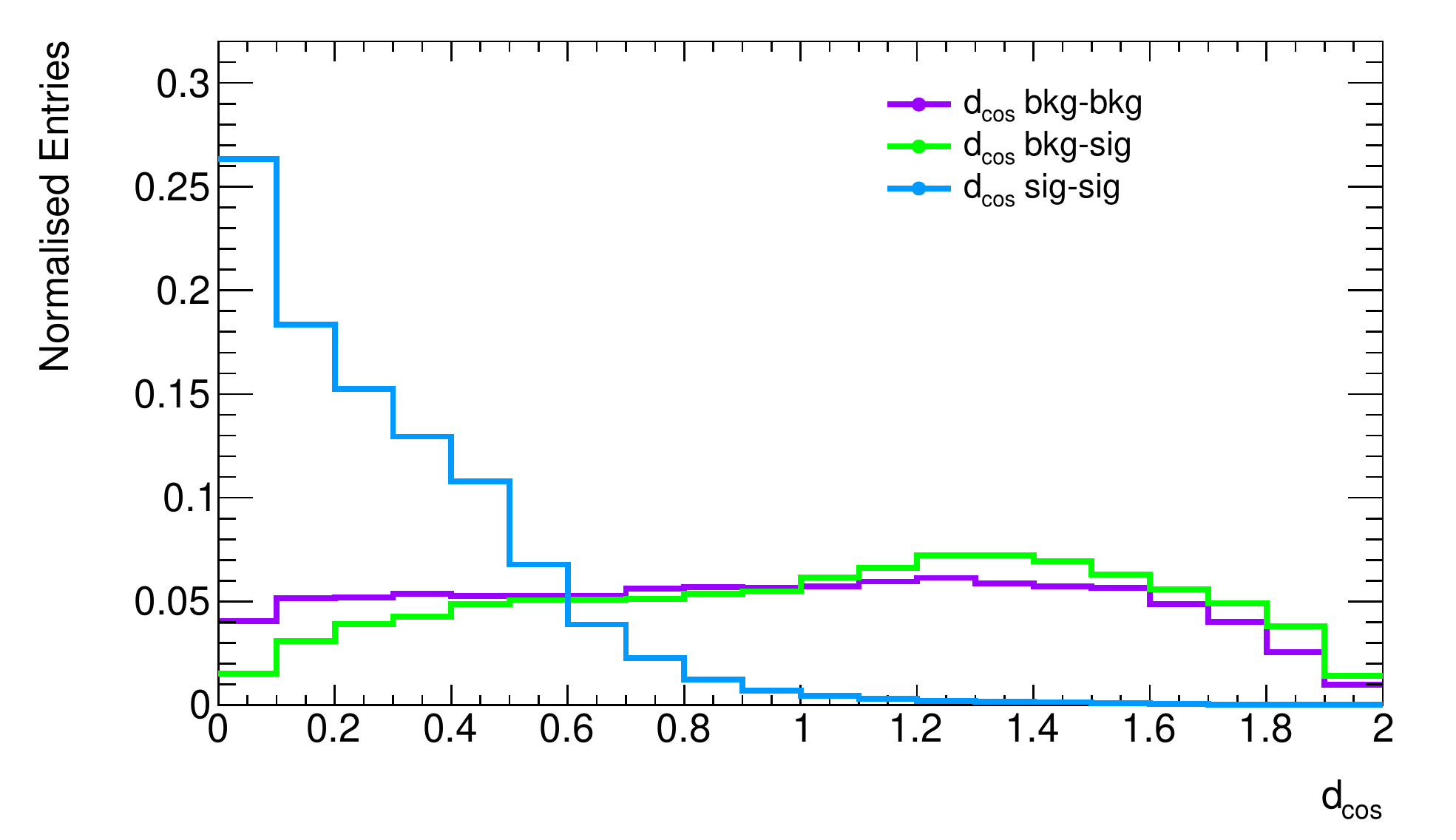}
\centering
\caption{cosine}
\label{6}
\end{subfigure} &
\begin{subfigure}[b]{0.45\textwidth}
\includegraphics[width=\textwidth]{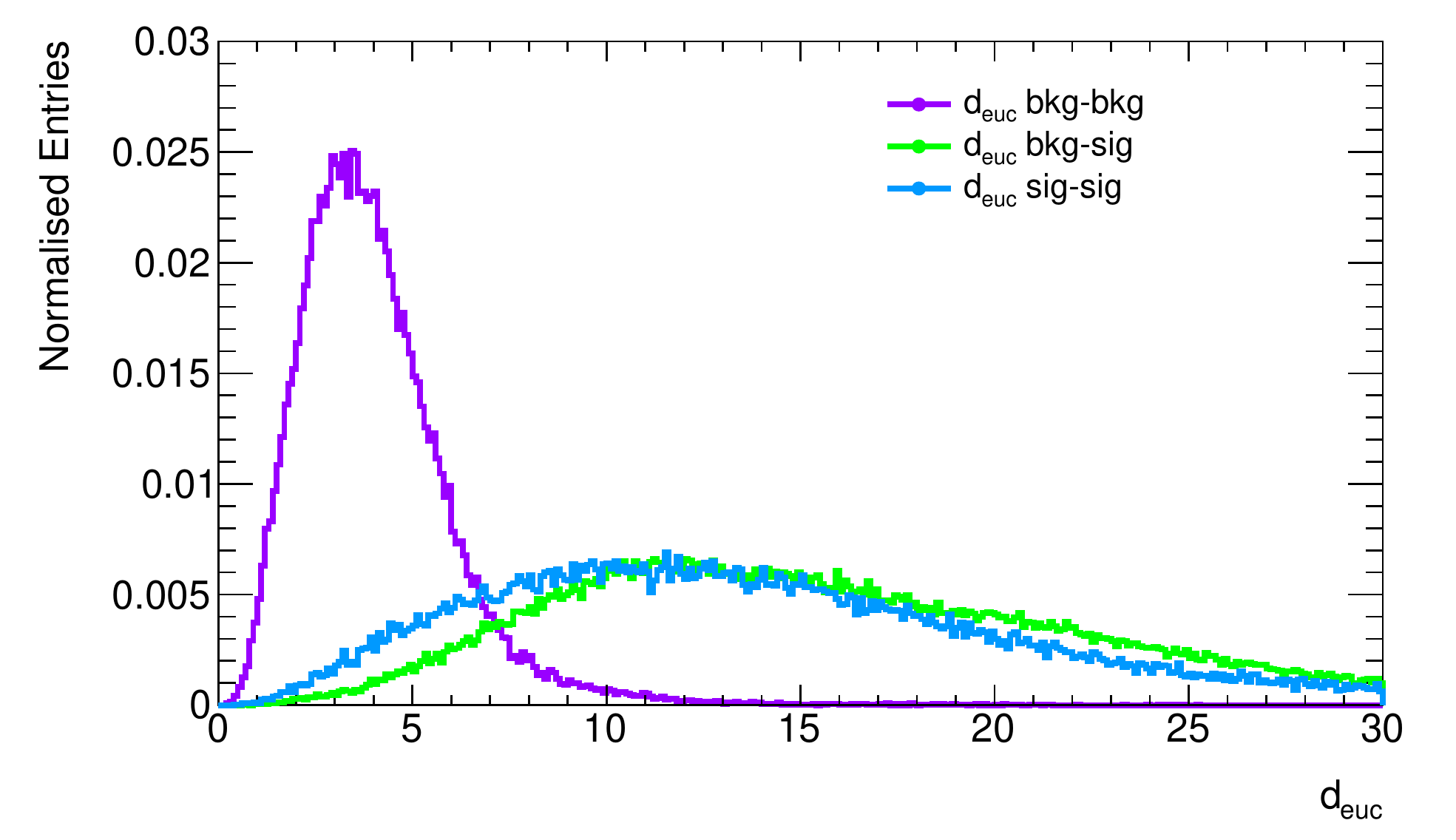}
\centering
\caption{Euclidean}
\label{7}
\end{subfigure} \\
\begin{subfigure}[b]{0.45\textwidth}
\includegraphics[width=\textwidth]{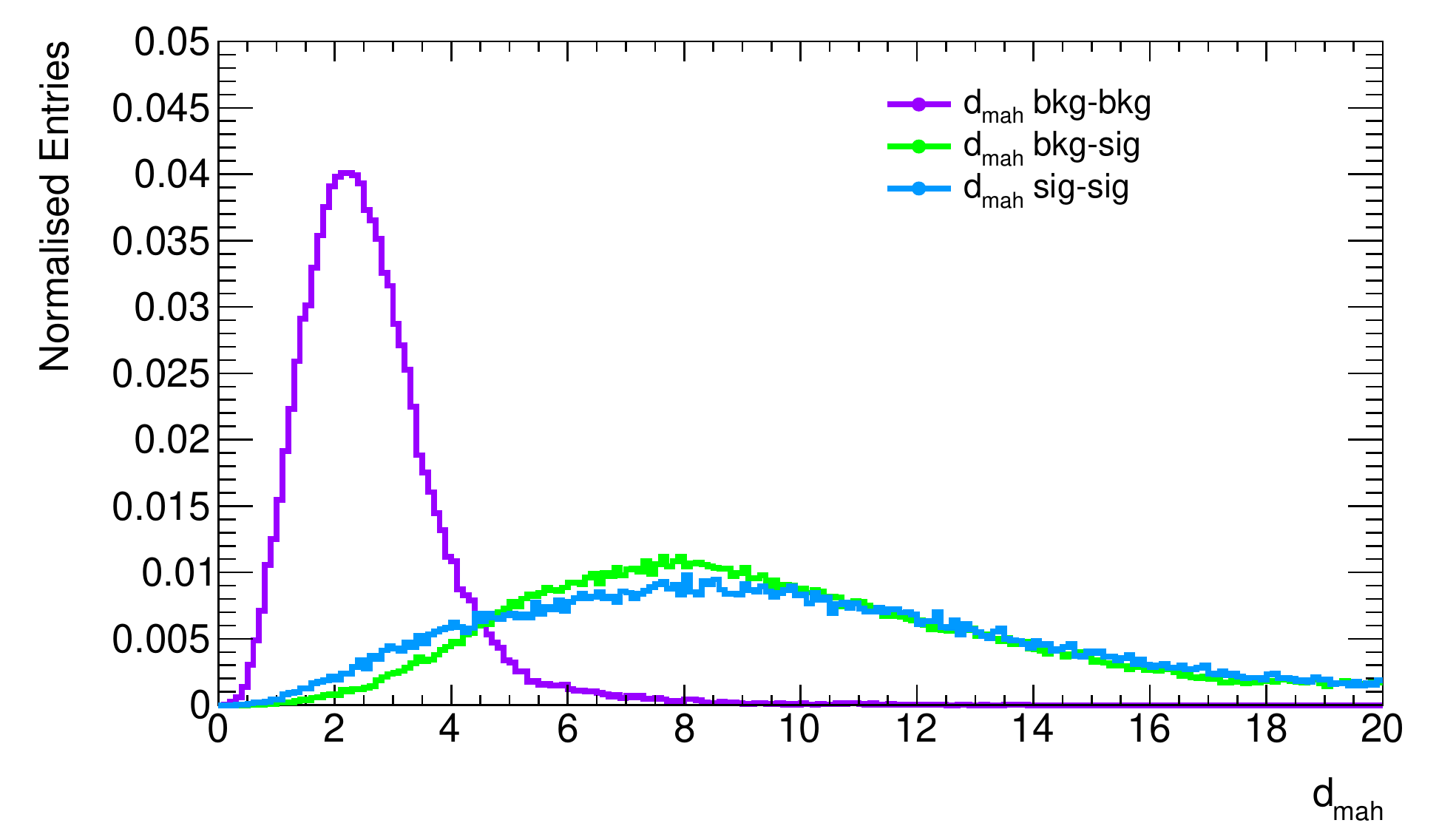}
\centering
\caption{Mahalanobis}
\label{8}
\end{subfigure} \\ 
\end{tabular}
\caption{\label{fig:ewdists} Distributions of the distance metrics used in our prototype electroweakino search.}
\end{figure}

When constructing the network for a given event sample these variables are scaled to equalise their weight in the analysis and account for their differing units. This ``median'' scaling is performed by subtracting the median of the variable's distribution in the background sample and normalising by the median absolute deviation (MAD). The MAD is calculated by subtracting the median of the background dataset from each point to produce a new dataset, then finding the median of the new dataset. A median scaling is chosen rather than a mean scaling to avoid being overly sensitive to tail effects. The distance metrics defined in Section~\ref{sec:network} are then calculated for a dataset containing 4000 of our simulated signal events and 4000 of our simulated SM events. Distributions of these distances, normalised to unit area, are shown for signal and background events in Figure~\ref{fig:ewdists}, where we have separated the distributions for signal-signal, signal-background and background-background distances.

%We note that that distances of signal events to signal events are shown in the ``signal'' histogram, distances of background events to background events are shown in the ``background'' histogram, and distances of signal events to background events (background-signal) events are shown in the signal (background) histogram, thus distances between signal and background events enter twice in each of the figures.

\FloatBarrier
Focussing on the signal-signal and background-background distributions, we see that these metrics exhibit reasonably large differences between the signal and background processes, indicating the presence of different typical scales at which events are expected to be clustered in the two cases. It is interesting to note that, in some cases,
the signal-signal and background-background roles are reversed; the Bray-Curtis, correlation
and cosine metrics all have the signal-signal distributions being more concentrated at small distances than the background-background distributions.
This raises the option of flipping the friendship condition to occur if the distance in kinematic space is greater than the linking length, although we continue to adopt the standard definition in the following analysis for consistency between distance metrics.
%So-called `enemy' networks can alo be built by reversing the requirement on linking length such that adequately distant events are linked, although we continue to primarily adopt the standard definition in the following analysis for consistency between distance metrics.
%Where an enemy network was used to calculate a network variable, a superscript $enemy$ is used.
%This already tells us that these two metrics are likely to prove the most interesting for the definition of graph networks. Interestingly, the Euclidean distance, which one might suggest as the simplest option, looks to be rather similar in the signal and background cases, although there is a small difference in behaviour. 
For each distance metric $d_{ij}$ between pairs of events $i$ and $j$, we define the adjacency matrix for the graph network by using the definition in Equation~\ref{eq:adj}. The linking length $l$ is chosen to be indicative of a characteristic scale of separation of background events, which differs from that of signal events. This then ensures that background events are well-connected in the networks, whilst signal events typically have fewer connections. Our final choices for each distance metric are summarised in Table~\ref{tab:ewlengths}.

\begin{table}[h!]
  \begin{center}
    \caption{Linking length values used for each distance metric for our prototype electroweakino analysis.}
    \label{tab:ewlengths}
    \begin{tabular}{c|c} 
      \textbf{Distance metric} & \textbf{Linking length} \\
      \hline
        $d_{\text{bray}}$ & 0.7 \\      
        $d_{\text{cheb}}$ & 4.8 \\
        $d_{\text{city}}$ & 12 \\  
        $d_{\text{corr}}$ & 0.6 \\ 
        $d_{\text{cos}}$ & 0.6 \\
        $d_{\text{euc}}$ & 6.4 \\
        $d_{\text{mah}}$ & 4.8 \\
    \end{tabular}
  \end{center}
\end{table}

After applying the event pre-selection, and with our definitions of ``friendship'' in place, we build networks for each distance metric and calculate, for each event, the local metrics defined in Section~\ref{sec:network}. Our analysis will proceed in two stages, designed to reflect how it could be performed in practice at the LHC:

\begin{itemize}
\item In this section, we use a signal-plus-background network to design signal regions that would be sensitive to our chosen electroweakino benchmark point, using a knowledge of which events in our simulated network are background events, and which are signal events. Within the ATLAS and CMS collaborations, this would correspond to using MC samples to design and optimise signal regions, using only one set of generated background events. Hypothetical significance results are obtained by placing selections on the original variables, and local network metrics, and counting the number of signal and background events in the resulting search region. For this procedure to be valid, it is essential that the background contribution to the signal-plus-background network metric distributions does not change significantly from the background distribution that would result from only having the background present in the network. We have checked this explicitly for all variables that are used to define our optimum analyses below~\footnote{As an alternative, one could split the background MC into two sets, and optimise the variable selections using both a signal-plus-background network, and a background-only network. We do not pursue that option in this paper, in order to reduce our simulation time.}.
\item In Section~\ref{sec:ewresults}, we develop a realistic example of an LHC exclusion test by simulating an independent set of background events that corresponds to the LHC data that would be obtained in the absence of a signal. These MC events are to be interpreted as ``mock LHC data'', and we build background-only networks from those events to represent the networks that would be obtained in such an LHC dataset. The yields in our search regions for the mock LHC data can then be compared to the yields expected from our signal-plus-background network analysis to determine the exclusion significance of our benchmark model (which now incorporates the effect of statistical fluctuations in the real LHC data). In practise, this test could be repeated on a variety of signal models, to generate exclusion limits in, for example, simplified model parameter planes.
\end{itemize}

%For the samples used in the electroweakino signal-plus-background network the expected yields for 150 fb$^{-1}$ are 7700 and 41 events for the $WZ$ background and electroweakino signal respectively. 
The metric calculations are computationally expensive to evaluate. For the electroweakino example, we used all events that pass the pre-selection (which amounts to just over 10,000 signal and 10,000 background events before weighting). We are confident that one could evaluate network metrics for $\mathcal{O}(100,000)$ network events with a suitable parallellisation of the network metric calculations, making the use of network variables after a looser pre-selection a realistic possibility within the ATLAS and CMS collaborations. With our chosen distance metrics, we have $7 \times 8 =$56 new variables in total, and we are also free to retain the original variables for extra kinematic selections.

\begin{figure}[htb!]
\begin{tabular}{cc}
\begin{subfigure}[b]{0.5\textwidth}
\includegraphics[width=\textwidth]{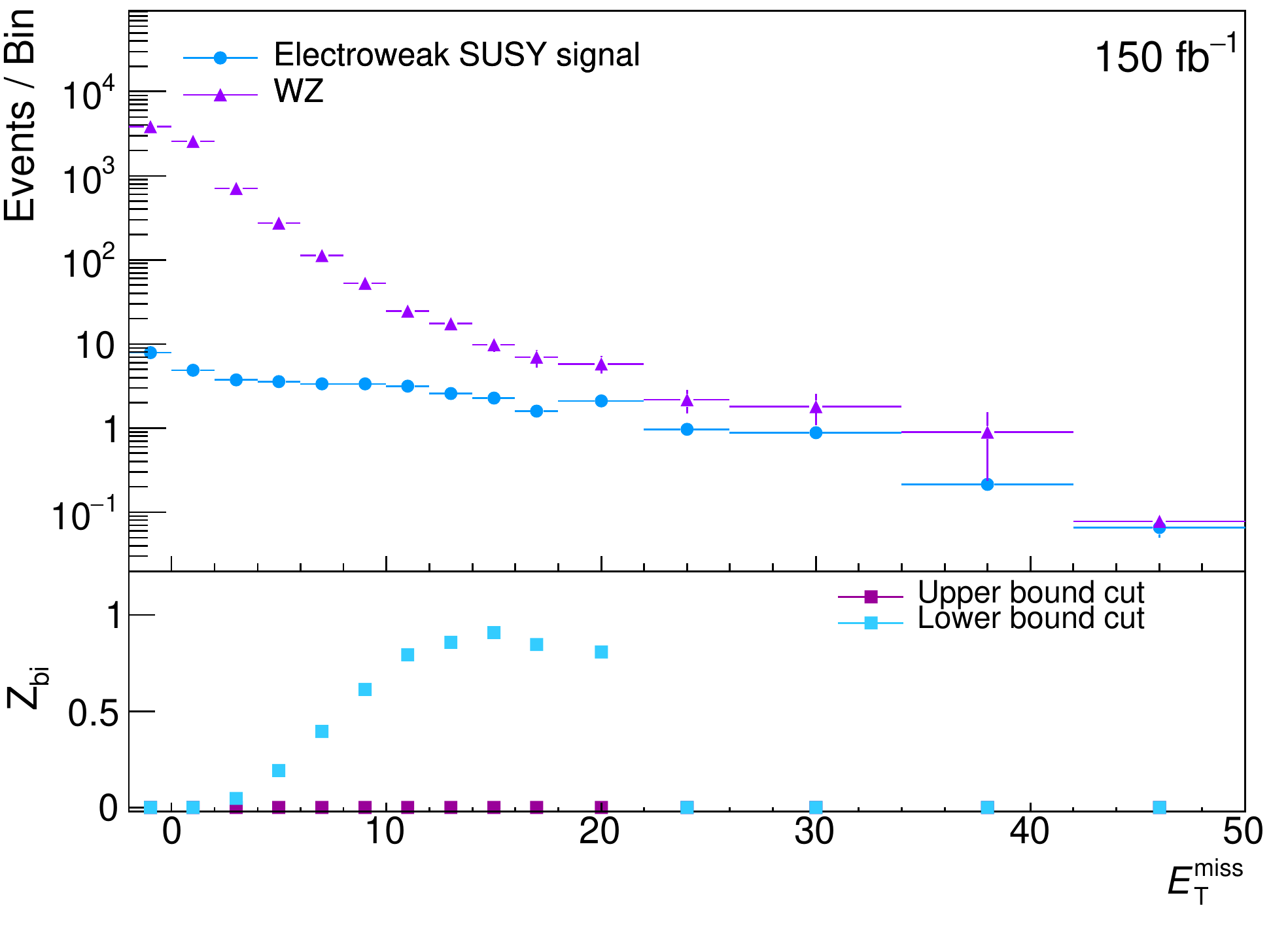}
\centering
\caption{$E_{\text{T}}^{\text{miss}}$}
\label{fig:ewvarsagain1}
\end{subfigure} &
\begin{subfigure}[b]{0.5\textwidth}
\includegraphics[width=\textwidth]{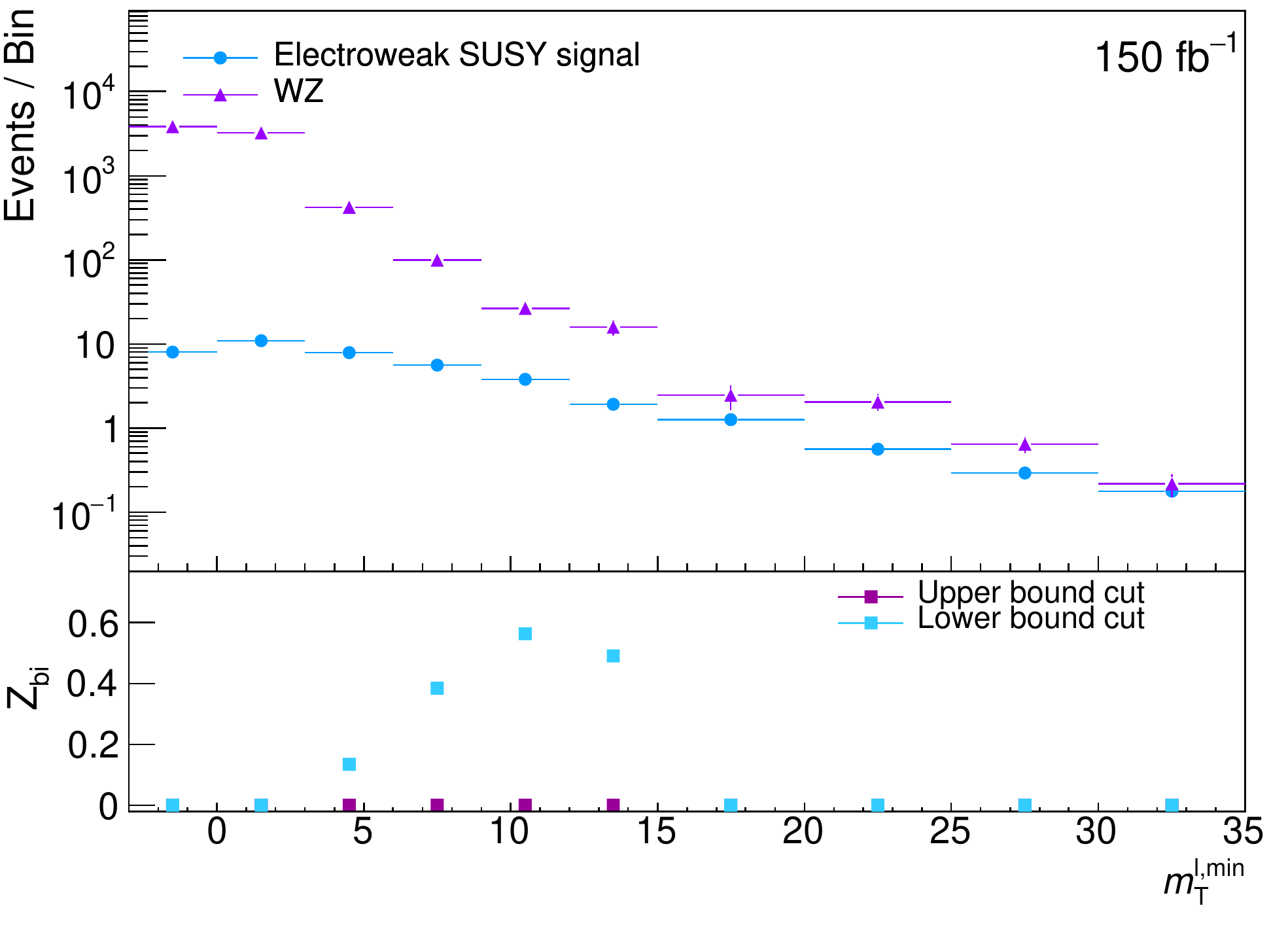}
\centering
\caption{$m_{\text{T}}^{l, \text{min}}$}
\label{fig:ewvarsagain2}
\end{subfigure} \\
\begin{subfigure}[b]{0.5\textwidth}
\includegraphics[width=\textwidth]{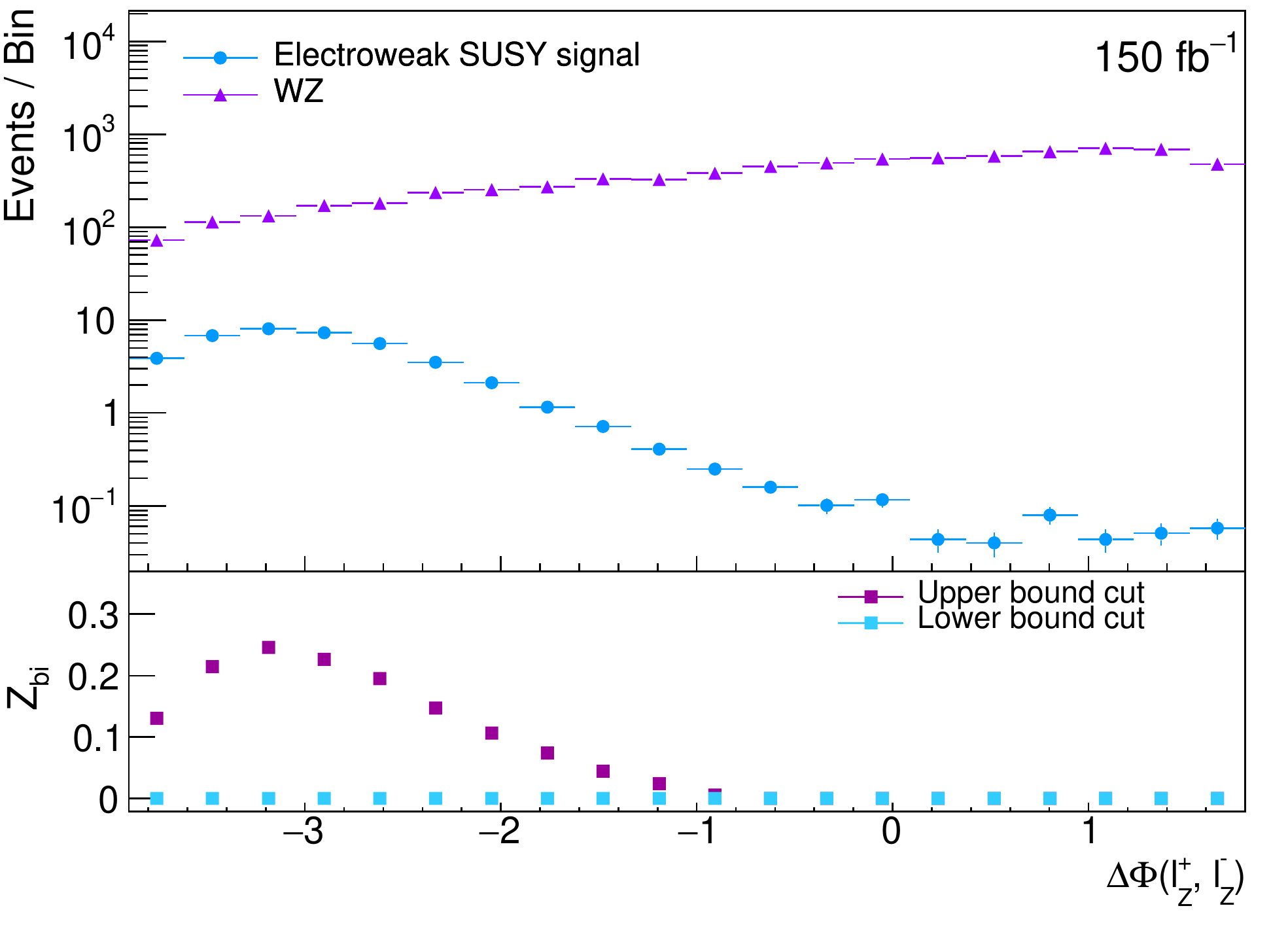}
\centering
\caption{$\Delta\Phi(l_{Z}^{+},l_{Z}^{-})$}
\label{fig:ewvarsagain3}
\end{subfigure} &
\begin{subfigure}[b]{0.5\textwidth}
\includegraphics[width=\textwidth]{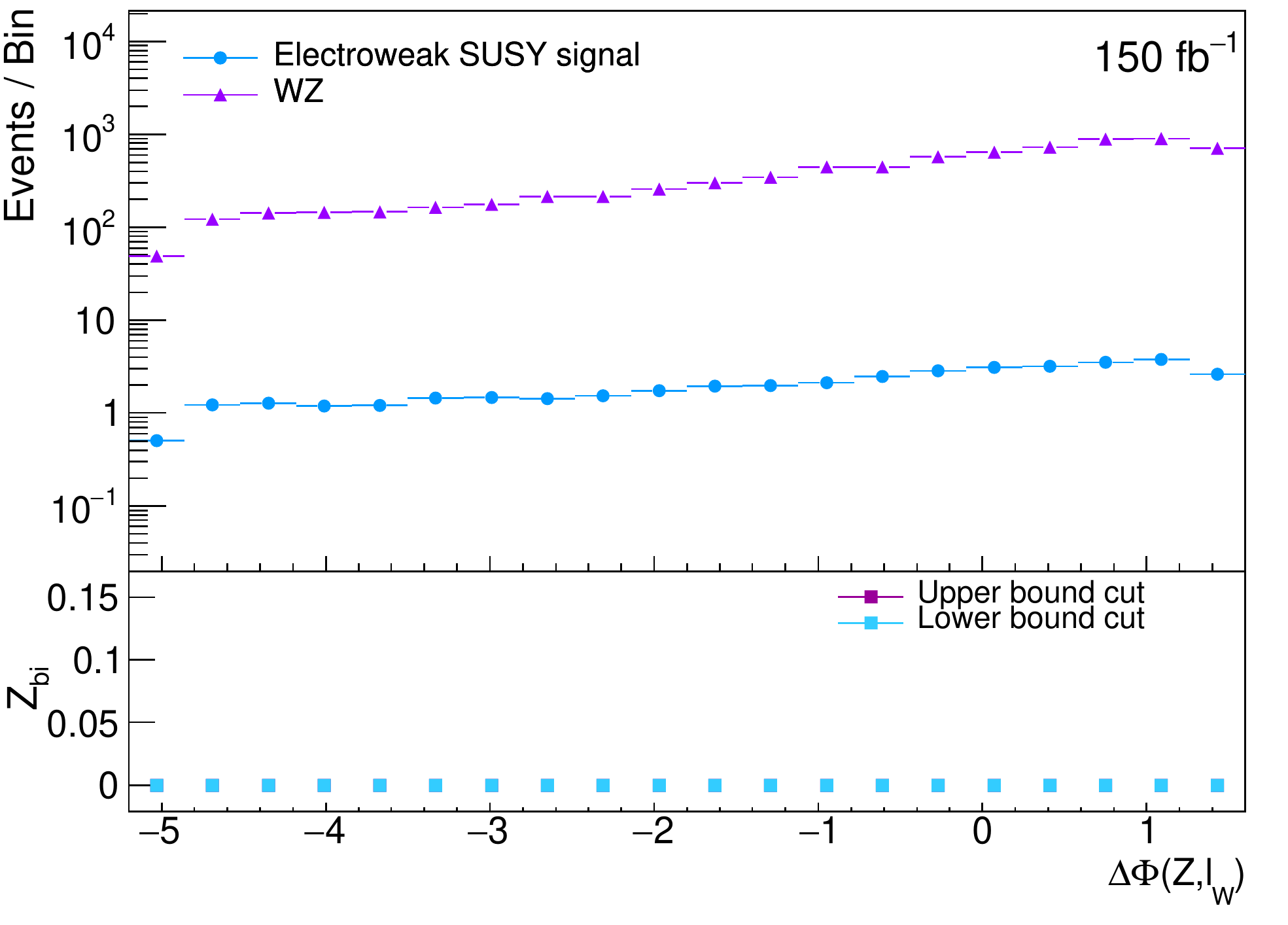}
\centering
\caption{$\Delta\Phi(Z,l_{W})$}
\label{fig:ewvarsagain4}
\end{subfigure} \\
\begin{subfigure}[b]{0.5\textwidth}
\includegraphics[width=\textwidth]{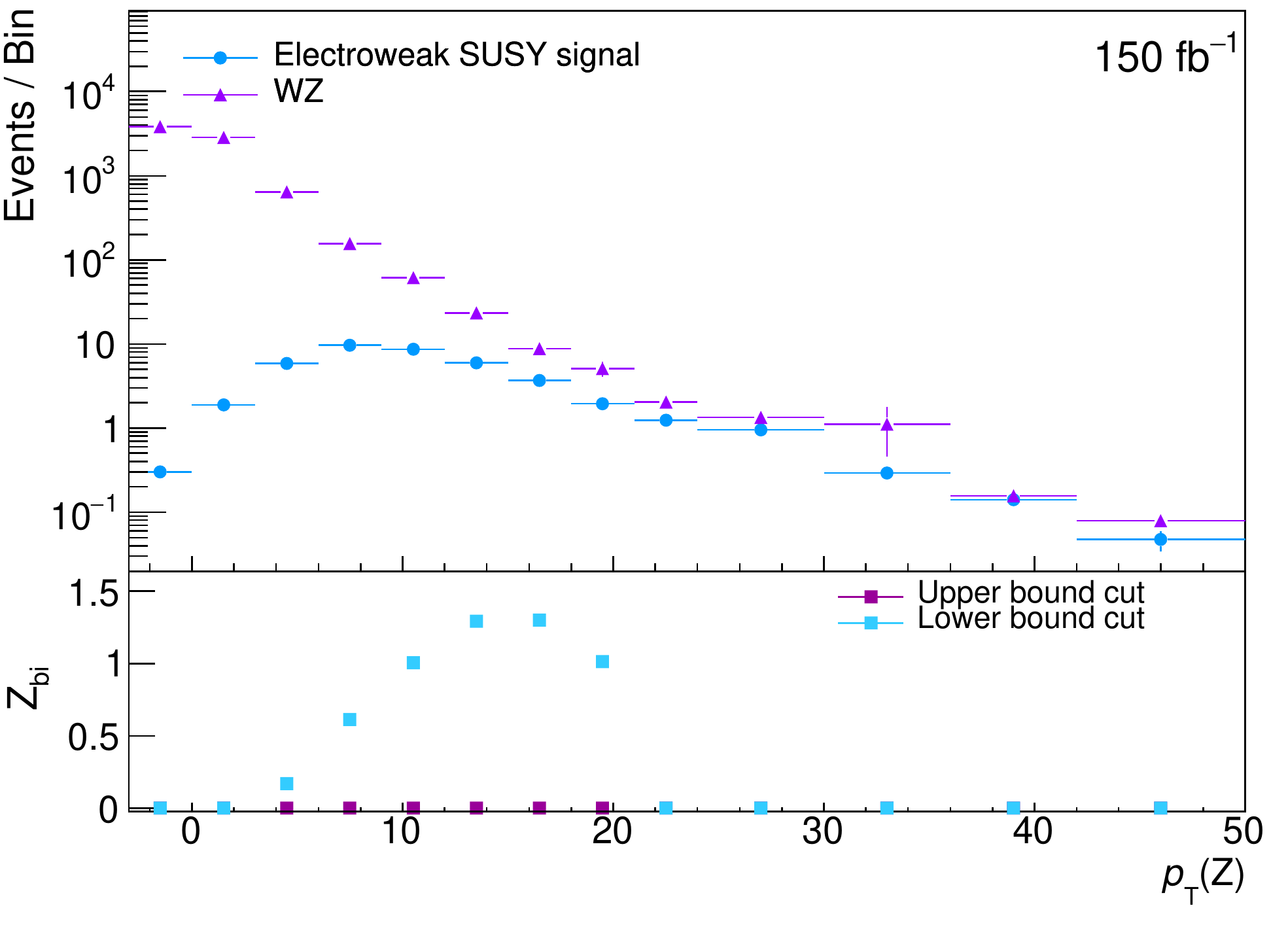}
\centering
\caption{$p_{\text{T}}(Z)$}
\label{fig:ewvarsagain5}
\end{subfigure} \\
\end{tabular}
\caption{\label{fig:ewvars} Event rates as a function of the kinematic variables used in our prototype electroweakino search, with the median scaling applied. Events in the overflow bin are not shown in the distribution but are included in the $Z_{bi}$ calculation.}
\end{figure}

After building the networks, we explored a variety of potential event selections that used either the network metrics alone, the original kinematic variables alone, or a combination of the original kinematic variables and network metrics. %This revealed that a further selection $p_{\text{T}}(Z)>160$~GeV provides additional signal-background discrimination that enhances the role of the network metrics. %Applying this cut instead before building the graph networks makes the network metrics less useful, 
Including tighter kinematic requirements in the pre-selection before building the graph network weakened the sensitivity of the network-driven searches, due to the increased kinematic similarity of the signal and background events that pass tighter kinematic selections.
The five kinematic variables are shown in Figure~\ref{fig:ewvars}. The lower panel of each figure shows the binomial significance $Z_{bi}$ %({\bf Suggest to change to $Z_{bi}$ unless i can find a reference that uses $Z_N$})
 for either an upper cut or lower cut on the variable at the value given on the horizontal axis, assuming a total systematic uncertainty of 15\% (chosen to be consistent with numbers quoted by the ATLAS run-2 searches in the 3-lepton channel~\cite{Aad:2019vvi,Aaboud:2018jiw}). This significance calculation also includes the statistical uncertainty of the background which is added to the systematic uncertainty in quadrature. When the signal or background weighted event yield drops below 3 events, the $Z_{bi}$ is set to 0. Differences in shape between the signal and SM distributions are clearly apparent, but no selection on a single kinematic variable is able to achieve an expected exclusion sensitivity at 95\% confidence level (corresponding to $Z_{bi}=1.64$) for our chosen benchmark point.

\FloatBarrier

\begin{figure}[h!]
\begin{tabular}{cc}
%\begin{subfigure}[b]{0.5\textwidth}
%\includegraphics[width=\textwidth]{figures/ref_paper_chebyshev_nsi_degree_4000-MCset1-fullsliced_All_stdkinset-eps-converted-to.pdf}
%\centering
%\caption{\degree{cheb}}
%\label{cheb:1}
%\end{subfigure} &
%\begin{subfigure}[b]{0.5\textwidth}
%\includegraphics[width=\textwidth]{figures/ref_paper_chebyshev_nsi_local_clustering_4000-MCset1-fullsliced_All_stdkinset-eps-converted-to.pdf}
%\centering
%\caption{\LC{cheb}}
%\label{cheb:2}
%\end{subfigure} \\
%\begin{subfigure}[b]{0.5\textwidth}
%\includegraphics[width=\textwidth]{figures/ref_paper_chebyshev_nsi_local_soffer_clustering_4000-MCset1-fullsliced_All_stdkinset-eps-converted-to.pdf}
%\centering
%\caption{\SC{cheb}}
%\label{cheb:3}
%\end{subfigure} & 
\begin{subfigure}[b]{0.5\textwidth}
\includegraphics[width=\textwidth]{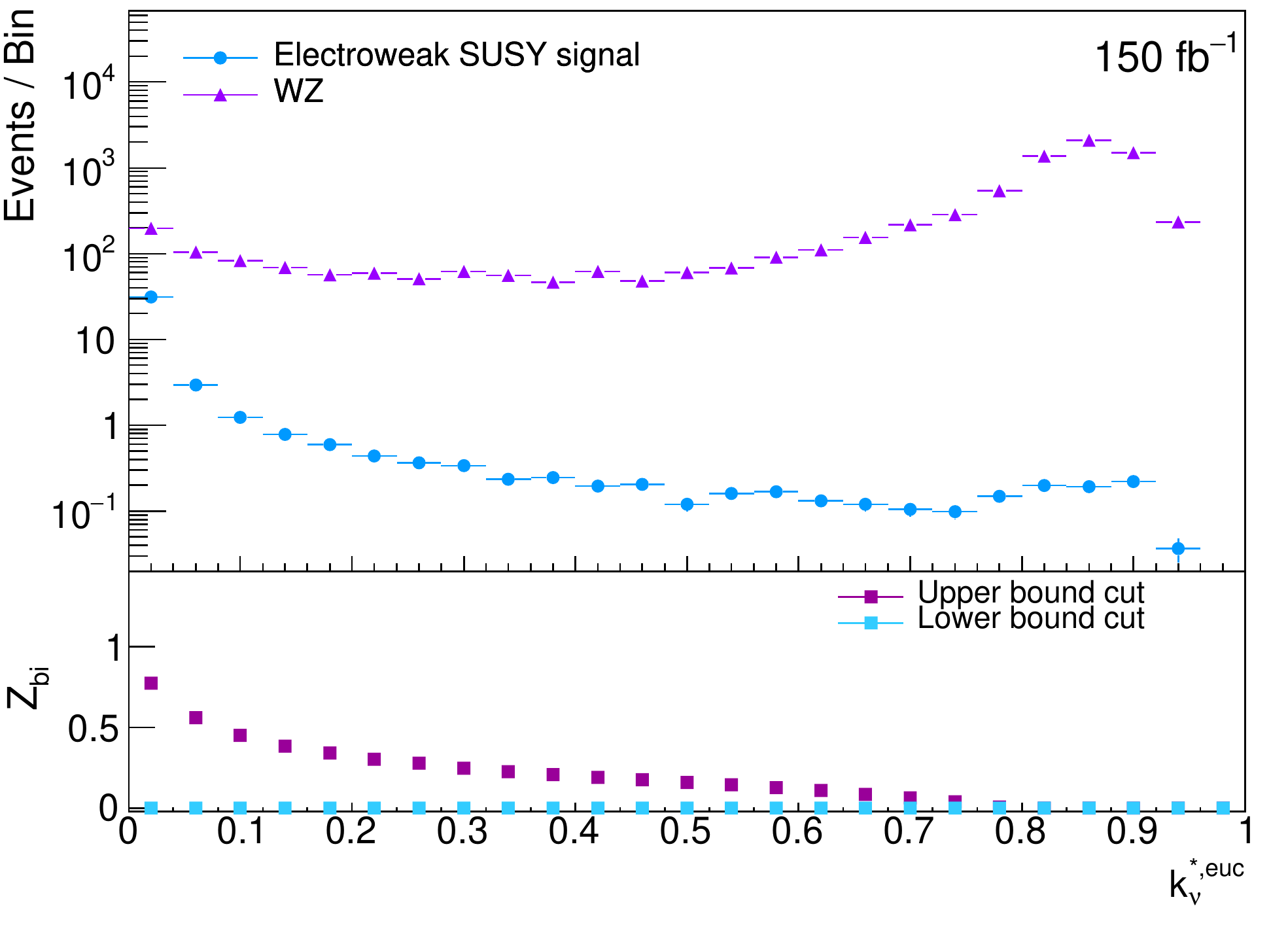}
\centering
\caption{\degree{euc}}
\label{euc:1}
\end{subfigure} &
%\begin{subfigure}[b]{0.5\textwidth}
%\includegraphics[width=\textwidth]{figures/ref_paper_euclidean_nsi_local_clustering_4000-MCset1-fullsliced_All_stdkinset-eps-converted-to.pdf}
%\centering
%\caption{\LC{euc}}
%\label{euc:3}
%\end{subfigure} &
%\begin{subfigure}[b]{0.5\textwidth}
%\includegraphics[width=\textwidth]{figures/ref_paper_euclidean_nsi_local_soffer_clustering_4000-MCset1-fullsliced_All_stdkinset-eps-converted-to.pdf}
%\centering
%\caption{\SC{euc}}
%\label{euc:4}
%\end{subfigure} \\
%\end{tabular}
%\caption{\label{fig:Networks:Electroweakinos:chebeuc} Event rates as a function of useful network metrics for our prototype electroweakino analysis calculated using $d_{\text{cheb}}$ and $d_{\text{euc}}$. Events in the overflow bin are not shown in the distribution but are included in the $Z_{bi}$ calculation.}
%\end{figure}
%
%
%\begin{figure}[h!]
%\begin{tabular}{cc}
\begin{subfigure}[b]{0.5\textwidth}
\includegraphics[width=\textwidth]{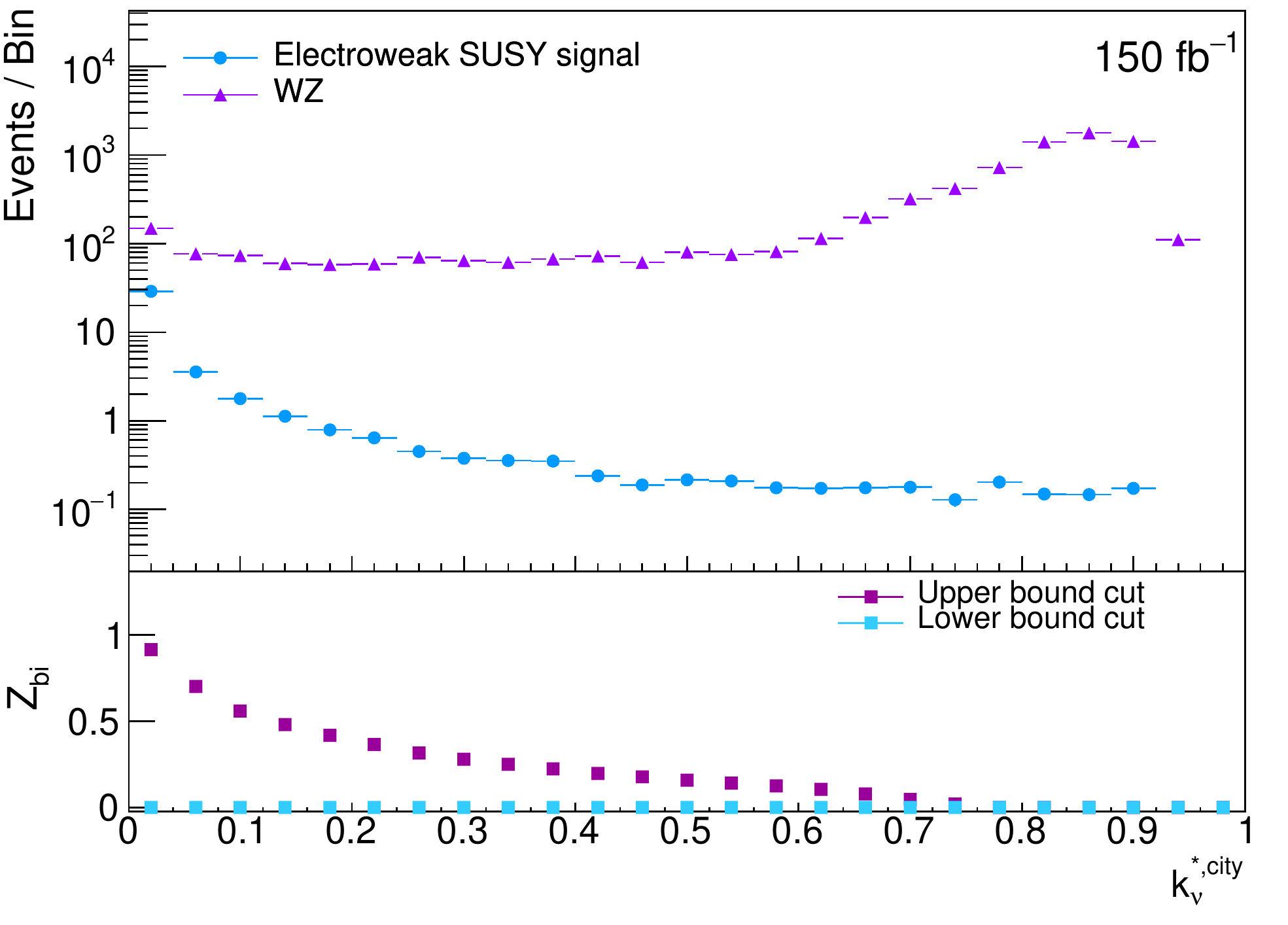}
\centering
\caption{\degree{city}}
\label{city:1}
\end{subfigure} \\
%\begin{subfigure}[b]{0.5\textwidth}
%\includegraphics[width=\textwidth]{figures/ref_paper_cityblock_nsi_average_neighbors_degree_4000-MCset1-fullsliced_All_stdkinset-eps-converted-to.pdf}
%\centering
%\caption{\ave{city}}
%\label{city:2}
%\end{subfigure} \\
%\begin{subfigure}[b]{0.5\textwidth}
%\includegraphics[width=\textwidth]{figures/ref_paper_cityblock_nsi_max_neighbors_degree_4000-MCset1-fullsliced_All_stdkinset-eps-converted-to.pdf}
%\centering
%\caption{\maxd{city}}
%\label{city:3}
%\end{subfigure} &
%\begin{subfigure}[b]{0.5\textwidth}
%\includegraphics[width=\textwidth]{figures/ref_paper_cityblock_nsi_local_soffer_clustering_4000-MCset1-fullsliced_All_stdkinset-eps-converted-to.pdf}
%\centering
%\caption{\SC{city}}
%\label{city:4}
%\end{subfigure}\\
%\begin{subfigure}[b]{0.5\textwidth}
%\includegraphics[width=\textwidth]{figures/ref_paper_cityblock_nsi_harmonic_closeness_4000-MCset1-fullsliced_All_stdkinset-eps-converted-to.pdf}
%\centering
%\caption{\HC{city}}
%\label{city:5}
%\end{subfigure} &
%\begin{subfigure}[b]{0.5\textwidth}
%\includegraphics[width=\textwidth]{figures/ref_paper_cityblock_nsi_exponential_closeness_4000-MCset1-fullsliced_All_stdkinset-eps-converted-to.pdf}
%\centering
%\caption{\EC{city}}
%\label{city:6}
%\end{subfigure}\\
\end{tabular}\caption{\label{fig:Networks:Electroweakinos:city} Event rates as a function of useful network metrics for our prototype electroweakino analysis calculated using $d_{\text{city}}$. Events in the overflow bin are not shown in the distribution but are included in the $Z_{bi}$ calculation.}
\end{figure}

\begin{figure}[h!]
\begin{tabular}{cc}
\begin{subfigure}[b]{0.5\textwidth}
\includegraphics[width=\textwidth]{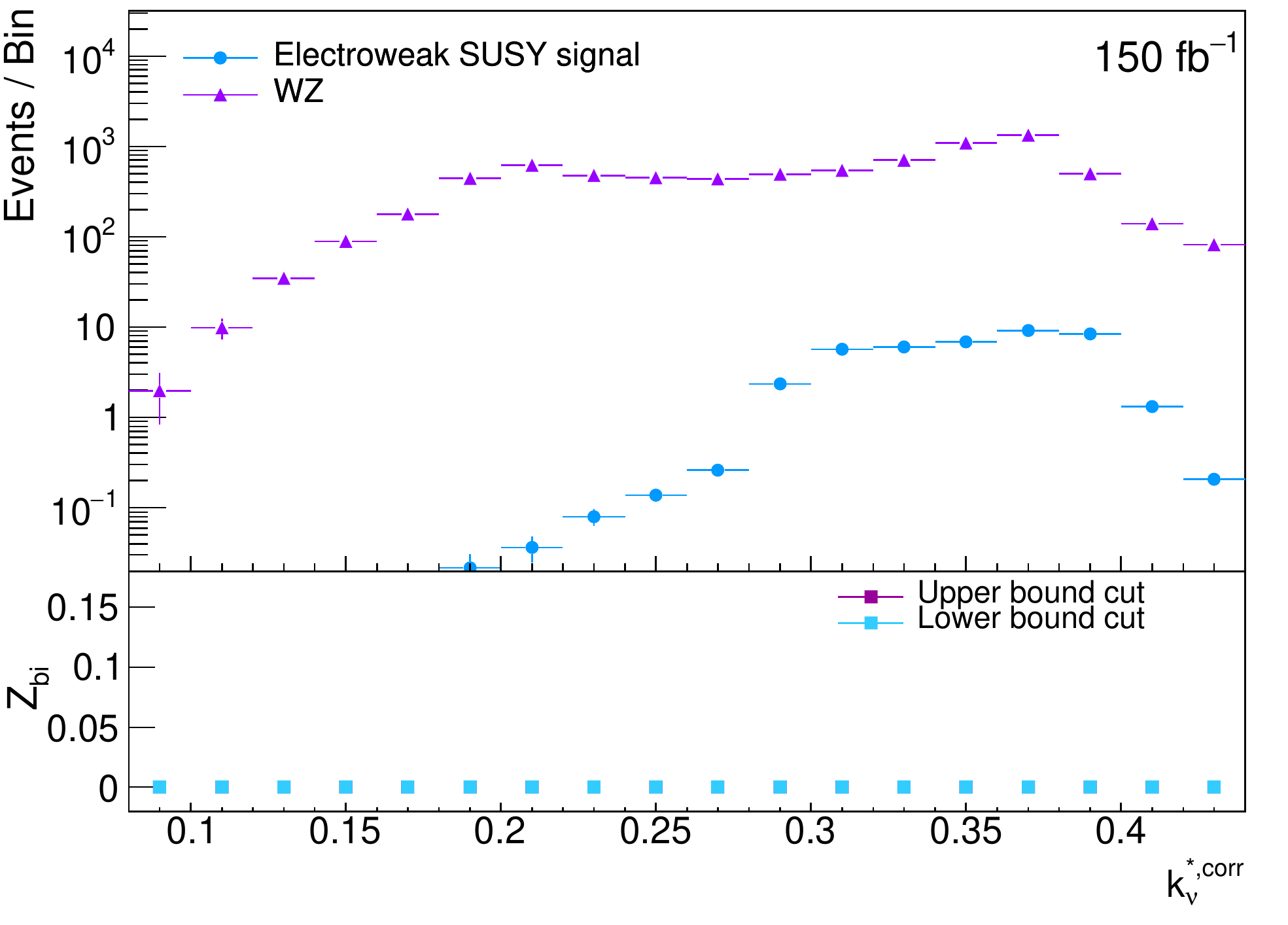}
\centering
\caption{\degree{corr}}
\label{corrcos:1}
\end{subfigure} &
\begin{subfigure}[b]{0.5\textwidth}
\includegraphics[width=\textwidth]{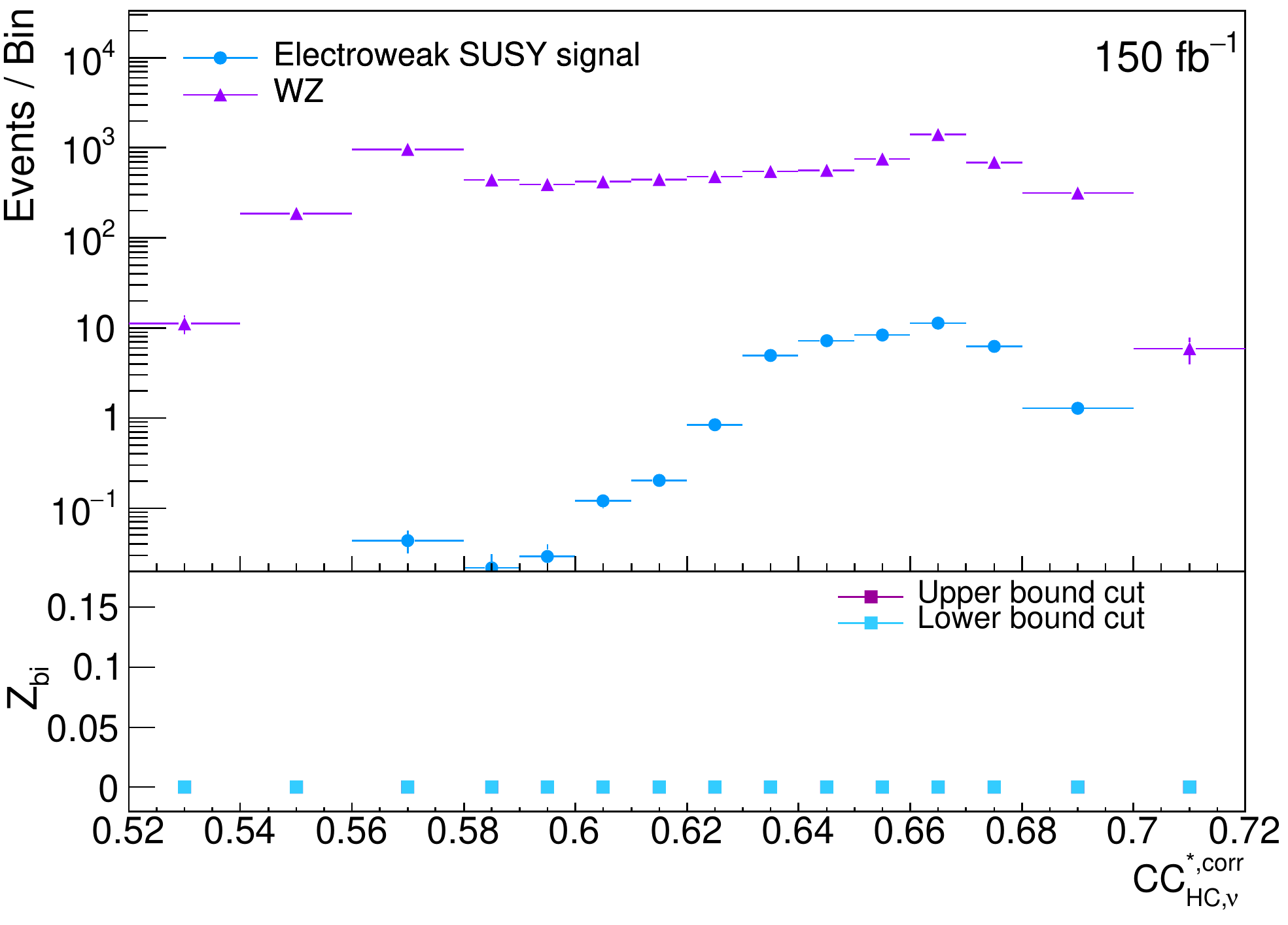}
\centering
\caption{\HC{corr}}
\label{corrcos:2}
\end{subfigure}\\
\begin{subfigure}[b]{0.5\textwidth}
\includegraphics[width=\textwidth]{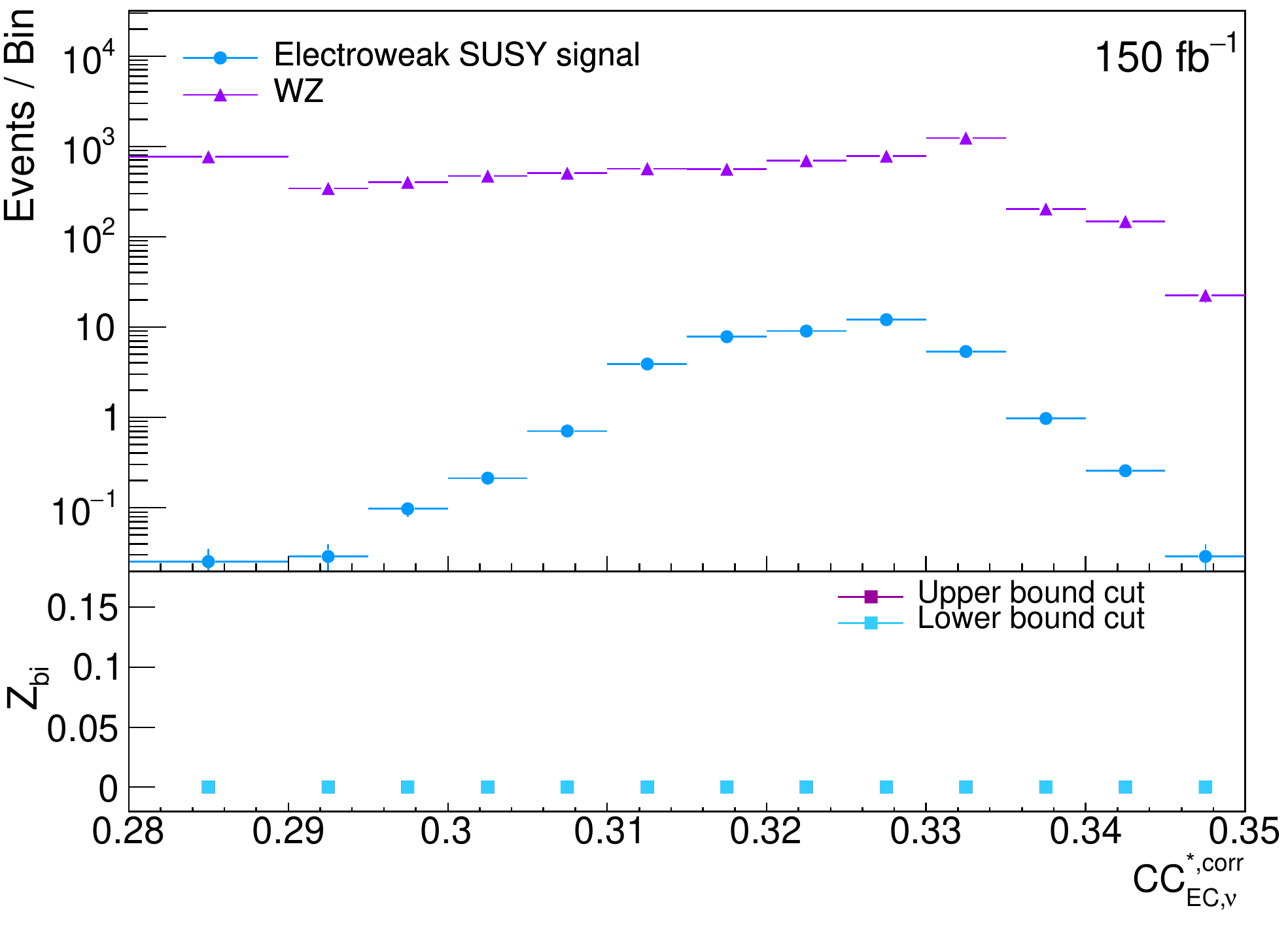}
\centering
\caption{\EC{corr}}
\label{corrcos:3}
\end{subfigure} &
\begin{subfigure}[b]{0.5\textwidth}
\includegraphics[width=\textwidth]{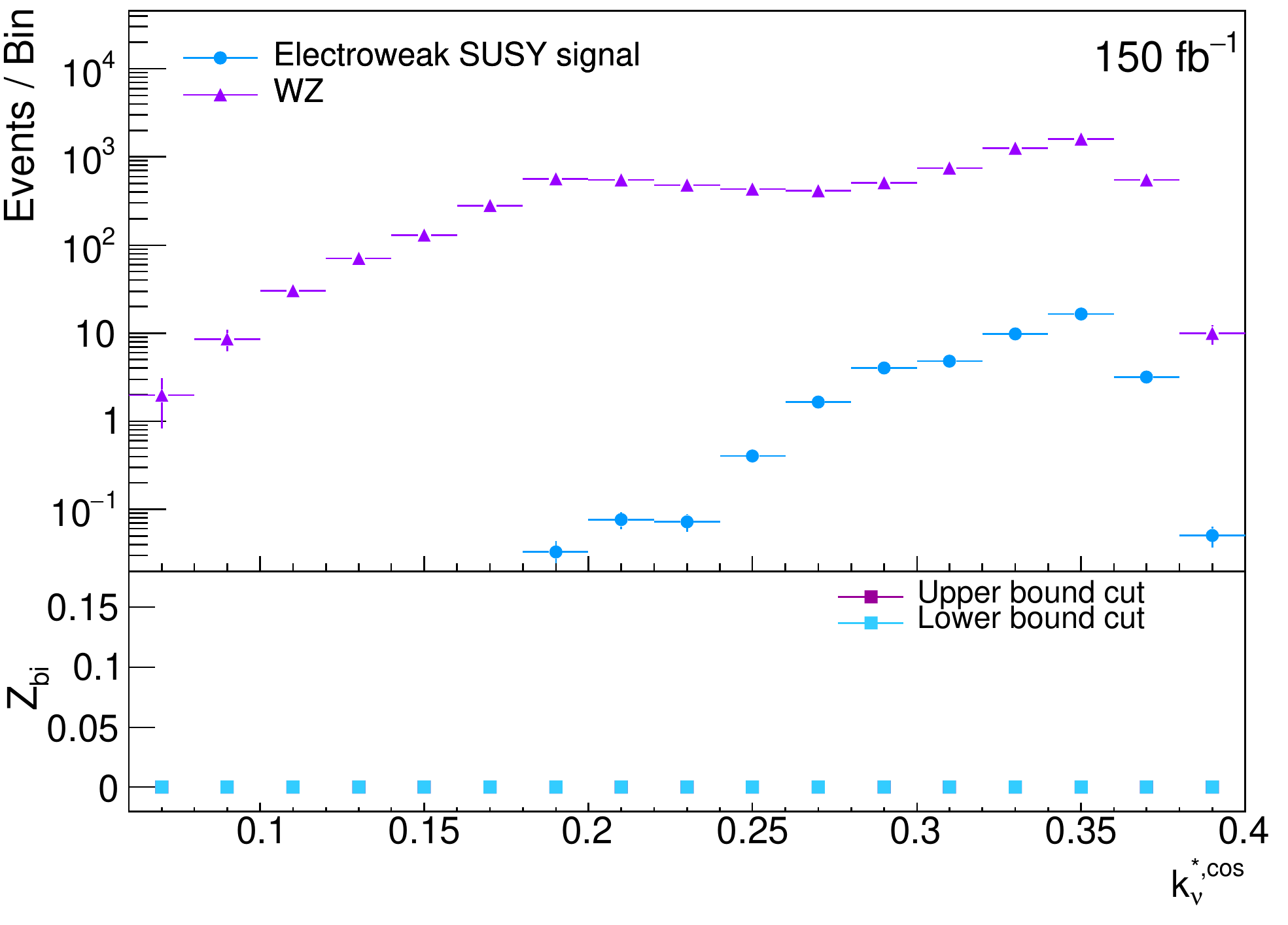}
\centering
\caption{\degree{cos}}
\label{corrcos:4}
\end{subfigure} \\
\begin{subfigure}[b]{0.5\textwidth}
\includegraphics[width=\textwidth]{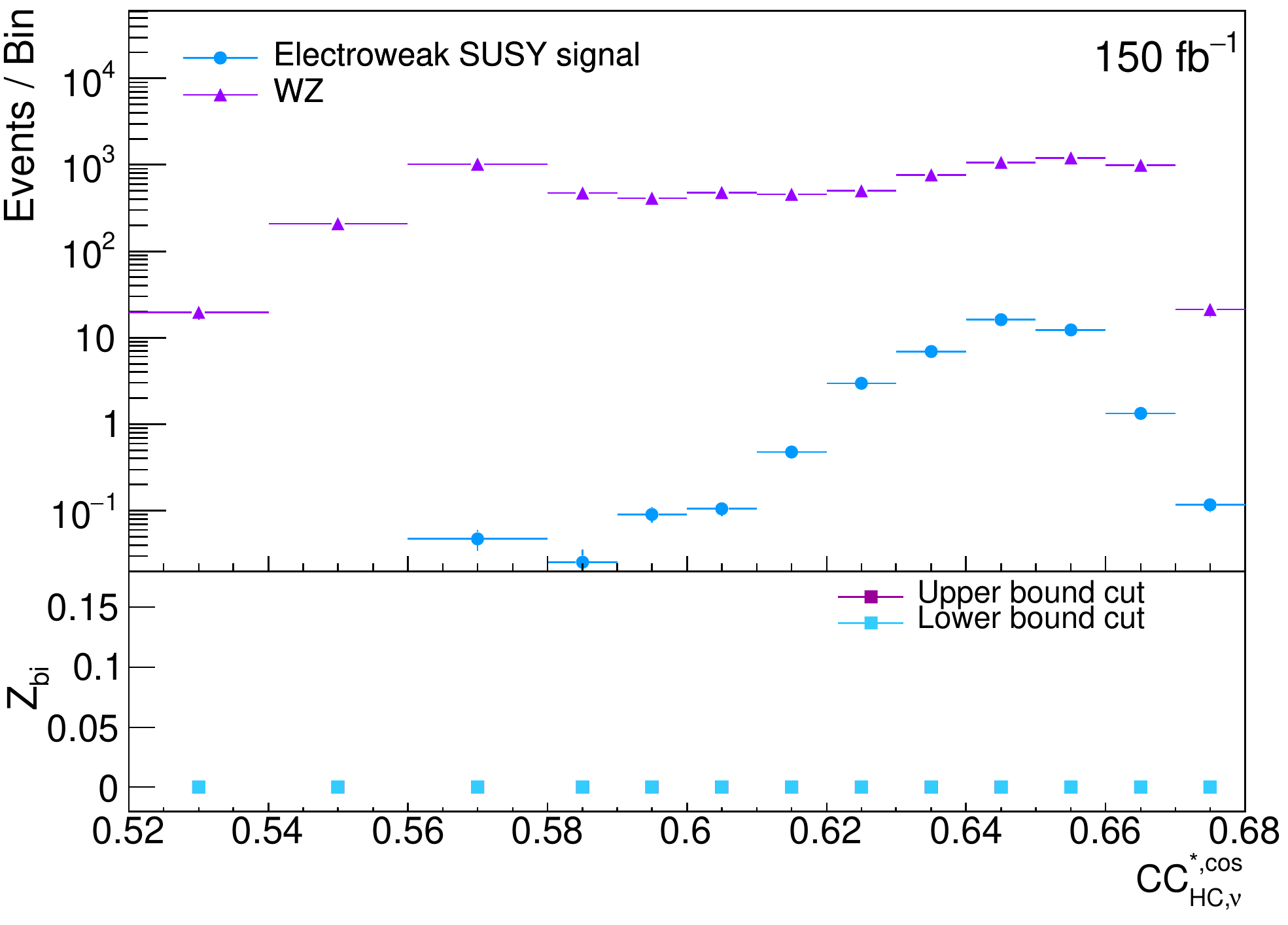}
\centering
\caption{\HC{cos}}
\label{corrcos:5}
\end{subfigure} &
%\begin{subfigure}[b]{0.5\textwidth}
%\includegraphics[width=\textwidth]{figures/ref_paper_cosine_nsi_average_neighbors_degree_4000-MCset1-fullsliced_All_stdkinset-eps-converted-to.pdf}
%\centering
%\caption{\ave{cos}}
%\label{corrcos:5}
%\end{subfigure} \\
\end{tabular}
\caption{\label{fig:Networks:Electroweakinos:corrcos} Event rates as a function of useful network metrics for our prototype electroweakino analysis calculated using $d_{\text{corr}}$ and $d_{\text{cos}}$. Events in the overflow bin are not shown in the distribution but are included in the $Z_{bi}$ calculation.}
\end{figure}

These distributions can be contrasted with the distributions for the network metrics in the signal-plus-background network. In Figures~\ref{fig:Networks:Electroweakinos:city}-\ref{fig:Networks:Electroweakinos:corrcos}, we show various useful network metrics for the Euclidean, Chebyshev, cityblock, Bray-Curtis and cosine distance metrics.
The supersymmetric events consistently show lower values of $k_{\nu}^{*}$ as well as the clustering and closeness variables.
This suggests that the supersymmetric events form fewer connections with other events than the Standard Model events, and that the events they do connect with are also sparsely connected. This arises from a combination of factors including the different typical lepton four-momenta expected in the SUSY case (given that the final state has a higher multiplicity), different signal distributions for variables which have well-defined kinematic endpoints for the background, and the fact that the smaller number of SUSY events limits the number of possible connections relative to background events. The rare and complex nature of the SUSY events for our benchmark model can be expected to be common among many BSM model processes, suggesting that these network metrics could also be useful for other cases. This includes the more complex supersymmetry models favoured by recent global fits that significantly depart from simplified model assumptions.

Taken as a set, the local network metric distributions offer a large set of variables that offer greater discrimination between signal and background than can be obtained with the original kinematic variables. %To test the robustness of this conclusion, we have checked that the distributions are indeed invariant under changing the number of MC events used to build the networks, and also that the network variable distributions remain the same if the poorly-sampled kinematic tails of the original variables are removed. These checks are presented in Appendix~\ref{app:ewmc}.

To find optimum kinematic selections for potential analyses, we first consider each network metric in turn. We determine which variables will give the highest significance based on a single upper or lower cut, considering many possible cut values for each variable. This is then iteratively repeated with the additional variables until the number of signal and background events passing the combined selections drops below 3 in each case. Some examples of the $Z_{bi}$ values and event yields obtained for promising search regions which provide $Z_{bi}>1.64$ are shown in Table~\ref{tab:Networks:Electroweakinos:zns}. This calculation includes both the statistical uncertainty on the event yields, and an assumed systematic uncertainty of 15\%. 

\FloatBarrier

\begin{table}[!h]
	\begin{center}
	\small
	\begin{tabular}{l|ccc}
		\hline
		Requirement	& $N_{signal}$ & $N_{background}$ & $Z_{bi}$ \\
		\hline
%$\degree{euc}<0.003, \LC{euc}<0.750$ & $11.46 \pm 0.20$ & $18.99 \pm 2.09 $ & 1.69 \\
%$\degree{euc}<0.003, \ave{cos}>0.311$ & $9.97 \pm 0.19$ & $14.13 \pm 2.00 $ & 1.70 \\
$\degree{euc}<0.003, \EC{corr}>0.324$ & $7.78 \pm 0.17$ & $6.53 \pm 1.29 $ & 1.96 \\
$\degree{euc}<0.003, \HC{corr}>0.656$ & $8.45 \pm 0.18$ & $7.52 \pm 1.45 $ & 1.96 \\
%$\degree{euc}<0.003, \LC{cheb}<0.590$ & $5.52 \pm 0.14$ & $4.39 \pm 0.79 $ & 1.79 \\
%$\degree{euc}<0.003, \SC{cheb}<0.663$ & $6.61 \pm 0.15$ & $5.75 \pm 1.04 $ & 1.84 \\
%$\degree{euc}<0.003, \LC{city}<0.665$ & $6.41 \pm 0.15$ & $5.38 \pm 1.24 $ & 1.74 \\
%$\ave{euc}<0.006, \HC{corr}>0.659$ & $5.93 \pm 0.15$ & $4.49 \pm 0.87 $ & 1.87 \\
%$\ave{euc}<0.006, \LC{cheb}<0.590$ & $4.38 \pm 0.13$ & $3.09 \pm 0.41 $ & 1.78 \\
%$\ave{euc}<0.006, \SC{cheb}<0.671$ & $5.49 \pm 0.14$ & $4.13 \pm 0.46 $ & 1.95 \\
%$\ave{euc}<0.006, \LC{city}<0.654$ & $4.77 \pm 0.13$ & $3.28 \pm 0.81 $ & 1.69 \\
%$\degree{cheb}<0.002, \LC{cheb}<0.598$ & $5.02 \pm 0.14$ & $4.05 \pm 0.46 $ & 1.81 \\
%$\degree{cheb}<0.002, \SC{cheb}<0.659$ & $5.36 \pm 0.14$ & $4.72 \pm 0.78 $ & 1.71 \\
%$\degree{cheb}<0.002, \LC{city}<0.654$ & $4.90 \pm 0.13$ & $3.52 \pm 0.81 $ & 1.70 \\
%$\degree{cheb}<0.002, \SC{city}<0.715$ & $5.66 \pm 0.14$ & $3.98 \pm 0.83 $ & 1.87 \\
%$\degree{city}<0.004, \LC{euc}<0.745$ & $10.79 \pm 0.20$ & $15.37 \pm 1.84 $ & 1.80 \\
%$\degree{city}<0.004, \LC{euc}<0.745 && \SC{euc}<0.737$ & $6.57 \pm 0.15$ & $4.49 \pm 0.78 $ & 2.10 \\
$\degree{city}<0.004, \degree{cos}>0.343$ & $5.86 \pm 0.15$ & $3.20 \pm 1.05 $ & 1.89 \\
%$\degree{city}<0.004, \ave{cos}>0.311$ & $9.85 \pm 0.19$ & $12.57 \pm 1.77 $ & 1.81 \\
$\degree{city}<0.004, \HC{cos}>0.646$ & $7.43 \pm 0.16$ & $5.67 \pm 1.42 $ & 1.89 \\
$\degree{city}<0.004, \degree{corr}>0.359$ & $7.34 \pm 0.16$ & $5.43 \pm 1.10 $ & 2.04 \\
$\degree{city}<0.004, \EC{corr}>0.323$ & $8.13 \pm 0.17$ & $7.15 \pm 1.44 $ & 1.93 \\
$\degree{city}<0.004, \HC{corr}>0.656$ & $8.17 \pm 0.17$ & $7.27 \pm 1.45 $ & 1.92 \\
		\hline
	\end{tabular}
	\caption{\label{tab:Networks:Electroweakinos:zns}
		Examples of binomial significances for search regions that provide exclusion sensitivity without reducing yields below 3 events. The errors only include the statistical error.
		}
	\end{center}
\end{table}

\afterpage{\FloatBarrier}

\subsection{Results of realistic electroweakino exclusion test}
\label{sec:ewresults}
\begin{figure}[h!]
\begin{tabular}{cc}
\begin{subfigure}[b]{0.5\textwidth}
\includegraphics[width=\textwidth]{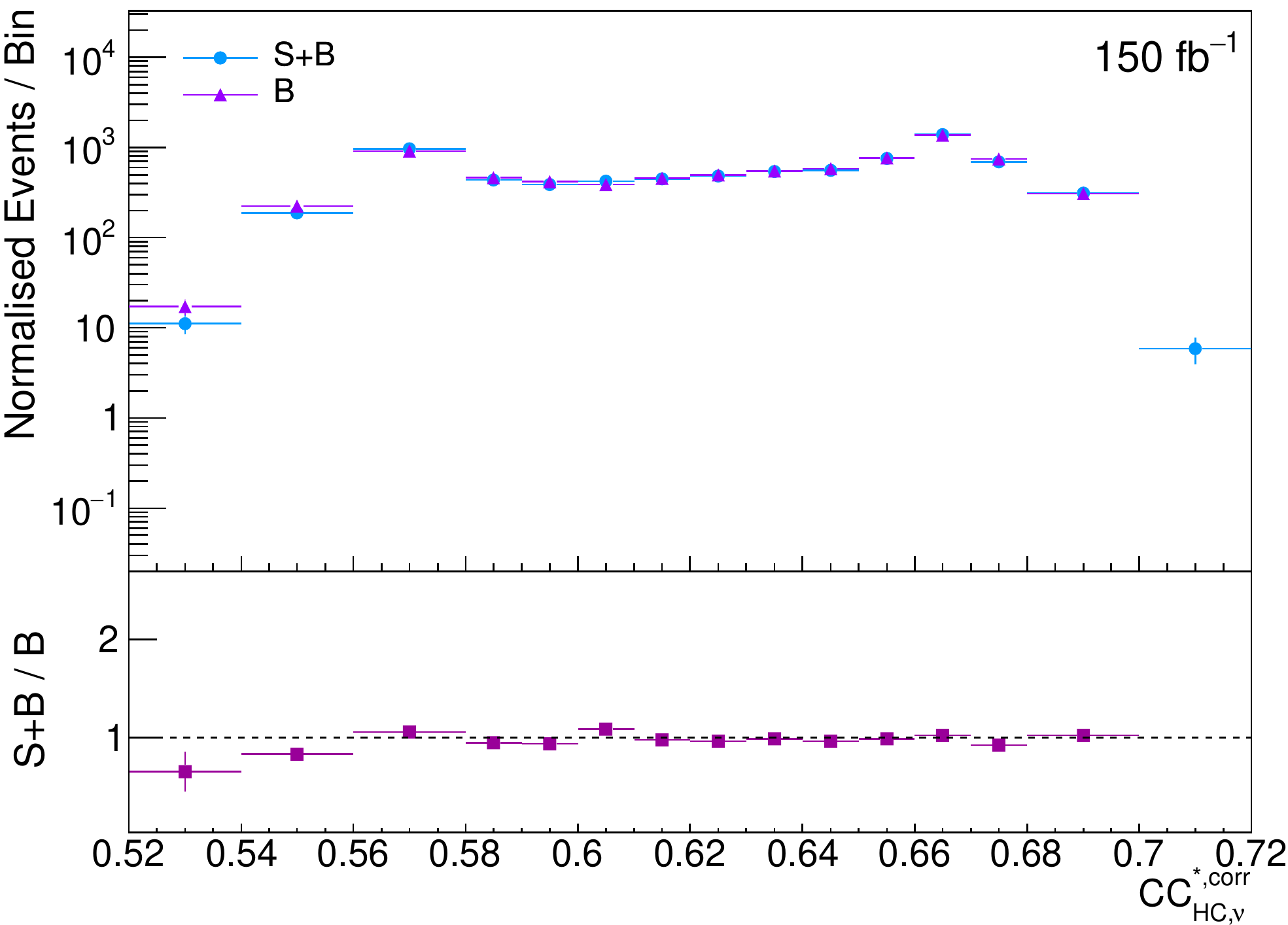}
\centering
\caption{\HC{corr}}
\label{1}
\end{subfigure} &
\begin{subfigure}[b]{0.5\textwidth}
\includegraphics[width=\textwidth]{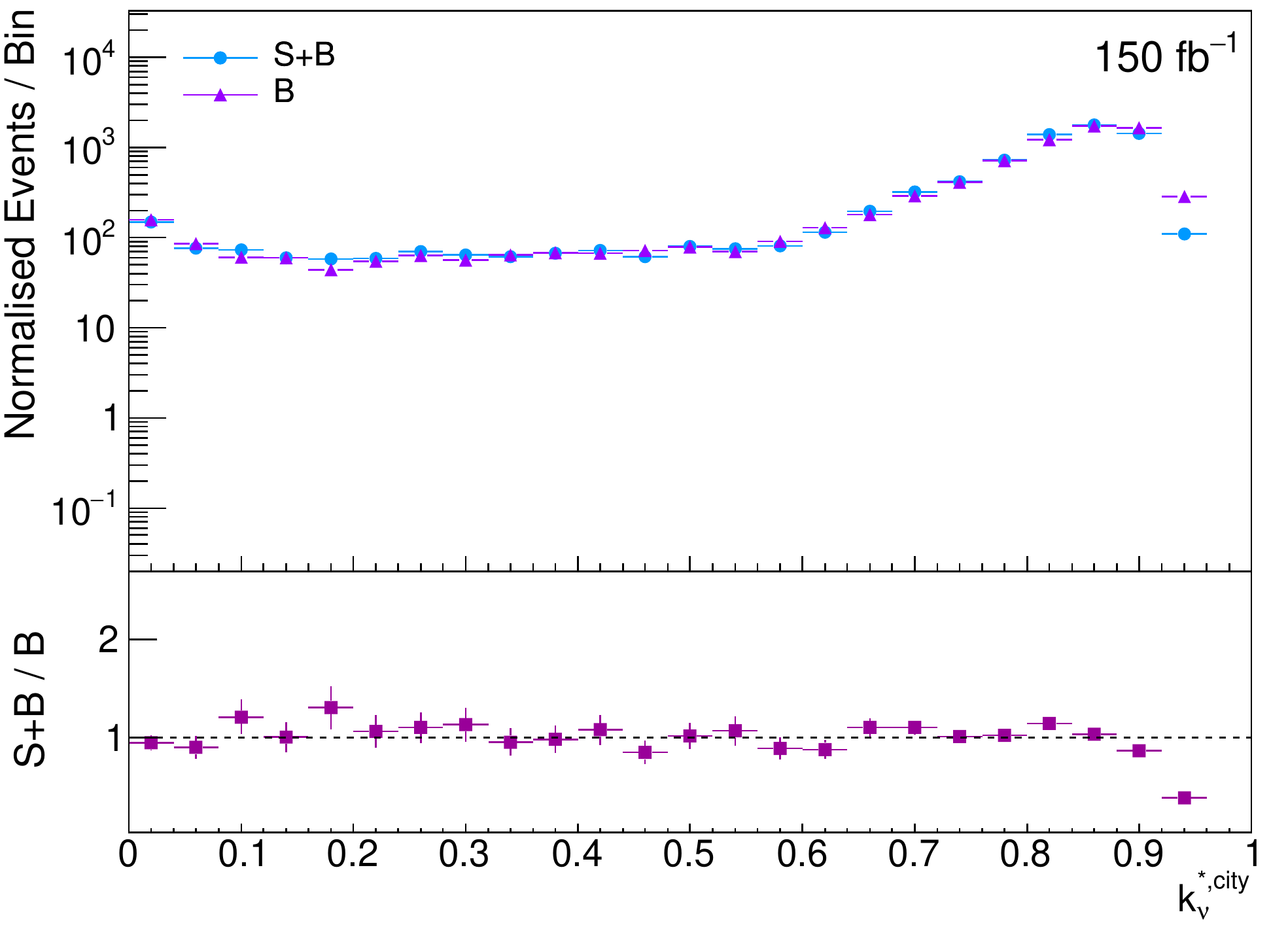}
\centering
\caption{\degree{city}}
\label{2}
\end{subfigure} \\
\centering
\end{tabular}
\caption{\label{fig:EW:results:SBvB} Event rates as functions of \SC{euc} and \degree{city} for $WZ$ background events calculated from either the signal-plus-backround or background-only network. These two networks use different sets of simulated $WZ$ events. Events in the overflow bin are not shown in the distribution.}
\end{figure}

Having defined signal regions using hypothetical signal-plus-background networks, we now compare the yields derived from the signal-plus-background networks with the background-only networks derived from our mock LHC dataset. This determines the exclusion sensitivity for our benchmark point that one would obtain with actual LHC data. The binomial significance is once again calculated with an error on the background yield that includes the statistical uncertainty plus an additional 15\% systematic uncertainty. Table~\ref{tab:Networks:Electroweakinos:results} compares the expected sensitivity of these regions with that obtained in some representative search regions constructed using only conventional kinematic variables. These selections are loosely inspired by the regions in the ATLAS 36.1fb$^{-1}$ search~\cite{Aaboud:2017aeu}, but generally have tighter requirements on $E_{\text{T}}^{\text{miss}}$ and $M_{T}^{l, \text{min}}$. We also impose slightly different pre-selection requirements (we require $n_{jets}<2$ whereas the ATLAS search region was inclusive in light jet multiplicity, but had ``binned'' regions considering the 0 and $>$0 light jet categories separately). The $Z_{bi}$ values differ from those of the previous section as expected, primarily due to statistical fluctuations in the assumed LHC dataset. 
As is demonstrated in Figure~\ref{fig:EW:results:SBvB}, which superimposes example background distributions taken from the background component of the signal-plus-background networks and the background-only networks, the general shape of the distributions are the same. This validates our previous procedure of designing search regions using signal-plus-background networks alone, and ultimately tells us that, for the chosen variables, the addition of rare signal events to the network does not subtantially alter the connection properties of background events. It remains the case that network methods provide exclusion sensitivity for the benchmark SUSY model, and they also outperform the analysis that uses cuts on the original kinematic variables. 

\begin{table}[!htb]
	\begin{center}
	\small
		\begin{tabular}{l|ccc}
			\hline
			Requirement(s)	& $N_{b-only}$ & $N_{s+b}$ & $Z_{bi}$ \\
			\hline
%$\degree{euc}<0.003, \LC{euc}<0.750 $ & 18.98 $\pm$ 2.09 & 30.45 $\pm$ 2.1 & 1.69 \\
%$\degree{euc}<0.003, \ave{cos}>0.311 $ & 8.4 $\pm$ 1.71 & 24.1 $\pm$ 2.01 & 3.15 \\
$\degree{euc}<0.003, \EC{corr}>0.324 $ & 7.27 $\pm$ 1.17 & 14.31 $\pm$ 1.3 & 1.76 \\
$\degree{euc}<0.003, \HC{corr}>0.656 $ & 8.43 $\pm$ 1.36 & 15.97 $\pm$ 1.46 & 1.73 \\
%$\degree{euc}<0.003, \LC{cheb}<0.590 $ & 3.39 $\pm$ 0.81 & 9.92 $\pm$ 0.8 & 2.24 \\
%$\degree{euc}<0.003, \SC{cheb}<0.663 $ & 4.73 $\pm$ 0.86 & 12.36 $\pm$ 1.05 & 2.32 \\
%$\degree{euc}<0.003, \LC{city}<0.665 $ & 4.52 $\pm$ 1.22 & 11.79 $\pm$ 1.25 & 2.05 \\
%$\ave{euc}<0.006, \HC{corr}>0.659 $ & 4.75 $\pm$ 0.88 & 10.42 $\pm$ 0.88 & 1.76 \\
%$\ave{euc}<0.006, \LC{cheb}<0.590 $ & 1.89 $\pm$ 0.28 & 7.46 $\pm$ 0.43 & 2.69 \\
%$\ave{euc}<0.006, \SC{cheb}<0.671 $ & 2.71 $\pm$ 0.34 & 9.62 $\pm$ 0.48 & 2.85 \\
%$\ave{euc}<0.006, \LC{city}<0.654 $ & 3.16 $\pm$ 0.97 & 8.05 $\pm$ 0.82 & 1.65 \\
%$\degree{cheb}<0.002, \LC{cheb}<0.598 $ & 3.15 $\pm$ 0.41 & 9.07 $\pm$ 0.48 & 2.33 \\
%$\degree{cheb}<0.002, \SC{cheb}<0.659 $ & 3.67 $\pm$ 0.45 & 10.08 $\pm$ 0.8 & 2.35 \\
%$\degree{cheb}<0.002, \LC{city}<0.654 $ & 3.4 $\pm$ 0.97 & 8.42 $\pm$ 0.82 & 1.66 \\
%$\degree{cheb}<0.002, \SC{city}<0.715 $ & 4.2 $\pm$ 1.0 & 9.64 $\pm$ 0.84 & 1.69 \\
%$\degree{city}<0.004, \LC{euc}<0.745 $ & 15.13 $\pm$ 1.71 & 26.15 $\pm$ 1.85 & 1.87 \\
$\degree{city}<0.004, \degree{cos}>0.343 $ & 3.48 $\pm$ 0.85 & 9.06 $\pm$ 1.06 & 1.90 \\
%$\degree{city}<0.004, \ave{cos}>0.311 $ & 8.32 $\pm$ 1.71 & 22.42 $\pm$ 1.78 & 2.88 \\
$\degree{city}<0.004, \HC{cos}>0.646 $ & 6.21 $\pm$ 1.29 & 13.1 $\pm$ 1.42 & 1.77 \\
$\degree{city}<0.004, \degree{corr}>0.359 $ & 6.08 $\pm$ 1.13 & 12.77 $\pm$ 1.11 & 1.79 \\
$\degree{city}<0.004, \EC{corr}>0.323 $ & 7.77 $\pm$ 1.46 & 15.29 $\pm$ 1.45 & 1.74 \\
$\degree{city}<0.004, \HC{corr}>0.656 $ & 7.17 $\pm$ 1.31 & 15.45 $\pm$ 1.46 & 2.01 \\
%$\degree{city}<0.004, \LC{cheb}<0.714 $ & 13.74 $\pm$ 1.67 & 26.14 $\pm$ 1.86 & 2.19 \\
%$\degree{city}<0.004, \LC{cheb}<0.714, \ave{cos}>0.3110$ & 4.42 $\pm$ 0.84 & 17.4 $\pm$ 1.15 & 3.74 \\
%$\degree{city}<0.004, \SC{cheb}<0.671 $ & 3.06 $\pm$ 0.38 & 11.64 $\pm$ 0.83 & 3.27 \\
%$\degree{city}<0.004, \LC{city}<0.703 $ & 7.72 $\pm$ 1.41 & 18.96 $\pm$ 1.76 & 2.54 \\
%$\degree{city}<0.004, \maxd{city}<750.000 $ & 31.17 $\pm$ 3.02 & 36.97 $\pm$ 2.44 & 0.58 \\
%$\degree{city}<0.004, \SC{city}<0.703 $ & 3.8 $\pm$ 1.0 & 8.83 $\pm$ 0.75 & 1.61 \\
%$\ave{city}<0.009, \LC{euc}<0.673 $ & 3.96 $\pm$ 0.74 & 8.65 $\pm$ 0.44 & 1.61 \\
%$\ave{city}<0.009, \SC{euc}<0.744 $ & 3.72 $\pm$ 0.76 & 8.73 $\pm$ 0.37 & 1.73 \\
$\ave{city}<0.009, \HC{corr}>0.659 $ & 4.4 $\pm$ 0.87 & 9.5 $\pm$ 0.85 & 1.63 \\
\hline
$\ptZ > 160$\,GeV, $\met>100$\,GeV, $\mtlmin>150$\,GeV & 93.93 $\pm$ 7.66 & 105.28 $\pm$ 7.55 & 0.47 \\
$\ptZ > 160$\,GeV, $\met>200$\,GeV, $\mtlmin>150$\,GeV & 28.55 $\pm$ 3.83 & 34.99 $\pm$ 3.62 & 0.64 \\
$\ptZ > 160$\,GeV, $\met>300$\,GeV, $\mtlmin>150$\,GeV & 7.05 $\pm$ 1.59 & 13.41 $\pm$ 1.78 & 1.48 \\
			\hline
		\end{tabular}
	\caption{\label{tab:Networks:Electroweakinos:results}
		Yields in our electroweakino search regions for our mock LHC data set ($N_{b-only}$) and our mock MC set ($N_{s+b}$). Also shown is the sensitivity of search regions using only the original kinematic variables. The errors quoted are statistical only, whilst the $Z_{bi}$ calculation uses a relative background uncertainty.}
	\end{center}
\end{table}
\FloatBarrier

\section{Case study 2: The search for stop quarks}
\label{sec:stop}
\subsection{Stop analysis design}
As a second demonstration of the application of network analysis to an LHC search, we consider the search for supersymmetric top quarks at the LHC. This is to determine if the apparent gains seen in the previous section are generic to all final states and signal topologies, or whether the same network metrics that work well in the electroweakino case fail in another example.

Before building the network we require there to be exactly one electron or muon, with transverse momentum $p_{\text{T}}>25$~GeV, at least 2 $b$-jets (with $p_{\text{T}} > 30$ GeV), and a missing transverse energy, $E_{\text{T}}^{\text{miss}}$, greater than 100 GeV. We have then examined a number of variables that are normally used in stop searches, choosing the following set that show good discrimination between the background and signal processes for our benchmark signal point:

\begin{itemize}
\item The leading jet transverse momentum $p_{\text{T}}^{j_1}$;
\item the value of the missing transverse momentum $E_{\text{T}}^{\text{miss}}$;
\item the minimum value of the transverse mass, $m_{\text{T}}^{b,\text{min}}$  formed by the two $b$-jets and missing transverse energy in the event. $m_T^{b}$ is defined as $\sqrt{2p_{\text{T}}^bE_{\text{T}}^{\text{miss}}\left[1-\cos(\Delta\phi)\right]}$, with $p_{\text{T}}^b$ signifying the tranverse momentum of each $b$-jet, and $\Delta\phi$ giving the difference in $\phi$ between the each $b$-jet and the missing transverse momentum;
\item the minimum value of the invariant mass formed by the lepton and each of the two $b$-jets in the event, $m_{bl}^{\text{min}}$. For the top pair production process, this has a maximum value, whilst signal events can have higher values;
\item the scalar sum of the moduli of the transverse momenta for the lepton and the 2 $b$-jets, which we refer to as $H_{\text{T}}$;
\item the asymmetric $m_{T2}$ value, $am_{\text{T2}}$ defined in Refs.~\cite{Barr:2009jv,Konar:2009qr,Bai:2012gs,Lester:1999tx,Lester:2014yga}.
\end{itemize}

\begin{figure}[ht!] 
\begin{subfigure}{0.48\textwidth}
\includegraphics[width=\linewidth]{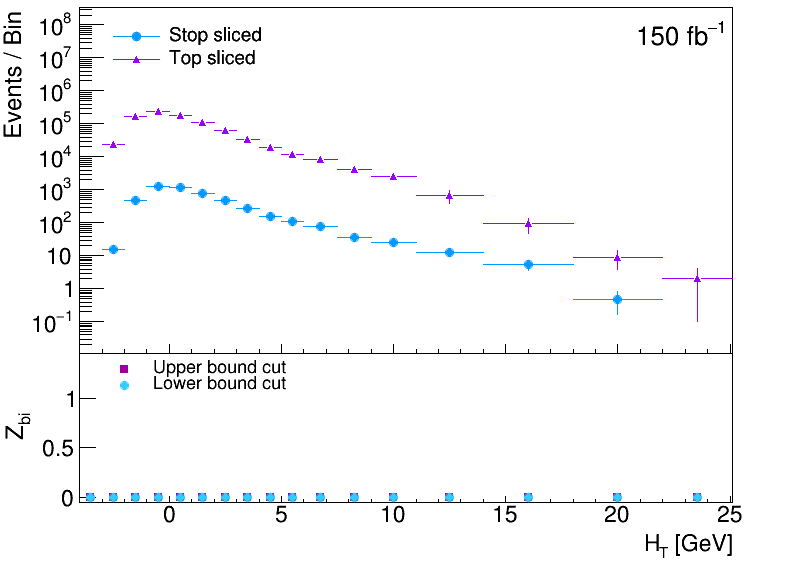}
\caption{$H_{\text{T}}$} \label{fig:a}
\end{subfigure}\hspace*{\fill}
\begin{subfigure}{0.48\textwidth}
\includegraphics[width=\linewidth]{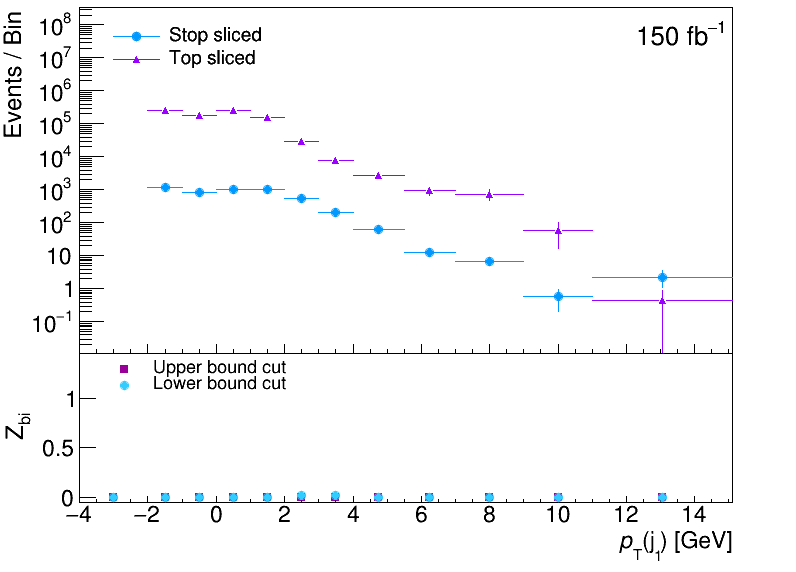}
\caption{$p_{\text{T}}(j_1)$} \label{fig:b}
\end{subfigure}

\medskip
\begin{subfigure}{0.48\textwidth}
\includegraphics[width=\linewidth]{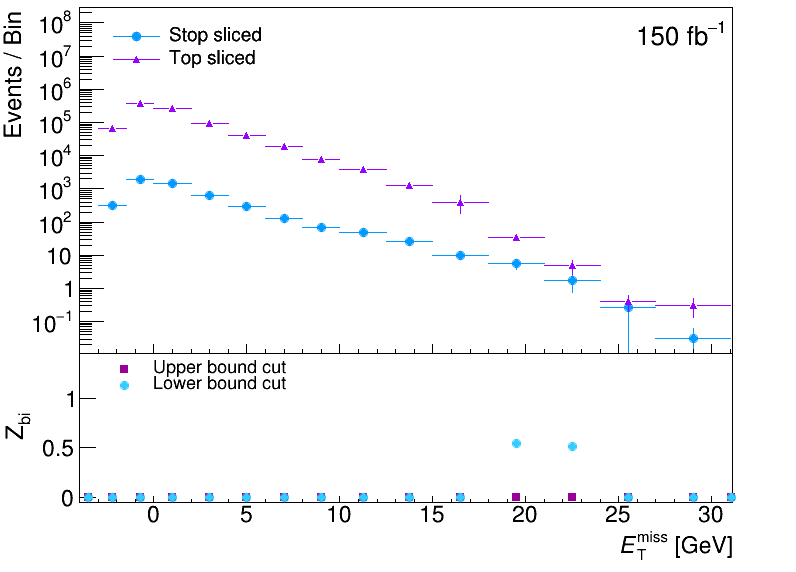}
\caption{$E_{\text{T}}^{\text{miss}}$} \label{fig:c}
\end{subfigure}\hspace*{\fill}
\begin{subfigure}{0.48\textwidth}
\includegraphics[width=\linewidth]{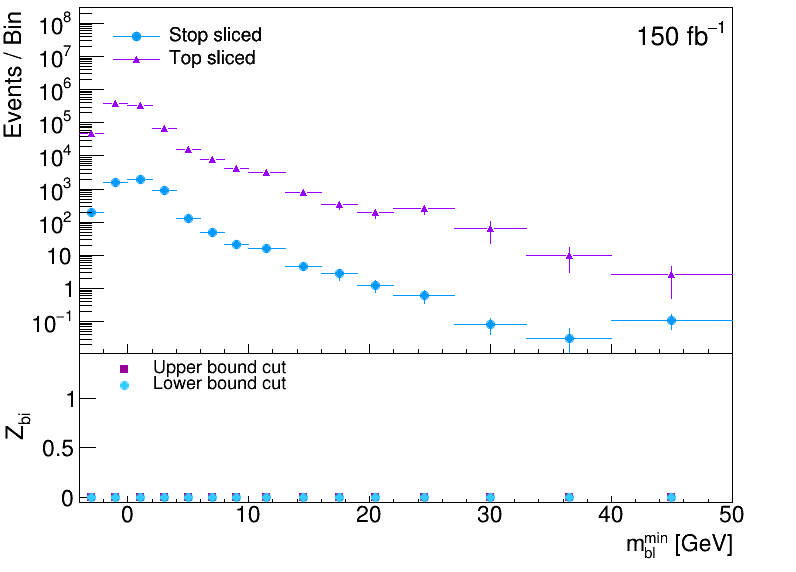}
\caption{$m_{bl}^{\text{min}}$} \label{fig:d}
\end{subfigure}

\medskip
\begin{subfigure}{0.48\textwidth}
\includegraphics[width=\linewidth]{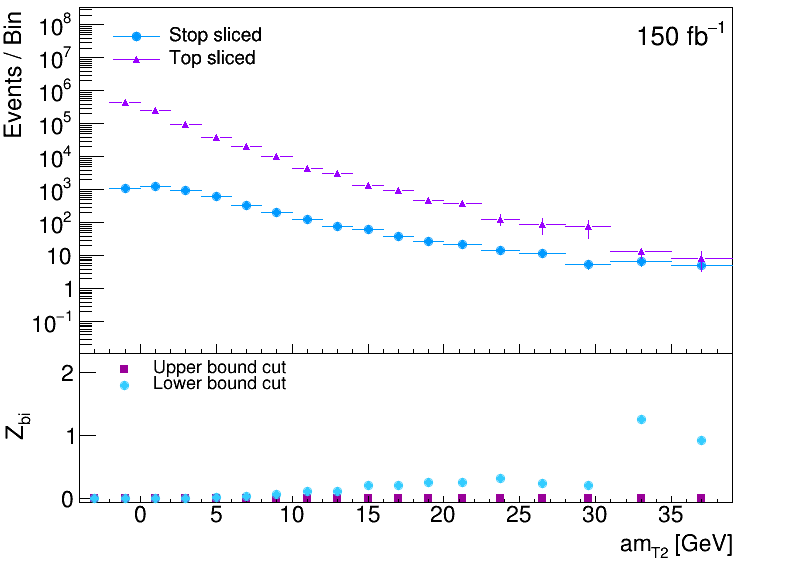}
\caption{$am_{\text{T2}}$} \label{fig:e}
\end{subfigure}\hspace*{\fill}
\begin{subfigure}{0.48\textwidth}
\includegraphics[width=\linewidth]{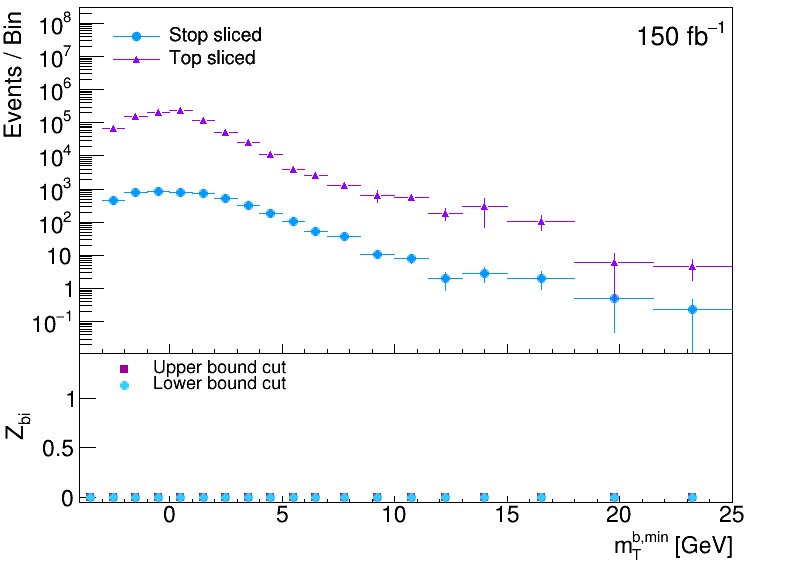}
\caption{$m_{\text{T}}^{b,\text{min}}$} \label{fig:f}
\end{subfigure}

\caption{Event rates as functions of the kinematic variables for the stop simplified model example that show the most difference between the signal and the background. Each variable has been scaled by the ``background median'' procedure described in the text. Events in the overflow bin are not shown in the distribution but are included in the $Z_{bi}$ calculation.} \label{fig:stopkin}
\end{figure}

\begin{figure}[htb!]
\begin{tabular}{cc}
%\begin{subfigure}[b]{0.45\textwidth}
%\includegraphics[width=\textwidth]{figures/braycurtis_stop.pdf}
%\centering
%\caption{}
%\label{1}
%\end{subfigure} &
%\begin{subfigure}[b]{0.4\textwidth}
%\includegraphics[width=\textwidth]{figures/canberra.pdf}
%\centering
%\caption{}
%\label{2}
%\end{subfigure}
%\begin{subfigure}[b]{0.45\textwidth}
%\includegraphics[width=\textwidth]{figures/chebyshev_stop.pdf}
%\centering
%\caption{}
%\label{3}
%\end{subfigure} \\
%\begin{subfigure}[b]{0.45\textwidth}
%\includegraphics[width=\textwidth]{figures/cityblock_stop.pdf}
%\centering
%\caption{}
%\label{4}
%\end{subfigure} &
\begin{subfigure}[b]{0.45\textwidth}
\includegraphics[width=\textwidth]{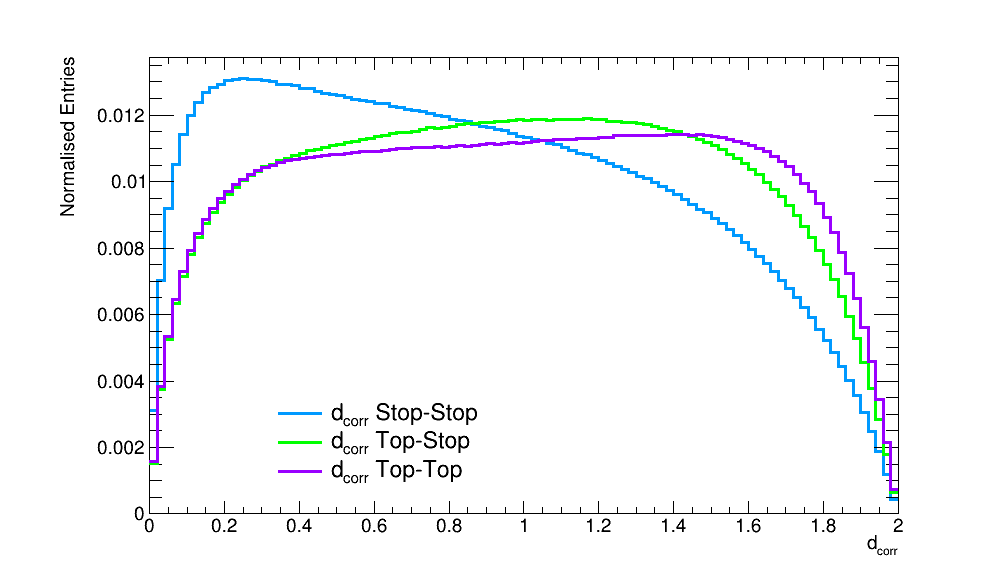}
\centering
\caption{correlation}
\label{5}
\end{subfigure} &
\begin{subfigure}[b]{0.45\textwidth}
\includegraphics[width=\textwidth]{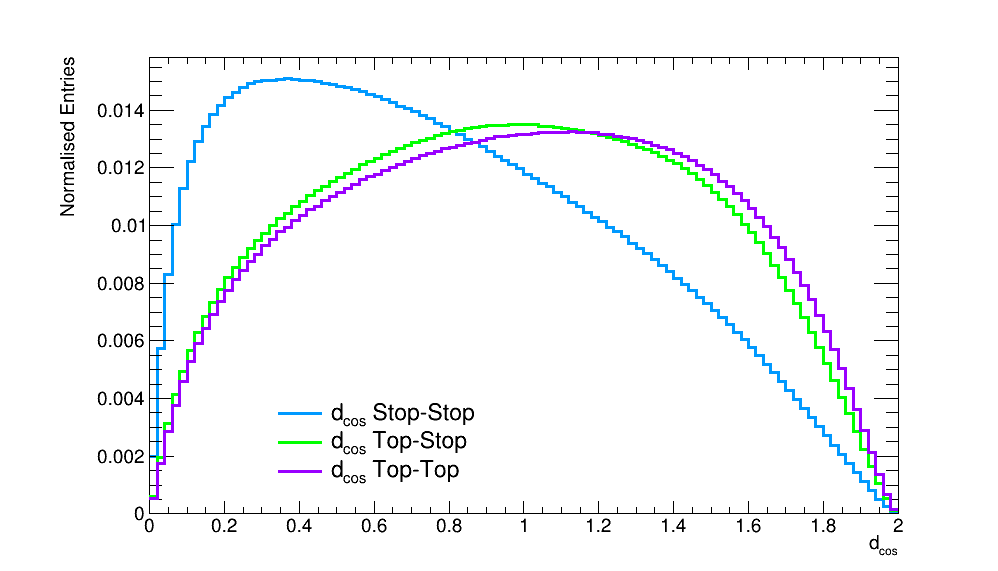}
\centering
\caption{cosine}
\label{6}
\end{subfigure} \\
\begin{subfigure}[b]{0.45\textwidth}
\includegraphics[width=\textwidth]{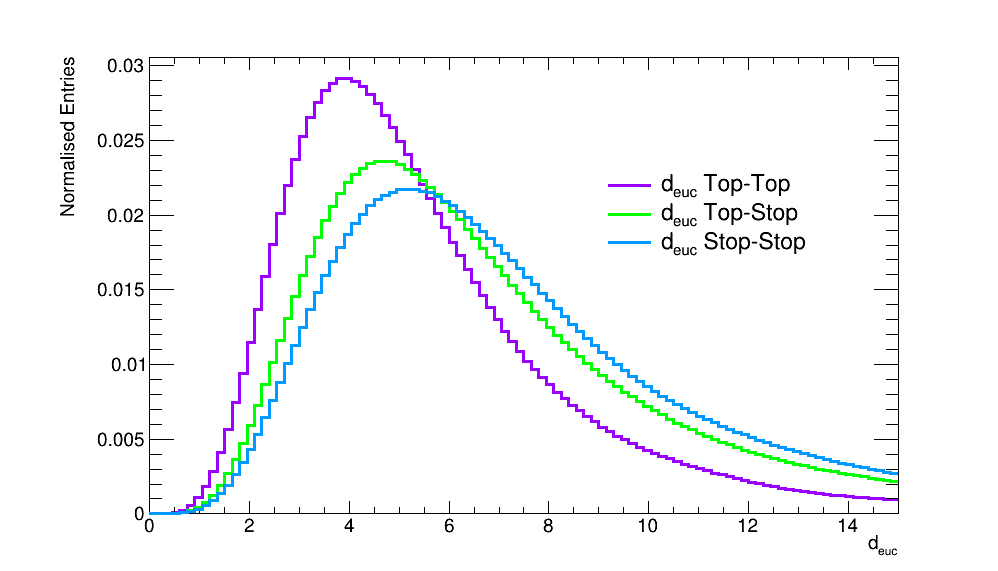}
\centering
\caption{Euclidean}
\label{7}
\end{subfigure} &
\begin{subfigure}[b]{0.45\textwidth}
\includegraphics[width=\textwidth]{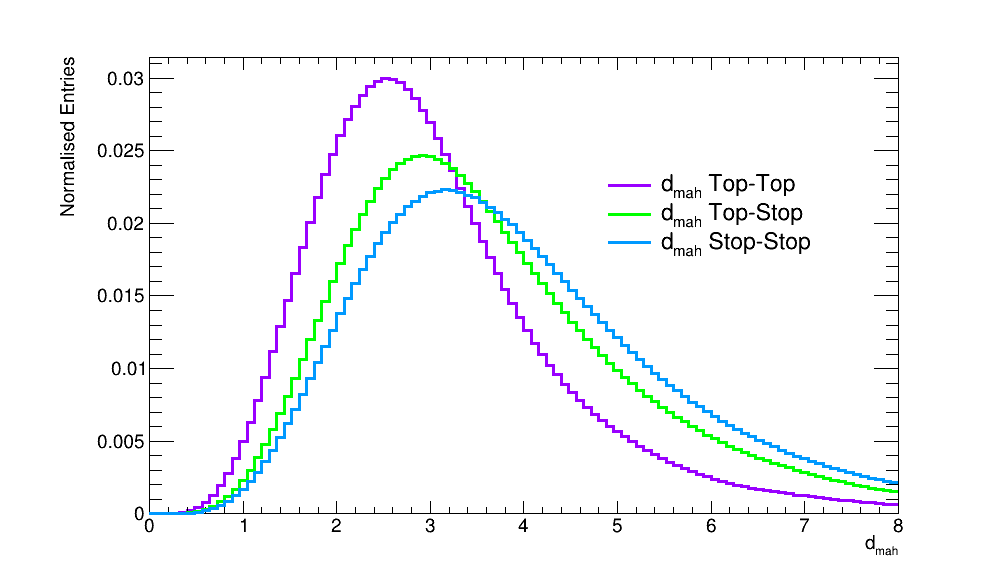}
\centering
\caption{Mahalanobis}
\label{8}
\end{subfigure} \\ 
\end{tabular}
\caption{\label{fig:stopdists} Distributions of the distance metrics we use in our prototype stop search, calculated using inclusive samples for simplicity.}
% We found that the distance distributions for the sliced samples (add to text) (used inclusive distances as a guide for the sliced distances, directly translates, avoid weighting the distance distributions ) 
\end{figure}

We show histograms of event rates as functions of these variables after the pre-selection in Figure~\ref{fig:stopkin}, with the samples scaled using the ``background median'' procedure defined previously. In all cases the signal lies substantially below the background across the entire range of the distribution. The variables are used in the calculation of our various distance metrics, exactly as we did in the previous section. Similar histograms of the distance metrics are provided for our stop example in Figure~\ref{fig:stopdists}. Based on these plots, we select the linking lengths given in Table~\ref{tab:stoplengths}, where we note that we have suppressed metrics that did not turn out to be useful for any local network metric. As in the electroweakino example, the correlation and cosine metrics have the signal being more concentrated at small distances than the background. We again retain the standard friendship condition when building our networks, where two nodes are linked by an edge if they are closer in distance than the linking length.

\FloatBarrier

\begin{table}[h!]
  \begin{center}
    \caption{Linking length values used for each distance metric for our prototype stop analysis.}
    \label{tab:stoplengths}
    \begin{tabular}{c|c} 
      \textbf{Distance metric} & \textbf{Linking length} \\
      \hline
      $d_{\text{corr}}$ & 0.7 \\
      $d_{\text{cos}}$ & 0.8 \\      
      $d_{\text{euc}}$ & 5.5 \\
      $d_{\text{mah}}$ & 4.0 \\
    \end{tabular}
  \end{center}
\end{table}

\begin{figure}[ht!] 

\begin{subfigure}{0.48\textwidth}
\includegraphics[width=\linewidth]{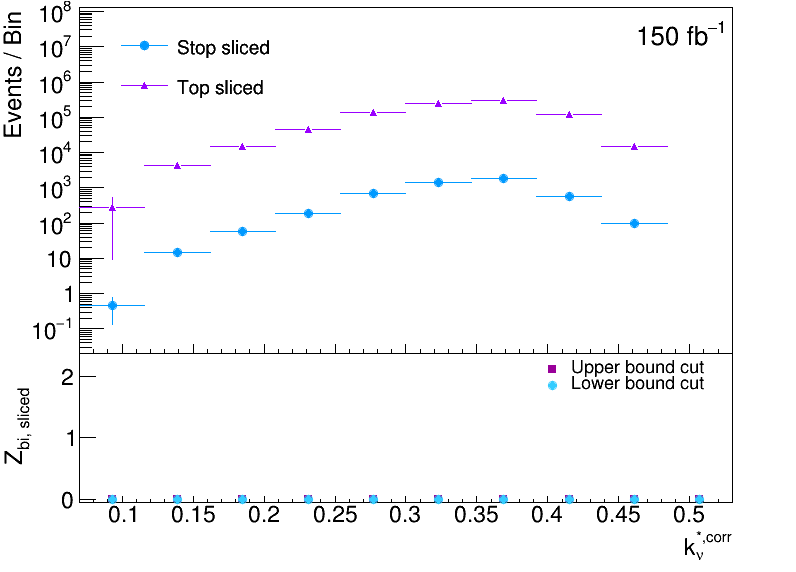}
\caption{$k_{\nu}^{*,\text{corr}}$} \label{fig:e}
\end{subfigure}\hspace*{\fill}
\begin{subfigure}{0.48\textwidth}
\includegraphics[width=\linewidth]{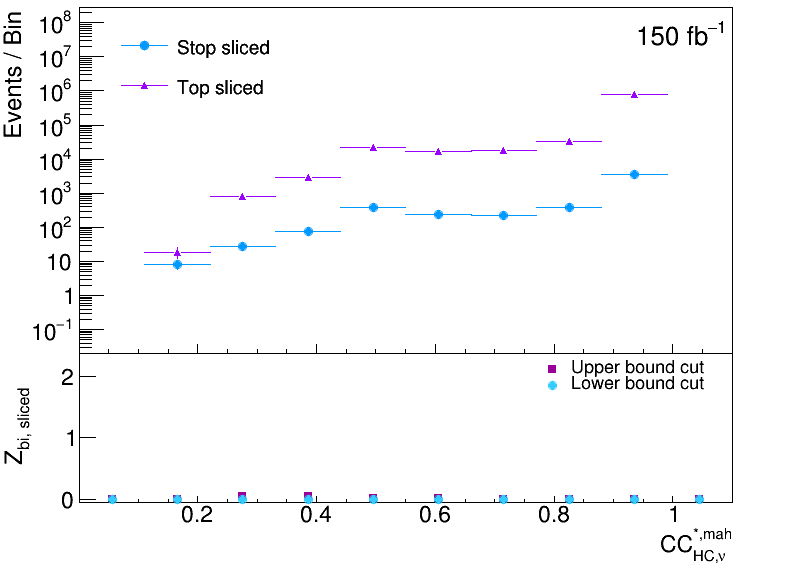}
\caption{$CC_{HC,\nu}^{*,\text{mah}}$} \label{fig:f}
\end{subfigure}

\medskip
\begin{subfigure}{0.48\textwidth}
\includegraphics[width=\linewidth]{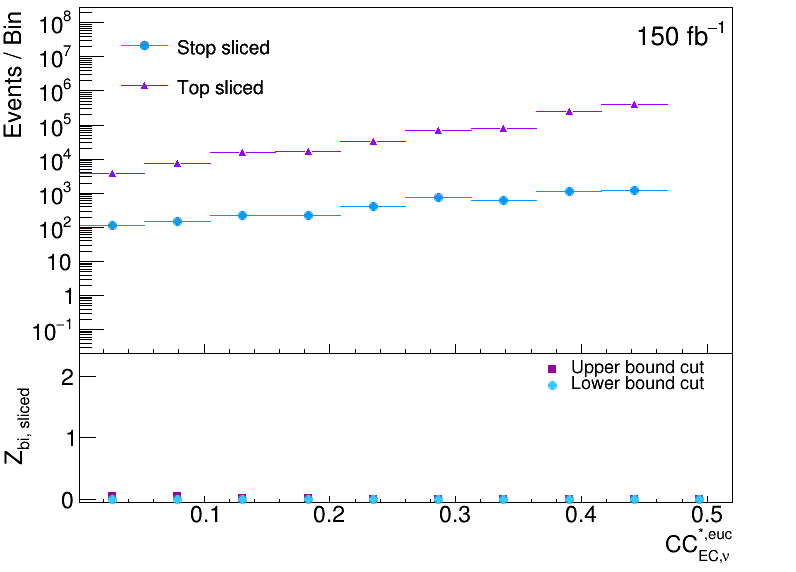}
\caption{$CC_{EC,\nu}^{*,\text{euc}}$} \label{fig:c}
\end{subfigure}\hspace*{\fill}
\begin{subfigure}{0.48\textwidth}
\includegraphics[width=\linewidth]{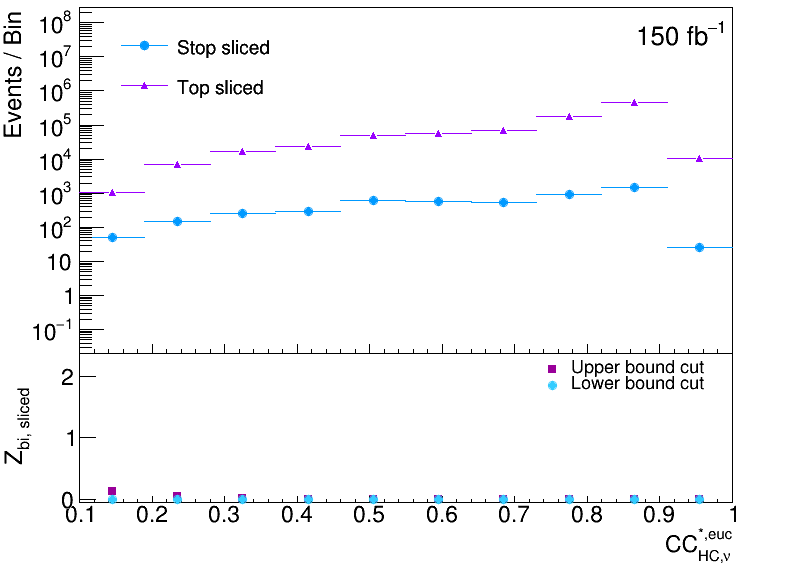}
\caption{$CC_{HC,\nu}^{*,\text{euc}}$} \label{fig:d}
\end{subfigure}

\medskip
\begin{subfigure}{0.48\textwidth}
\includegraphics[width=\linewidth]{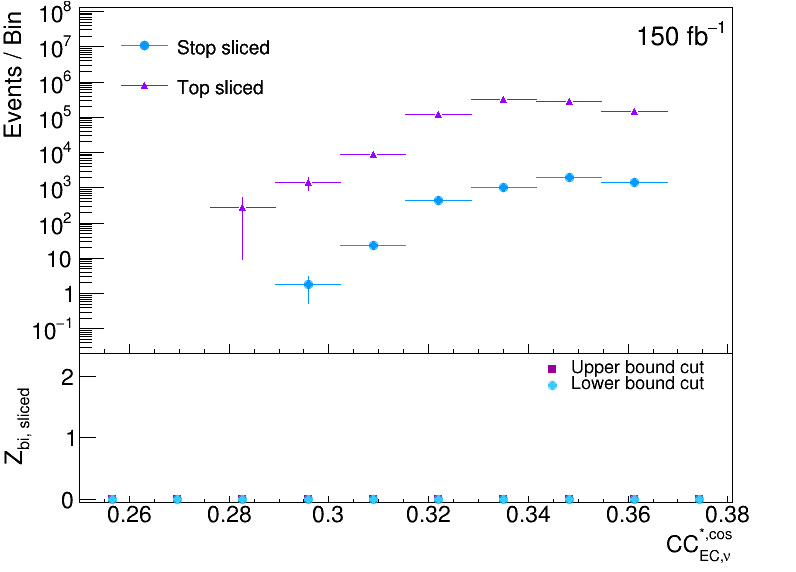}
\caption{$CC_{EC,\nu}^{*,\text{cos}}$} \label{fig:b}
\end{subfigure}
\begin{subfigure}{0.48\textwidth}
\includegraphics[width=\linewidth]{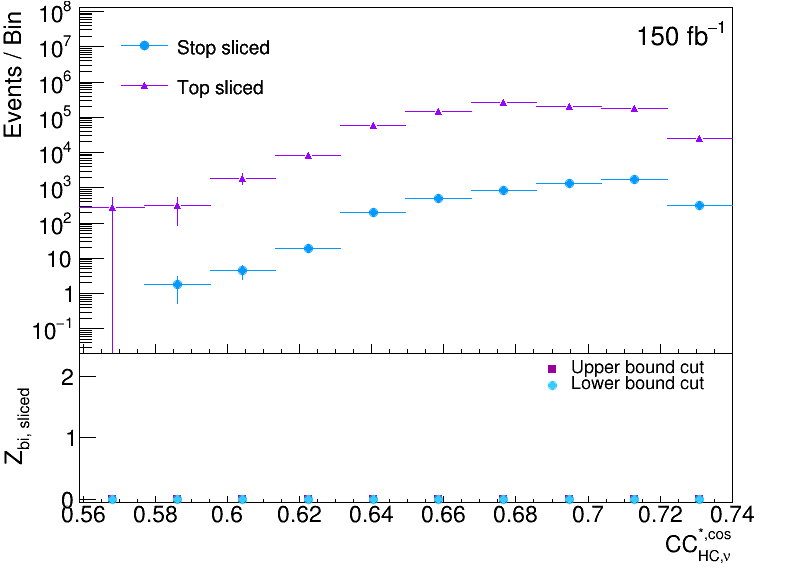}
\caption{$CC_{HC,\nu}^{*,\text{cos}}$} \label{fig:b}
\end{subfigure}

%\medskip
%\begin{subfigure}{0.48\textwidth}
%\includegraphics[width=\linewidth]{mahalanobis_nsi_degree_stop.png}
%\caption{$k_{\nu}^{*,\text{mah}}$} \label{fig:e}
%\end{subfigure}\hspace*{\fill}
%\begin{subfigure}{0.48\textwidth}
%\includegraphics[width=\linewidth]{correlation_nsi_closeness_stop.png}
%\caption{$CC_{C,\nu}^{*,\text{corr}}$} \label{fig:f}
%\end{subfigure}

\caption{Event rates as functions of the network variables for the stop simplified model example. Events in the overflow bin are not shown in the distribution but are included in the $Z_{bi}$ calculation.} \label{fig:stopnetwork}

\end{figure}

In Figure~\ref{fig:stopnetwork}, we show the network metrics that were used in the previous example, and that are considered robust under the theoretical reasoning described in Appendix~\ref{app:nsi}. Although the distributions show modest differences in shape in some cases, there is not enough separation of the signal and background distributions to render these particular local network metrics useful in stop searches. We found no combination of selections on these variables plus the original kinematic variables that gave sensitivity for exclusion at the LHC. It remains possible that different choices of the original kinematic variables used to build the network might change this picture, but it is clear that the use of local network metrics does not automatically give sensitivity to BSM physics signals. 

\begin{figure}[ht!] 
\begin{subfigure}{0.48\textwidth}
\includegraphics[width=\linewidth]{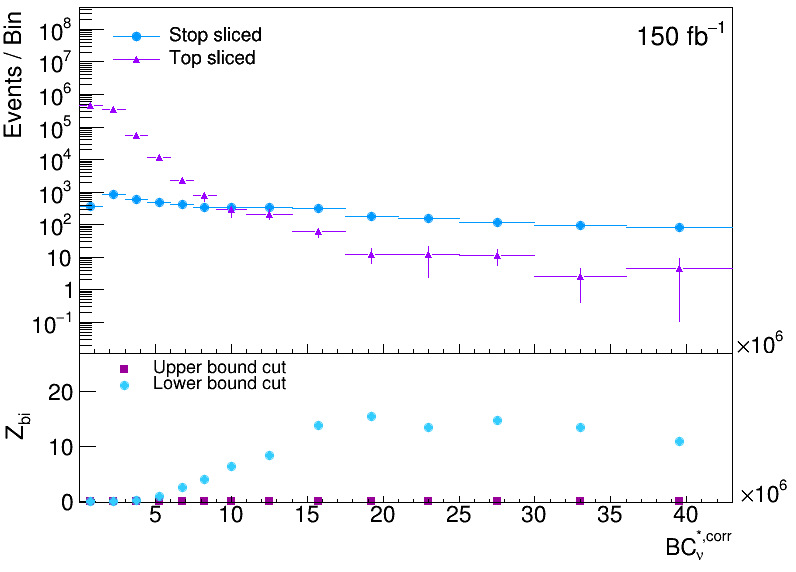}
\caption{$BC_{\nu}^{*,\text{corr}}$} \label{fig:a}
\end{subfigure}\hspace*{\fill}
\begin{subfigure}{0.48\textwidth}
\includegraphics[width=\linewidth]{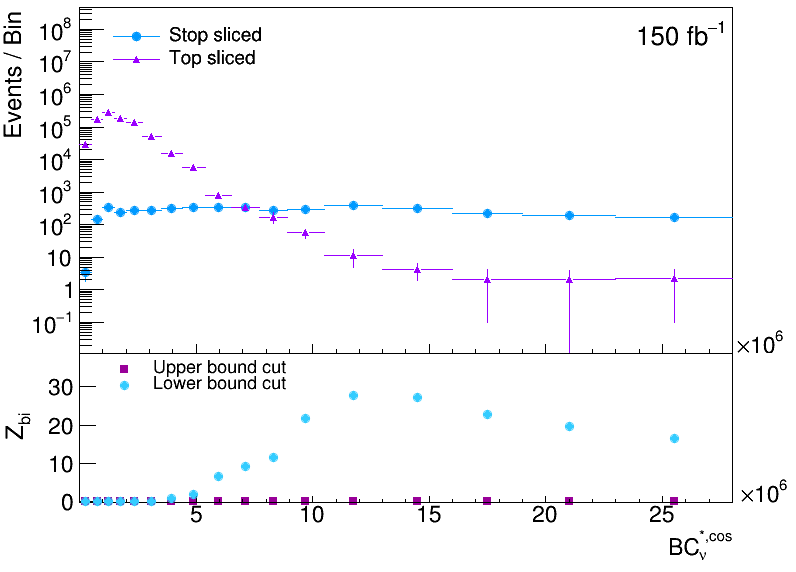}
\caption{$BC_{\nu}^{*,\text{cos}}$} \label{fig:b}
\end{subfigure}

\medskip
\begin{subfigure}{0.48\textwidth}
\includegraphics[width=\linewidth]{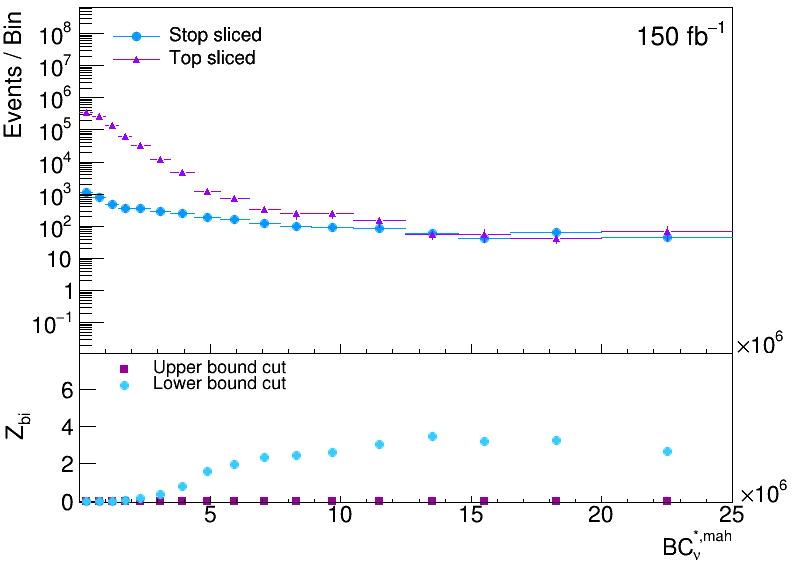}
\caption{$BC_{\nu}^{*,\text{mah}}$} \label{fig:c}
\end{subfigure}\hspace*{\fill}
\begin{subfigure}{0.48\textwidth}
\includegraphics[width=\linewidth]{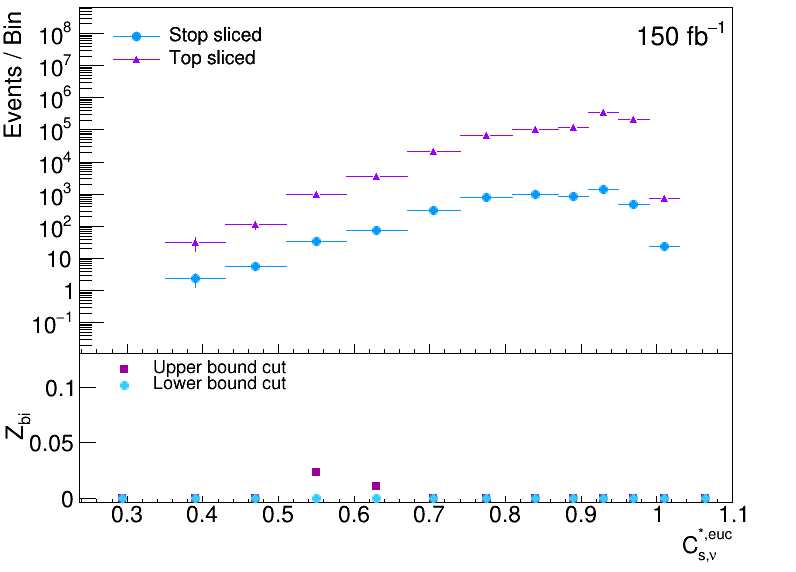}
\caption{$C_{s,\nu}^{*,\text{euc}}$} \label{fig:d}
\end{subfigure}

\medskip
\begin{subfigure}{0.48\textwidth}
\includegraphics[width=\linewidth]{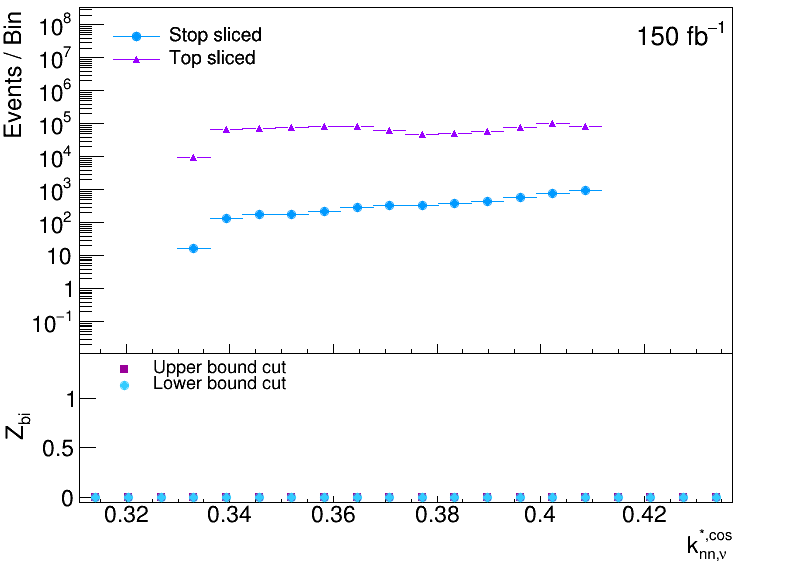}
\caption{$k_{nn,\nu}^{*,\text{cos}}$} \label{fig:e}
\end{subfigure}\hspace*{\fill}
\begin{subfigure}{0.48\textwidth}
\includegraphics[width=\linewidth]{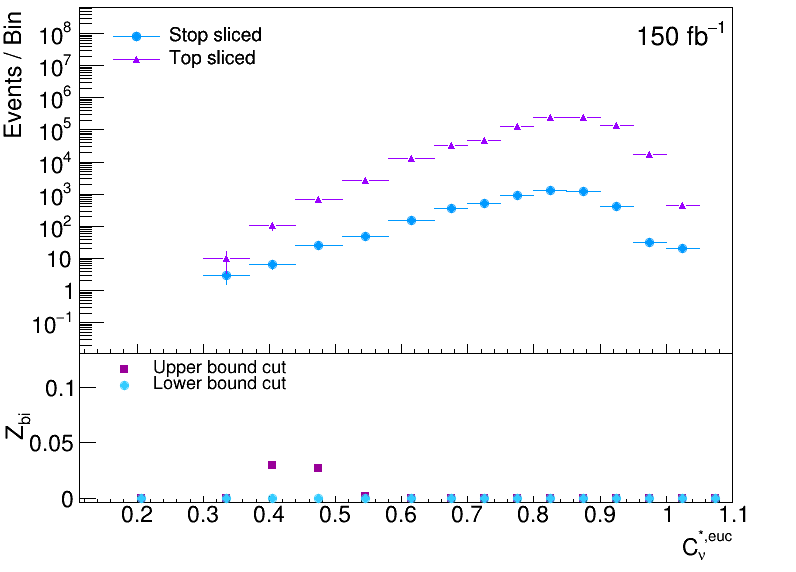}
\caption{$C_{\nu}^{*,\text{euc}}$} \label{fig:f}
\end{subfigure}

\caption{Event rates as functions of the network variables for the stop simplified model example that show the most difference between the signal and the background. Events in the overflow bin are not shown in the distribution but are included in the $Z_{bi}$ calculation.} \label{fig:stopnetworkbetweenness}

\end{figure}

Before concluding this section, it is worth noting that some network metric distributions show much greater signal-background separation (see Figure~\ref{fig:stopnetworkbetweenness}). For example, the betweenness measures typically fall off much faster for top events than for stop events. This is particularly true for networks built using the correlation and cosine distance metrics, for which the betweenness centrality distributions for the background do not show evidence of a flatter tail which is apparent in the case of the Mahalanobis betweenness. More modest discrimination comes from the Euclidean local clustering, the Euclidean local soffer clustering and the cosine average neighbors degree. Although we currently recommend caution on the use of these metrics (due to the potential violation of the n.s.i external connectivity assumption detailed in Appendix~\ref{app:nsi}), we feel that further investigation of the robustness of these metrics in LHC use cases is a strong avenue for future work.

%Whilst single cuts on the correlation or cosine betweenness centrality variables would seem to be the best motivated approach for exclusion and discovery, we also found optimum exclusion in the case that several variables are used at the same time. We followed a similar procedure to that of the previous section, with the exception that we modified the criterion on the number of events that must remain after applying all selections to take into account the change in the event weights. Table~\ref{tab:Networks:Stop:zns-disco} shows the analyses that delivered the highest binomial significance values.

\FloatBarrier

\FloatBarrier

\section{Discussion}
\label{sec:discussion}
For our electroweakino example, we have shown that the network variables provide exclusion potential for a benchmark point that has just been excluded by a three lepton ATLAS search performed using 139 fb$^{-1}$ of data recorded by the ATLAS experiment. %, whilst for the stop case we find that the metrics investigated do not provide any advantage over existing techniques. 

%we provide comfortable exclusion of a benchmark point in the 1 lepton final state that has just been excluded by a zero lepton ATLAS search utilising 139 fb$^{-1}$ of data.

So far, we have compared the efficacy of the network variables with the original kinematic variables by comparing prototype cut-and-count analyses, and by discussing the shape of the various variables. In order to more comprehensively compare the effectiveness of the graph network variables to other common search techniques, the sensitivity of a Boosted Decision Tree (BDT) is considered. We examine the electroweakino case, where a BDT is trained using the same kinematic variables as those used to define the distance metrics. Additionally, the same preselection is applied to the events as for the graph network.

\begin{figure}[h!]
\centering
\includegraphics[width=0.8\textwidth]{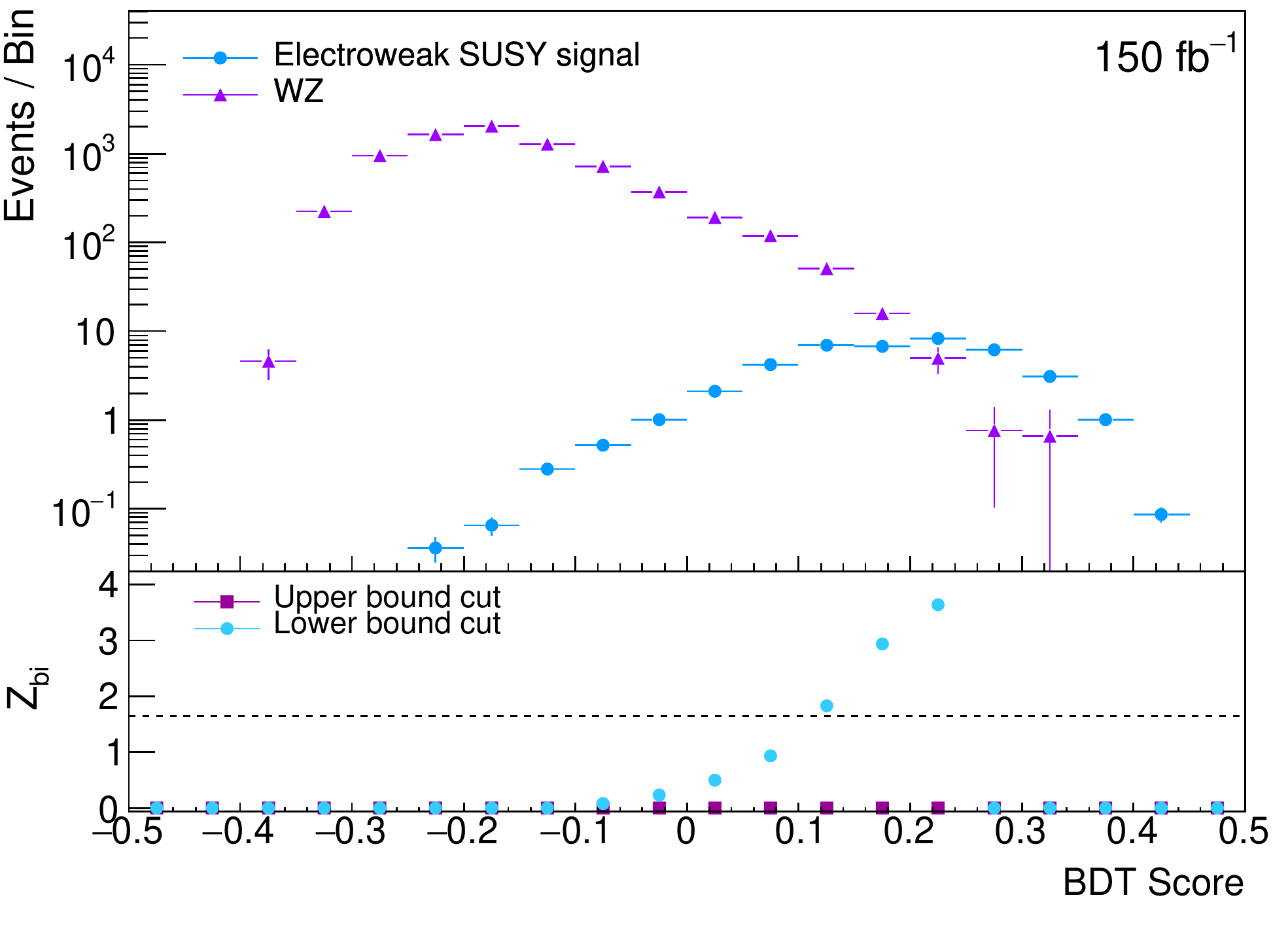}
\caption{\label{fig:bdtew} Distributions of the BDT score for electroweakino signal and WZ background events, in the case where the BDT is only trained on kinematic variables. In the lower panel the $Z_{bi}$ is also shown for cumulative upper or lower bound cuts on the score. Events in the overflow bin are not shown in the distribution but are included in the $Z_{bi}$ calculation. Note that we deliberately set $Z_{bi}$ to zero in the lower figure when the number of signal or background events falls below three (before weighting).}
\end{figure}

The ROOT package TMVA is used for the BDT training~\cite{hoecker2007tmva}, using its default settings.
To avoid a drop in sample size, the SUSY signal and $WZ$ background samples are randomly split into two sets, such that two BDTs are produced, each trained on one set and tested on the other.
The BDT output distribution for the full samples is shown in Figure~\ref{fig:bdtew}, along with the values of $Z_{bi}$ for upper and lower cuts on the BDT output variable, akin to the procedure used in the main body Figures.
The resulting max $Z_{bi}$ is $3.63$, which is slightly higher than the network variable cuts proposed in Table~\ref{tab:Networks:Electroweakinos:zns}.
Whilst the BDT performs admirably to exclude the signal that it was trained on, one would expect the BDT to do a worse job than the graph network at generalising to other signal points not trained on.

\begin{figure}[h!]
\centering
\includegraphics[width=0.8\textwidth]{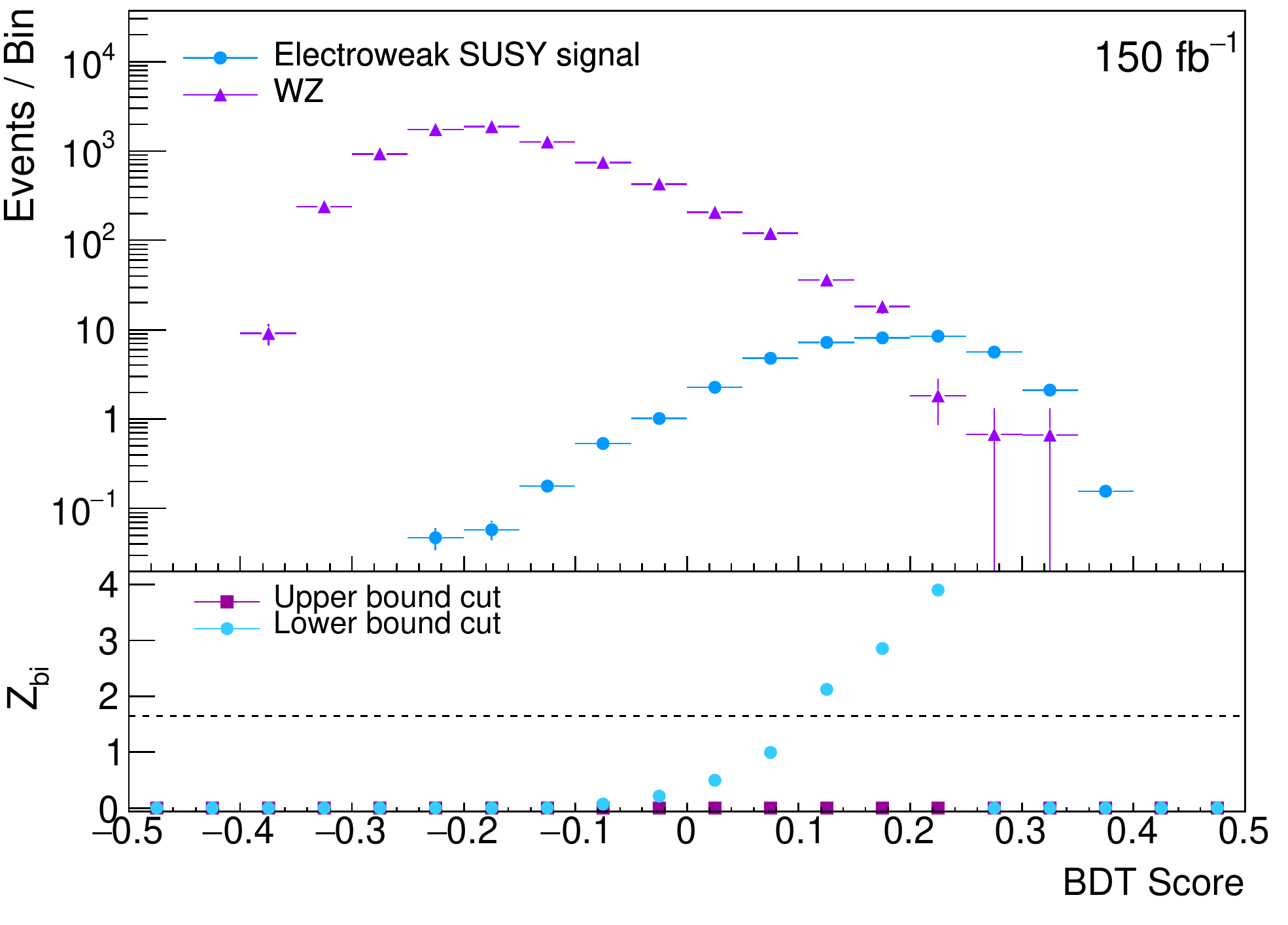}
\caption{\label{fig:bdtew_network} Distributions of the BDT score for electroweakino signal and WZ background events, in the case where \degree{euc} is added to the BDT. In the lower panel the $Z_{bi}$ is also shown for cumulative upper or lower bound cuts on the score. Events in the overflow bin are not shown in the distribution but are included in the $Z_{bi}$ calculation. Note that we deliberately set $Z_{bi}$ to zero in the lower figure when the number of signal or background events falls below three (before weighting).}
\end{figure}

It is also interesting to study whether the BDT performance can also be improved using the graph network variables investigated in the main body.
The same setup as described for the first BDT is used, with the addition of only one promising network variable: \degree{euc}.
The BDT output in this case can be seen in Figure~\ref{fig:bdtew_network}, where a clear improvement in $Z_{bi}$ is observed, reaching up to $3.98$. 
 
There is much remaining to be explored in future work, starting with the fact that our study is currently limited by the computational complexity of local network metric calculations for large networks. As more signal and background events are considered, the number of edges in the network grows to be very large, hampering our current parallelised calculations in the {\tt pyunicorn} package.  It would also be interesting to study how our prototype analyses change when choosing different linking lengths for each distance metric when defining the graph networks. A significant reduction in computational complexity could be achieved with shorter linking lengths, since this would make all of the networks more sparse. The cost, however, is that it will be harder to discriminate background from signal events, since both signal and background events will often have a low degree in the network. 

%The betweenness centrality is the most expensive calculation, and we note that there is room for both a smarter parallelisation strategy, and the use of fast approximations such as the graph neural network approach presented in Ref.~\cite{2019arXiv190510418F}.

Our graph network technique also offers the possibility of unsupervised event detection in the LHC data, since one can look for local network metric shapes that differ from those expected in a purely Standard Model network. A similar analysis would also be possible with \emph{global} network metrics. In both cases, this is only feasible if the number of events in a given final state in the full LHC dataset is small enough that network computations can be performed within a reasonable timescale. In the case of the ``supervised'' analyses presented above, one of the functions of the pre-selection is to reduce the number of events that must be processed to a manageable level, with the consequence that some model-dependence is introduced through the choice of pre-selection. Fast approaches to calculating n.s.i local network metrics would, however, provide a very powerful way to search for new physics in an agnostic fashion, complementing previous signal-model-independent approaches such as those presented in Refs.~\cite{DeSimone:2018efk,DAgnolo:2018cun,Farina:2018fyg,Heimel:2018mkt,Hajer:2018kqm,Kuusela:2011aa,Aaltonen:2007dg,CMS:2008gya,ATLAS:2017irs,Choudalakis:2011qn,Abbott:2000fb,Aktas:2004pz,Aaron:2008aa,Asadi:2017qon,Aaltonen:2008vt,CMS:2011fra,Cerri:2018anq,Blance:2019ibf,Roy:2019jae} (plus further techniques for resonances presented in Refs.~\cite{Collins:2018epr,Collins:2019jip} that do not rely on a background model).

\section{Conclusions}
\label{sec:conclusions}
We have shown that graph network analysis can be used to define powerful variables for discriminating signal and background processes at the LHC. By building graph networks using a variety of distance metrics, we have developed prototypical LHC analyses that use local metrics of the graphs as new event attributes, alongside the original kinematic variables that the networks were built from. This permits the mapping of the topological structure, in kinematic space, of different contributions to the LHC dataset.

The results indicate that the graph network technique offers comfortable exclusion potential for an electroweakino production scenario that is hard to exclude using conventional kinematic variables. Furthermore, the use of network metrics improved or exceeded the performance of a BDT trained on only the original kinematic variables. At the very least, our results indicate that the technique provides a promising alternative to current search methods, motivating a deeper analysis that could explore different distance metrics, linking lengths and pre-selections. Better scalability of the technique will require faster computational approaches for node-splitting invariant network metric calculations.

In the case of a stop production scenario in which the kinematics of the signal process closely resemble those of the top background, the metrics that we argue to be robust under the assumptions of node-splitting invariance did not prove to be sensitive to exclusion of the scenario. However, betweenness centrality measures calculated with the correlation, cosine and mahalanobis distances show much flatter distributions for the signal than for the background, which might be exploited in future work. This would require either that the robustness of these metrics under the external connectivity assumption of node splitting invariance could be verified, or that alternative forms of the metrics could be derived that account for violations of this assumption. 

%% Acknowledgments
%\begin{acknowledgments}
\acknowledgments
This work is partly supported by the Australian Research Council Discovery Project DP180102209. Martin White would like to thank Jobst Heizig and Jonathan Donges for helpful advice regarding the {\tt pyunicorn} package, and on the implementation and behaviour of the n.s.i betweenness centrality. Our perspective on unsupervised LHC searches has benefitted from interaction with the {\tt Dark Machines} research collective (\url{https://darkmachines.org}).

%\end{acknowledgments}

%%%%%%%%%%%%%%%%%%%%%%%%%%%%%%%%%%%%%%%%%%%%%%%%%%%%%%%%%%%%%%%%%%%%%%%%
%######################################################################%
%#                        APPENDICES                                  #%
%######################################################################%
%%%%%%%%%%%%%%%%%%%%%%%%%%%%%%%%%%%%%%%%%%%%%%%%%%%%%%%%%%%%%%%%%%%%%%%%

%% Appendices (use * to suppress numbering for a single appendix)
\appendix

\section{Justifying the robustness of n.s.i. network metrics}

\label{app:nsi}
We have performed a variety of checks that the n.s.i network metrics used in our study are indeed safe under reweighting the events. First, in \ref{emptest} we study the behaviour of network variables when different-sized MC samples are used to construct networks with different node-weight distributions. Second, in \ref{theotest}, theoretical consideration is given to the simplifying assumptions involved in using n.s.i. network metrics, their applicability to the present example (\ref{theotest1}), and whether this could distort the network metric distributions (\ref{theotest2}).

\subsection{Empirical Tests}
\label{emptest}

First, the behaviour of the network variables was studied when different numbers of MC background events are used to generate the signal-plus-background network. This is designed to test empirically on the MC simulated data that the crucial concept of \emph{node splitting invariance} is indeed robust.

\begin{figure}[htb!]
\begin{tabular}{cc}
\begin{subfigure}[b]{0.45\textwidth}
\includegraphics[width=\textwidth]{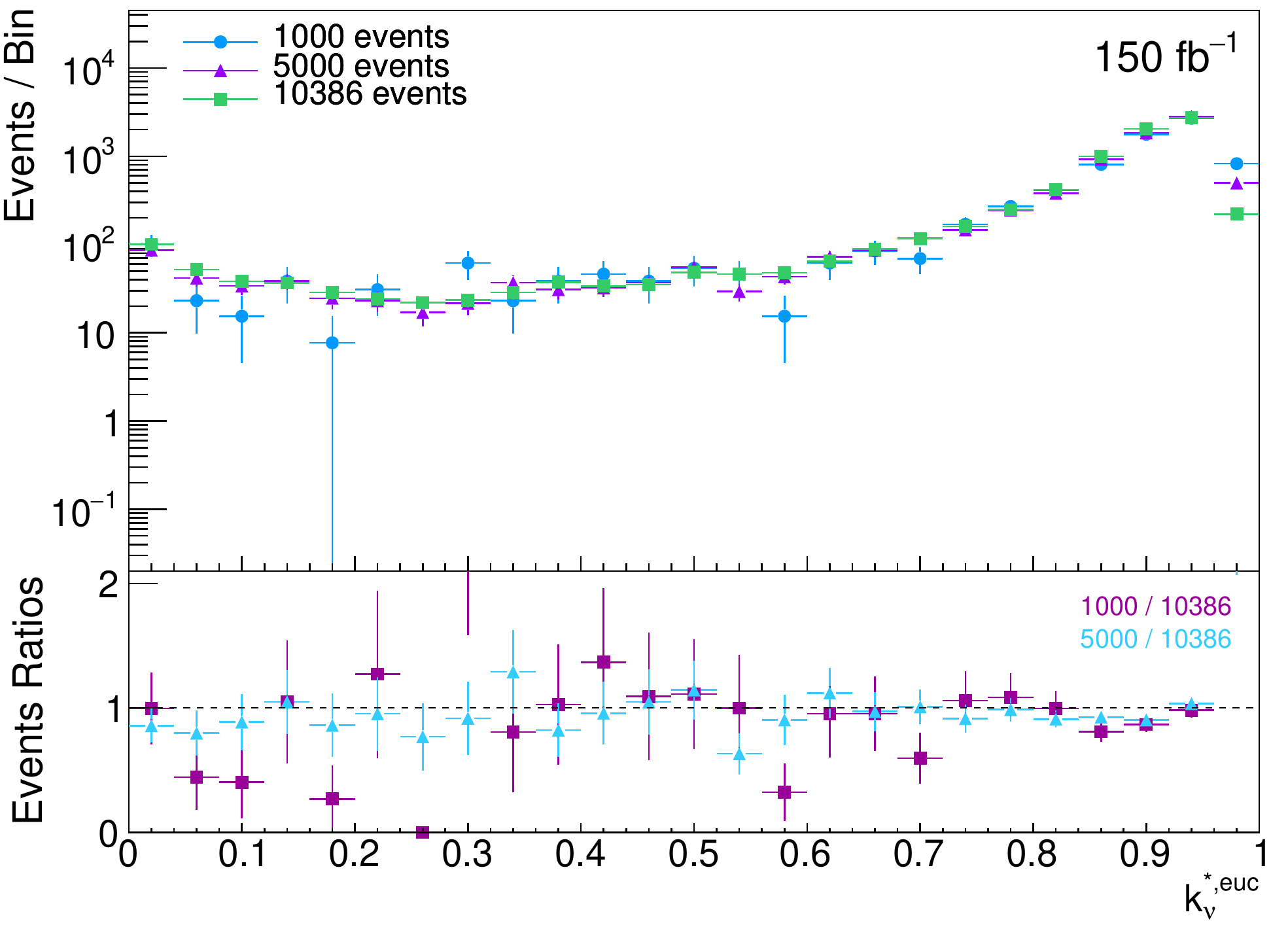}
\centering
\caption{\degree{euc}}
\label{11}
\end{subfigure} &
\begin{subfigure}[b]{0.45\textwidth}
\includegraphics[width=\textwidth]{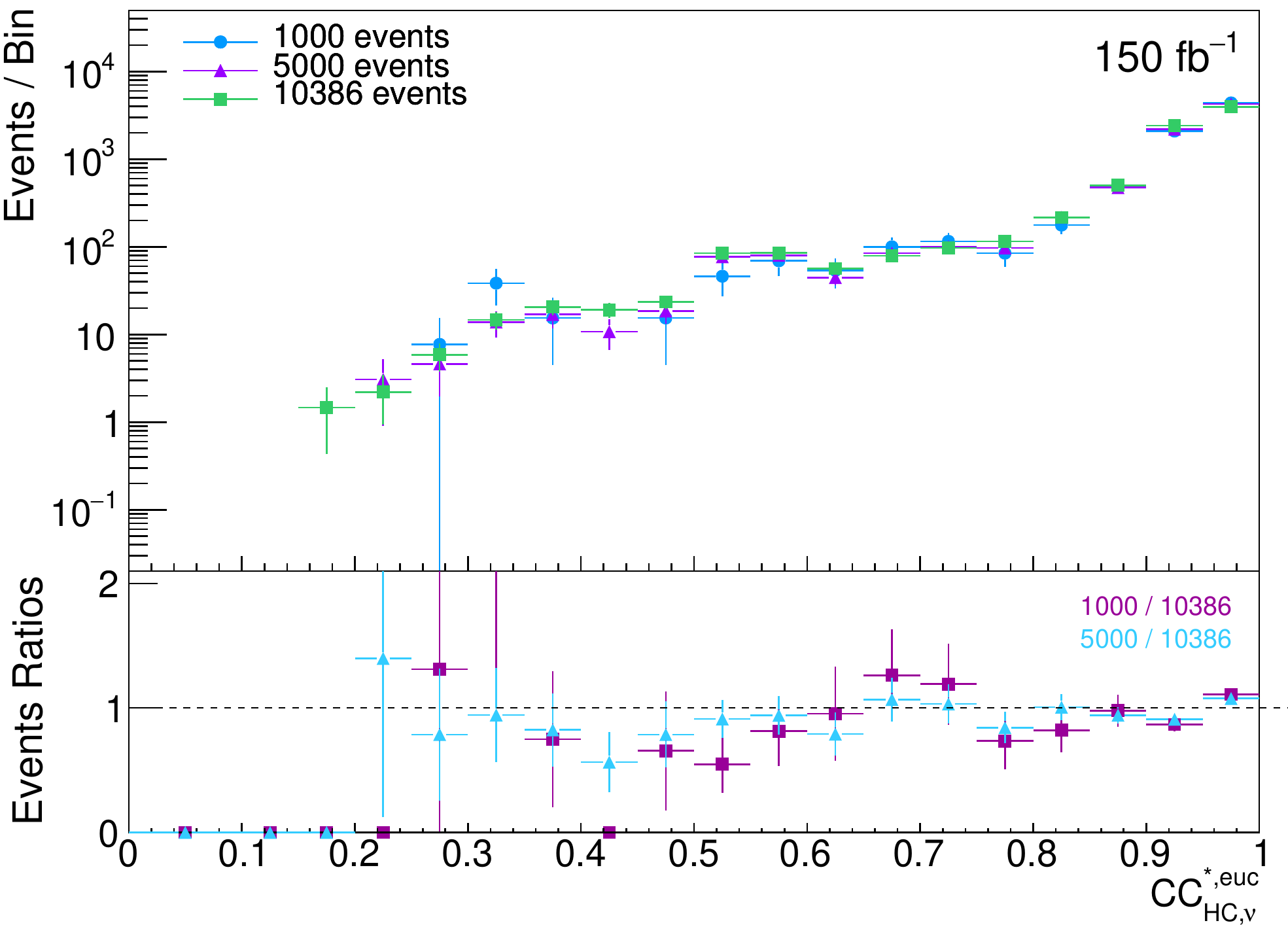}
\centering
\caption{\HC{euc}}
\label{14}
\end{subfigure} \\
\begin{subfigure}[b]{0.45\textwidth}
\includegraphics[width=\textwidth]{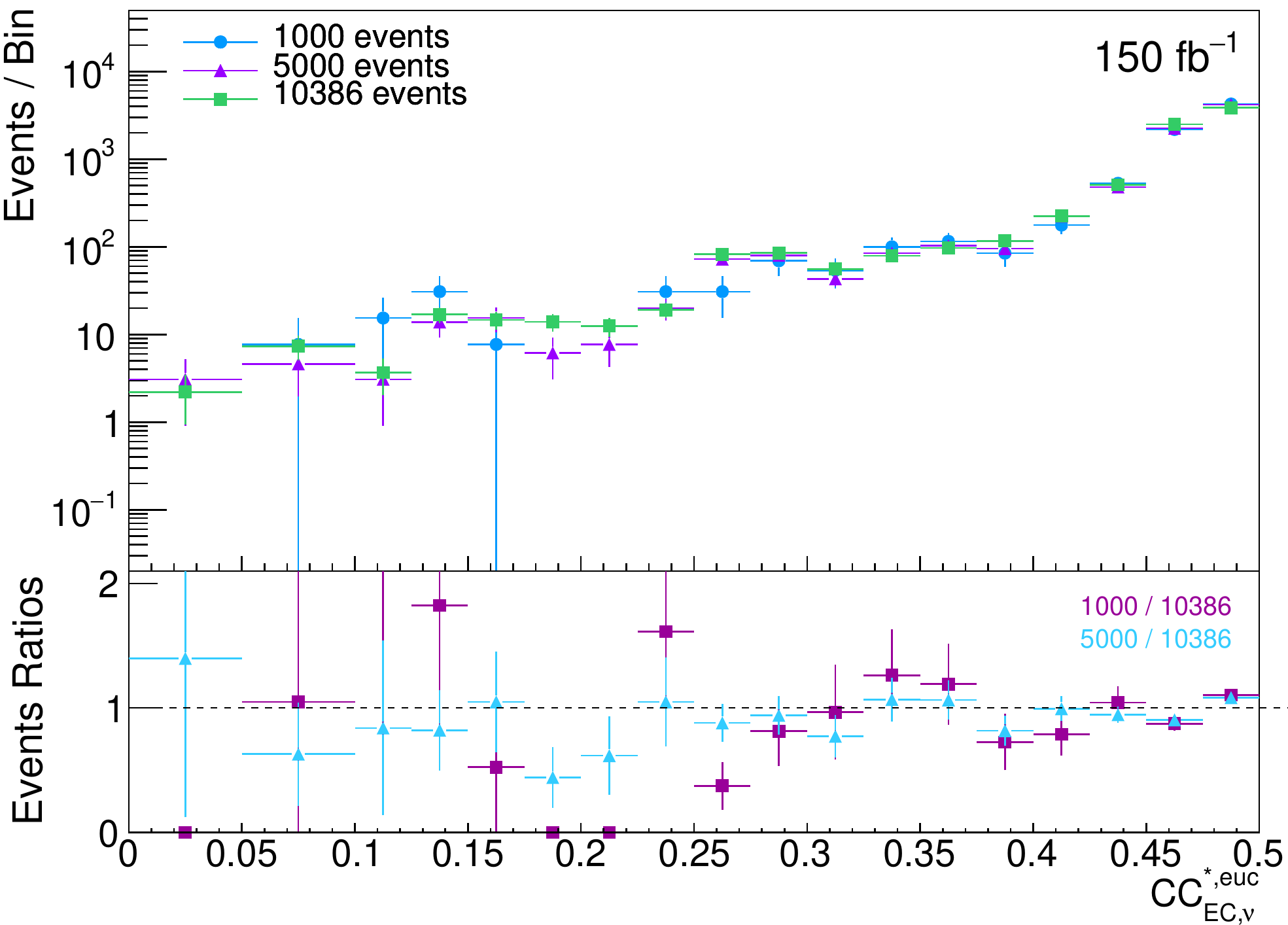}
\centering
\caption{\EC{euc}}
\label{15}
\end{subfigure} &
\begin{subfigure}[b]{0.45\textwidth}
\includegraphics[width=\textwidth]{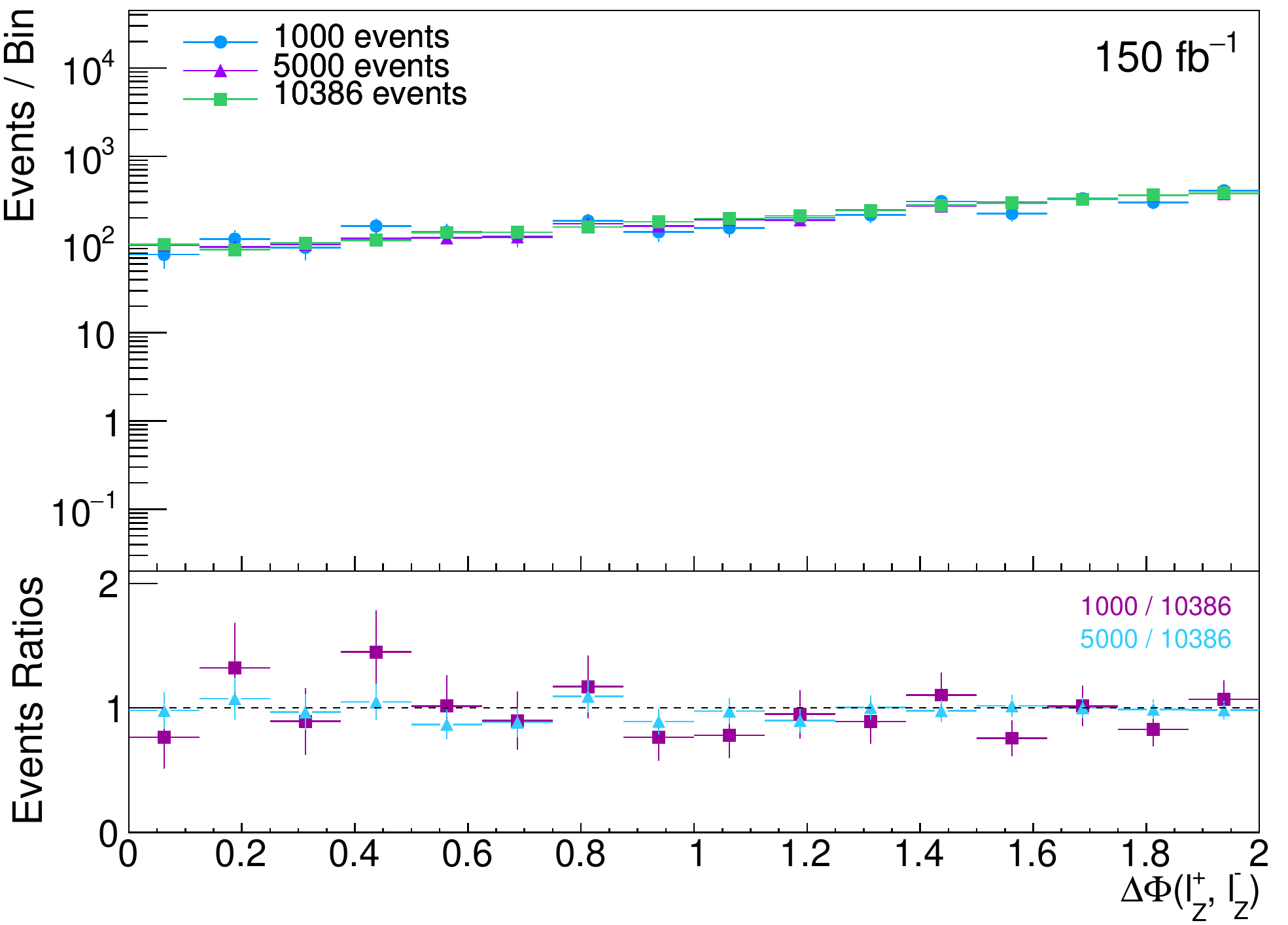}
\centering
\caption{\dPhiZlZl}
\label{110}
\end{subfigure} \\
\end{tabular} 
\caption{\label{fig:appEWtails:numbg}Event rates for the electroweakino case study as a function of the network metrics calculated using $d_{\text{euc}}$, calculated using different background sample sizes. Events in the overflow bin are not shown in the distribution.}
\end{figure}

We first show local network metrics for the electroweakino case with 11,197 signal events and 10,486 background events --- the full number of background events with a three lepton final state obtained from an inclusive generated set of 1,000,000 events (where ``inclusive'' means that no slicing was applied). In Figure~\ref{fig:appEWtails:numbg} distributions for some example network variables calculated from these samples are shown, in comparison to those calculated from networks built from samples using 1,000 or 5,000 background events.
From these plots it is clear that larger event samples produce less sparse tails and smoother distributions, however the bulk shapes of the distributions remain the same. 

\FloatBarrier

%%%%%%%%%%%%%%%%%%%%%%%%%%%%%%%%%%%%%%%%%%%%%%%%%%%%%%%%%%%%%%%%%%%%%%%%%%%%%%%%%%%%%%%%%%%%%%%%%%%%%%%%%%%%%%%%%%%%%%%%%%%%%%%%
Having investigated node-splitting invariance for inclusive MC events, we now move on to assess its validity on sliced events. The slicing procedure we describe in Section~\ref{sec:mc} not only ensures that the tail regions in the original kinematic variables are reliable, but also increases the statistical strength of the tails in the network metric distributions. The improvement is clear when we compare the network metric distributions after slicing our samples in \HT\ with the same metrics calculated from inclusive samples. The slicing procedure improves the modelling of the tails of the existing shapes of the inclusive network distributions, whilst giving a similar number of events when one integrates the tail (in the inclusive samples, the distributions stop early and have a sizable fluctuation in the final bin). 

\begin{figure}[ht!] 
\begin{subfigure}{0.48\textwidth}
\includegraphics[width=\linewidth]{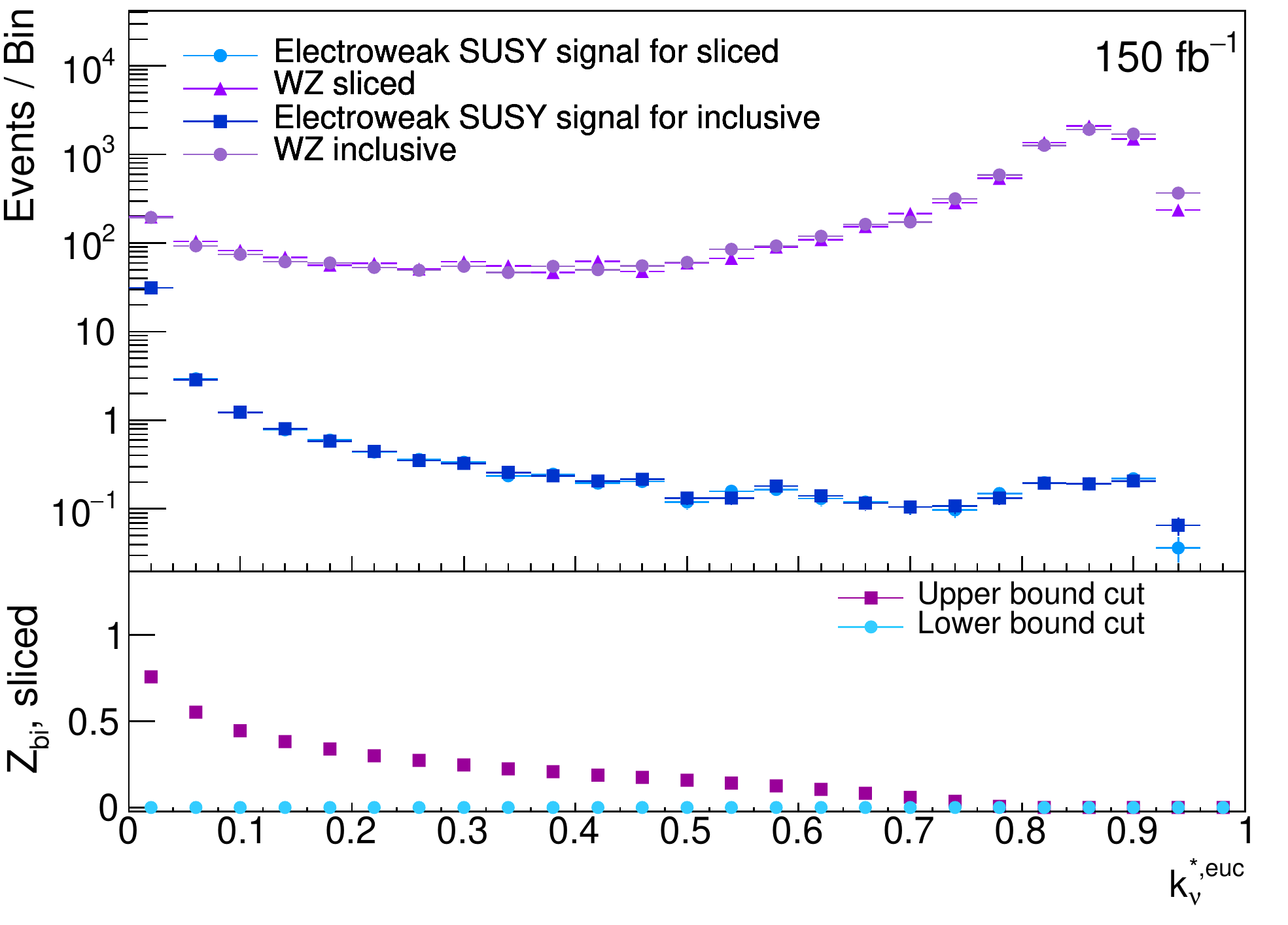}
\caption{\degree{euc}} \label{fig:ewoverlay:a}
\end{subfigure}
\begin{subfigure}{0.48\textwidth}
\includegraphics[width=\linewidth]{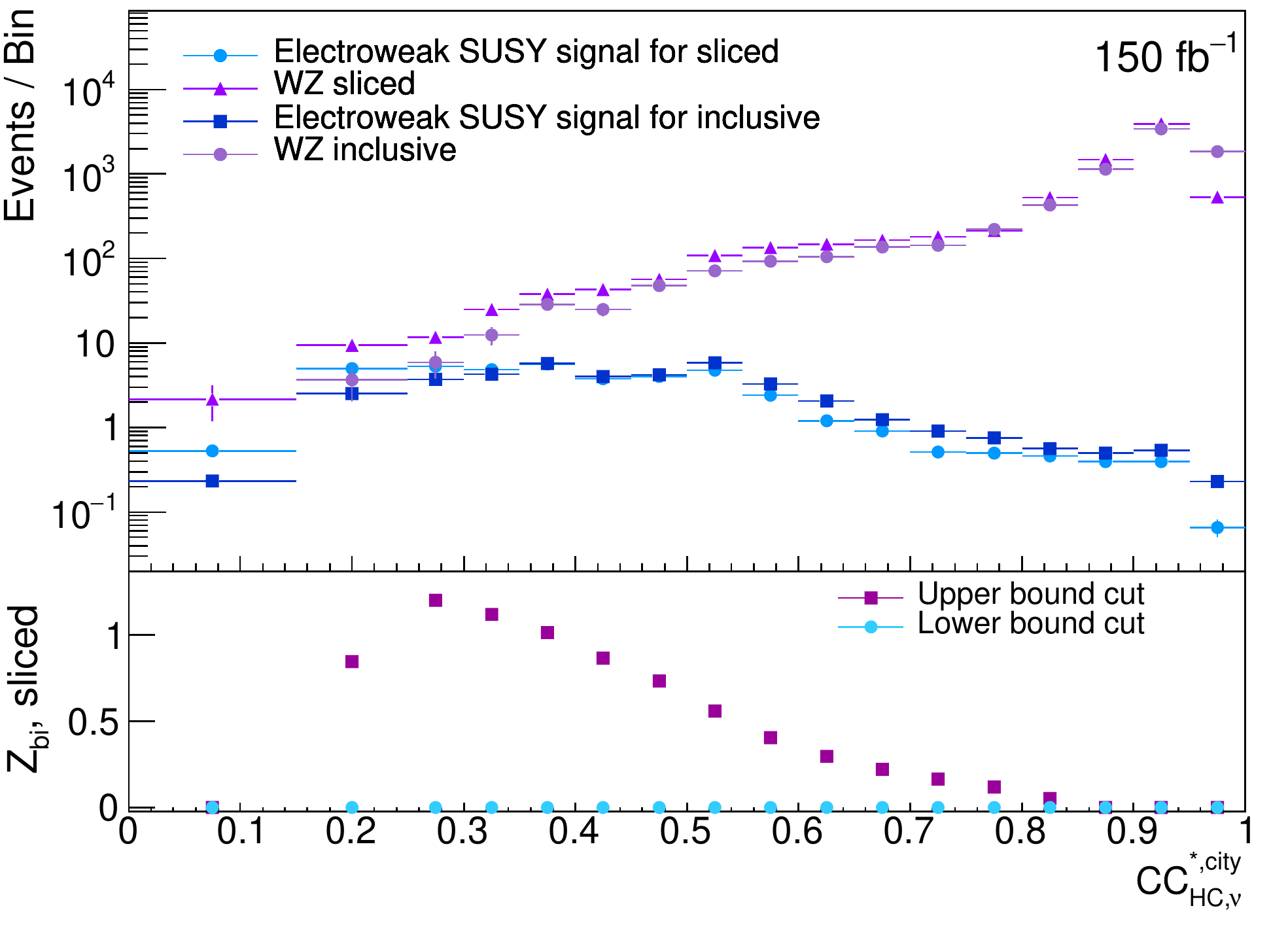}
\caption{\LC{euc}} \label{fig:ewoverlay:b}
\end{subfigure}
\caption{Event rates as functions of \degree{euc} and \HC{city} for the electroweakino simplified model example, comparing the network variables calculated from inclusive $WZ$ samples with those calculated from $WZ$ samples sliced in $p_T$. Events in the overflow bin are not shown in the distribution but are included in the $Z_{bi}$ calculation. In both cases the SUSY signal sample is the same, and the legend indicates the associated $WZ$ sample its network was built with. \label{fig:ewoverlays}}
\end{figure}

Although slicing requires that tail events are given very low weights, and therefore increases the range of weights applied to nodes in our ‘node-split’ networks, the shapes of the n.s.i network variable distributions are indeed robust under these weights. This suggests that for the network metrics considered node-splitting is robust under a wide range of node weights. Considering Figure~\ref{fig:ewoverlays}, which shows \degree{euc} and \HC{city} for either an inclusive $WZ$ sample or a $WZ$ sample sliced in \pt, we conclude that the tail behaviour of these metrics is a reliable feature of these distributions under the range of node weights introduced by slicing. In both cases an inclusive SUSY signal sample is used, and the small changes in the network distributions in the signal are caused by the small changes in the background.

While the results presented so far are promising, they do not provide conclusive evidence that node splitting invariance is robust enough to allow the MC generated node weighted network to provide an accurate comparison to an LHC event generated network. The range of node weights considered in the electroweakino analysis is very wide, and in the absence of a fully unweighted MC set to compare it to, we will appeal to theoretical arguments for the robustness of the chosen n.s.i. network metrics to artificial distortions based on the node weights. 

\FloatBarrier

\subsection{Theoretical Considerations}
\label{theotest}

\subsubsection{Accuracy of n.s.i. assumptions in this context}
\label{theotest1}

The node-splitting-invariance (n.s.i.) approach to defining complex network metrics on a node-weighted network was developed in Ref.~\cite{nsi}. In this paper, Heitzig et. al. use an axiomatic approach to derive alternative versions of common network metrics for node-weighted networks, where a node's weight is proportional to the size of the underlying domain of interest that the node represents. The n.s.i. approach requires that each node's value for a particular network metric will be unchanged whenever the operation of 'node-splitting' is performed on one or more nodes. Node-splitting means that any node $s$ with weight $w_s$ is split into two nodes $s'$ and $s''$ (weights $w_{s'}$ and $w_{s''}$), where $w_s = w_{s'} + w_{s''}$, $s'$ and $s''$ are linked to one another, and $s'$ and $s''$ are also linked to an identical set of neighbours (the same set as $s$ was linked to). If $G$ is the network which contained node $s$, and $G'$ is the refined network which contains $s'$ and $s''$, then all n.s.i. network metrics $\theta^*$ must satisfy 

\begin{equation}
    \theta^*(v \in G) = \theta^*(v \in G') \text{ for } v \ne s, s', s''
\end{equation}

and 

\begin{equation}
    \theta^* (s) = \theta^* (s') = \theta^* (s'')
\end{equation}

\begin{figure}[h!]
\centering
\includegraphics[scale=0.55]{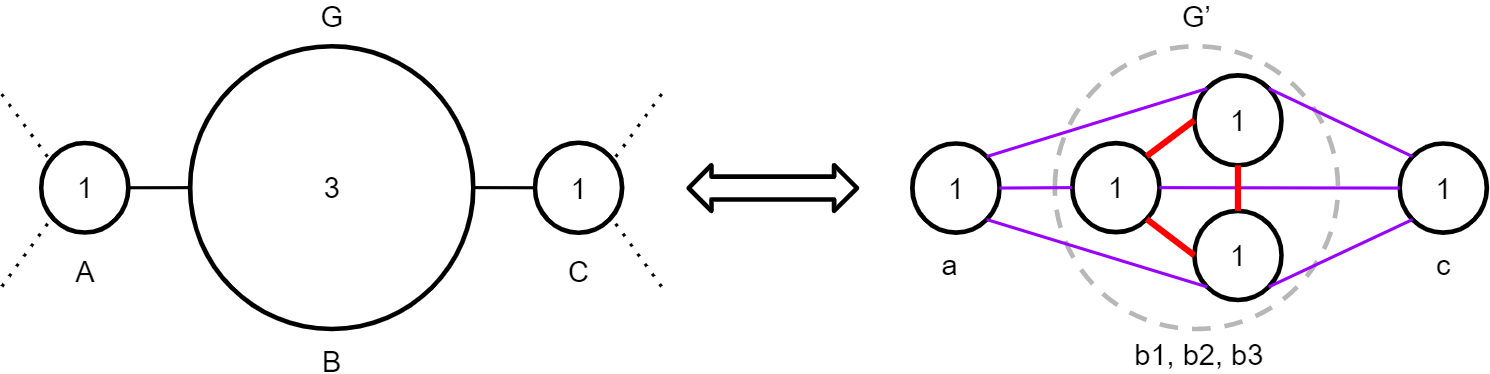} \hspace{10mm}
\caption{\label{fig:nodesplitting} The principle of node splitting invariance means that a node-weighted graph G is equivalent to a refined graph G’ where all nodes have been split into a number of unweighted nodes proportional to the weight. Here node B is split into b1, b2 and b3. The nodes b1, b2 and b3 are assumed to have i) full internal connectivity (red links) and ii) identical external connectivity (purple links).}
\end{figure}

Crucially, as illustrated in Figure \ref{fig:nodesplitting}, n.s.i. assumes that a weighted node can be taken to represent a group of identical nodes which possess i) full internal connectivity (shown in red) and ii) identical external connectivity (purple). In the context of a collision event network, we are concerned with how well a lower-resolution node-weighted network $G_{lr}$ built from a weighted sample of Monte Carlo events can be used to approximate a larger, higher-resolution network $G_{hr}$, built from an ideal larger unweighted sample of collision events. We would only expect n.s.i. metrics to provide a good approximation to the network metrics measured on the higher-resolution network if tight groups of almost identical nodes with i) almost full internal connectivity and ii) almost identical external connectivity are actually to be found in the higher-resolution network. A network containing differently sized clusters of tightly connected identical nodes possesses a highly specific topology, and so if the assumptions i) and ii) of the n.s.i. approach are not accurately reflective of the structure of a higher-resolution network, then applying n.s.i. network metrics may produce distortions in the values of certain local network metrics. This will be discussed further in the next section.

\begin{figure}[h!]
\centering
\includegraphics[scale=1]{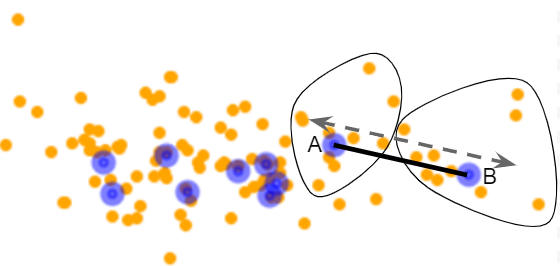} \hspace{10mm}
\caption{\label{fig:subsamples} Visualizing in 2-D the accuracy of representing a larger, higher-resolution sample (yellow dots, used to construct $G_{hr}$) by a smaller, weighted sample (blue circles, used to construct $G_{lr}$). Blue nodes A and B have circled around them approximately the 10 closest yellow nodes that would be assigned to them uniquely. The linking length (grey dotted line) is long enough to connect A and B, and long enough to connect any two points assigned to the same blue node, ensuring full internal connectivity. However, some of the yellow nodes assigned to A will not be linked to some of those assigned to B, therefore the assumption of identical external connectivity does not hold.}
\end{figure}

First, we briefly address the question of whether the n.s.i. assumptions of i) full internal connectivity and ii) identical external connectivity realistically apply in for the networks constructed in this paper. If a weighted MC generated node-weighted network $G_{lr}$ is broken down into an equivalent unweighted network via repeated application of the operation of node splitting, would the resultant $G_{lr}'$ provide an accurate approximation of $G_{hr}$, a higher-resolution network built from a larger, unweighted event sample? Since our networks are constructed using geometric proximity between events in a multidimensional kinematic space, the answer to this question will depend on the distribution of events in this kinematic space, and the chosen linking length. Figure  \ref{fig:subsamples} shows a 2-dimensional visualisation of how a smaller sample (10 blue circles) may be taken to represent a larger higher-resolution sample from the same distribution (100 yellow dots). If we were to build a node-weighted network from the blue circles, and compare this to an unweighted network built from the yellow dots, with the same linking length in both cases, would we find that each weighted node from a particular blue dot corresponds to a group of 10 fully internally connected and identically externally connected nodes in the higher-resolution network? If the linking length is long enough, then it is clear that the assumption of full internal connectivity will almost perfectly hold. In the electroweakino case study, for example, the networks constructed had link densities ranging from $0.213$ to $0.492$. With such a high link density it is extremely unlikely that a group of proximate nodes from $G_{hr}$ that are most accurately represented by the same weighted node from $G_{lr}$ would not all be linked to each other. 

However, the assumption of identical external connectivity is less safe. As illustrated in Figure  \ref{fig:subsamples}, two blue circles may fall within the linking length of each other, but only some of their nearby yellow dots fall within the linking length of each other. If the data is naturally tightly clustered, for example if the yellow dots all formed tight balls centred on the blue circles, then the assumption of identical external connectivity will almost perfectly hold. However without strong evidence that this is the case in the datasets considered, we choose to use only n.s.i. network metrics that are robust if the assumption of identical external connectivity does not hold. In the next section we provide theoretical justification for why the n.s.i. degree and n.s.i. closeness, harmonic closeness and exponential closeness centralities are robust under these conditions. 

\subsubsection{Justifying the choice of network metrics used}
\label{theotest2}

For a complex network constructed based on geometric proximity in a multidimensional kinematic space, we interpret the weight of a node as representing approximately the density of nodes in a nearby region. As discussed above, with a long enough linking length, which we argue is used in this analysis, all potential nodes that are represented by a single weighted node will be mutually connected (full internal connectivity). Being in close proximity the nodes will share many external neighbors, but their external connections will not be identical. Assuming identical external connectivity may, for some network metrics (e.g. betweenness centrality), cause a node-weight dependent bias in the network metric distributions, by treating larger-weighted nodes like unweighted nodes which have an unrealistically high number of identical twins. However, we argue that the n.s.i. degree and the three variants of the n.s.i. closeness centrality are safe to use in this context. The conclusions reached are in agreement with preliminary empirical tests we have performed, but further work is required to investigate under what conditions (e.g. network configuration, network size, distribution of node weights, link density) and to what extent imperfection in the n.s.i. approximation renders any of the network metrics unreliable. This is left as an important line of future research, and could lead to the future use of metrics we discount such as the betweenness centrality. 

Suppose we have some node $v \in G_{lr}$ which can be taken to represent most closely the nodes $v_1, ..., v_{w_v} \in G_{hr}$. If we assume that the set of neighbors of a weighted node $v$ in $G_{lr}$ is to a good approximation representative of the set of neighbors of $v_1, ..., v_w$ in $G_{hr}$, then the n.s.i. degree of $v$ will be approximately the average of the degree of all $v_1, ..., v_{w_v}$. The approximation will be less accurate if $v_1, ..., v_{w_v}$ have a sparse, non-symmetric or otherwise non-trivial structure, but this is primarily a problem in using a smaller, weighted Monte Carlo sample to approximate a finer-grained data set, and is not compounded by the n.s.i. approximation.

The n.s.i. closeness centrality of a node $u$ measures the average length $d_{ui}$ of shortest paths from $u$ to any other node in the network $i$. The three versions of closeness centrality differ in how they perform this average; closeness centrality sums the length of the shortest paths from $u$ to each other node in the network, and takes the inverse of this sum; harmonic closeness centrality sums the inverses of the distances of shortest paths from $u$ to every node in the network; exponential closeness centrality sums the inverses of $2^{d_{ui}}$. In all cases node weights along each path are multiplied together to approximate the number of shortest paths being traversed (see \cite{nsi} for precise formulae). Since all three of these closeness centralities take into account only the length of these shortest paths (not their uniqueness, as discussed below), we argue that they are not significantly distorted by the assumption of identical external connectivity. If we reconfigure an underlying higher-resolution network $G_{hr}$ to conform with n.s.i. assumptions, by removing some external links from some nodes in a group, and replacing them with external links identical to other external links in the group, this will not cause a significant change in the closeness centrality of any of the nodes in the cluster. So long as full internal connectivity holds, any of the new external links assigned to a node in order to satisfy external connectivity would have had a path length of 2, and now have this path length reduced to 1. Due to the fact that the network is constructed via geometric proximity, we expect that any of the broken links will likely still be connected by a path length of 2. Therefore, the assumption of identical external connectivity will only cause some minor changes in the average path lengths starting from a node, and these changes will to some extent cancel out. 

In contrast, an incorrect assumption of identical external connectivity may cause higher-weighted nodes to have a lower betweenness centrality. This is because unlike closeness centrality, betweenness depends also on the uniqueness of shortest paths passing through a node. In calculating betweenness centrality, the contribution of a shortest path from $a$ to $b$ which passes through $v$ is always scaled by the inverse of the total number of shortest paths that pass from $a$ to $b$. While nearby nodes in a higher-resolution network may have a diverse and distinct set of shortest paths passing through each of them, n.s.i. forces all nodes in the same group to lie along an identical set of shortest paths (see the identical external purple links for $b1, b2, b3$ in Figure \ref{fig:nodesplitting}). Thus n.s.i. betweenness centrality may be systematically lower for larger-weight nodes, if the assumption of identical external connectivity does not actually hold.

Finally, we note that improvements in the method of construction of our node-weighted network could lead to increased accuracy of the n.s.i. assumption of identical external connectivity. For example, we could generate a node-weighted network by taking a very large sample of Monte Carlo events, and clustering events together into a larger weighted event only where a tight cluster of events naturally exists. This would ensure that a weighted node accurately represents the density of a very local area of the kinematic space. Efficient algorithms such as k-means clustering could be used to generate a set of representative weighted nodes from a larger sample. Methods considered in \cite{Komiske:2019jim} to find the smallest representative set of events could also be applied. Building a more robust node-weighted network in this way may significantly increase the power of this analysis by permitting the use of important topological network measures omitted here, particular betweenness centrality and clustering coefficient.

\FloatBarrier

\bibliography{Network}
\bibliographystyle{JHEP}

%%%%%%%%%%%%%%%%%%%%%%%%%%%%%%%%%%%%%%%%%%%%%%%%%%%%%%%%%%%%%%%%%%%%%%%%
\end{document}